\documentclass[aps,prd,showpacs,twocolumn,floatfix,nofootinbib]{revtex4-1}
\usepackage{graphicx}
\usepackage{amsmath,amssymb,siunitx}

\begin{document}
\title{Dynamical mass ejection from black hole-neutron star binaries}
\author{
Koutarou Kyutoku$^{1,2}$,
Kunihito Ioka$^{3,4}$,
Hirotada Okawa$^{5,6}$,
Masaru Shibata$^5$,
and
Keisuke Taniguchi$^{7,8}$
}
\affiliation{
$^1$Department of Physics, University of Wisconsin-Milwaukee, P.O. Box
413, Milwaukee, Wisconsin 53201, USA\\
$^2$Interdisciplinary Theoretical Science (iTHES) Research Group, RIKEN,
Wako, Saitama 351-0198, Japan\\
$^3$Theory Center, Institute of Particles and Nuclear Studies, KEK,
Tsukuba 305-0801, Japan\\
$^4$Department of Particle and Nuclear Physics, the Graduate University
for Advanced Studies (Sokendai), Tsukuba 305-0801, Japan\\
$^5$Yukawa Institute for Theoretical Physics, Kyoto University, Kyoto
606-8502, Japan\\
$^6$Advanced Research Institute for Science and Engineering, Waseda
University, Okubo 3-4-1, Shinjuku, Tokyo, 169-8555, Japan\\
$^7$Graduate School of Arts and Sciences, University of Tokyo, Komaba,
Meguro, Tokyo 153-8902, Japan\\
$^8$Department of Physics, University of the Ryukyus, Nishihara, Okinawa
903-0213, Japan
}
\date{\today}

\begin{abstract}
 We investigate properties of material ejected dynamically in the merger
 of black hole-neutron star binaries by numerical-relativity
 simulations. We systematically study the dependence of ejecta
 properties on the mass ratio of the binary, spin of the black hole, and
 equation of state of the neutron-star matter. Dynamical mass ejection
 is driven primarily by tidal torque, and the ejecta is much more
 anisotropic than that from binary neutron star mergers. In particular,
 the dynamical ejecta is concentrated around the orbital plane with a
 half opening angle of \ang{10}--\ang{20} and often sweeps out only a
 half of the plane. The ejecta mass can be as large as $\sim 0.1
 M_\odot$, and the velocity is subrelativistic with $\sim 0.2$--$0.3c$
 for typical cases. The ratio of the ejecta mass to the bound mass (disk
 and fallback components) is larger, and the ejecta velocity is larger,
 for larger values of the binary mass ratio, i.e., for larger values of
 the black-hole mass. The remnant black hole-disk system receives a kick
 velocity of $O(100)$ \si{\km\per\second} due to the ejecta linear
 momentum, and this easily dominates the kick velocity due to
 gravitational radiation. Structures of postmerger material, velocity
 distribution of the dynamical ejecta, fallback rates, and gravitational
 waves are also investigated. We also discuss the effect of ejecta
 anisotropy on electromagnetic counterparts, specifically a
 macronova/kilonova and synchrotron radio emission, developing analytic
 models.
\end{abstract}

\maketitle

\section{Introduction} \label{sec:intro}

Coalescences of black hole-neutron star binaries are one of the most
promising gravitational-wave sources for ground-based
laser-interferometric detectors \cite{adligo2010,virgo2015,kagra2012},
along with those of binary neutron stars and binary black holes. The
sensitivity of these detectors will reach a sufficiently high level in
the coming years to detect gravitational waves from compact binary
coalescences more often than once a year
\cite{ligovirgo2010,dominik_bombfhbp2015}. The first direct detection of
gravitational waves must have a dramatic impact on fundamental
physics. Furthermore, gravitational waves from binaries involving
neutron stars will tell us neutron-star properties like the radius,
compactness, and tidal deformability. Knowledge of neutron-star
properties will allow us to constrain the equation of state of nuclear-
and supranuclear-density matter, and therefore gravitational waves will
also give us valuable information on nuclear physics.

Simultaneous detection of electromagnetic radiation from compact binary
mergers, i.e., electromagnetic counterparts to gravitational waves, is
eagerly desired \cite{metzger_berger2012,piran_nr2013}. It will support
gravitational-wave detection and enhance scientific returns from each
coalescence event. For example, source localization on the celestial
sphere is much more accurate with electromagnetic instruments than with
gravitational-wave detector networks
\cite{nissanke_kg2013}. Gravitational-wave data analysis benefits from
accurate localization by solving degeneracy between the sky location and
other amplitude parameters such as the luminosity distance. Accurate
localization of the source is also indispensable to find its host galaxy
and to determine the cosmological redshift. By combining this
information, the luminosity distance-redshift relation will be derived
without relying on the cosmic distance ladder,\footnote{See also
Ref.~\cite{messenger_read2012} for an alternative approach free from
electromagnetic observation.} and we will obtain a novel method to test
cosmological models \cite{schutz1986}. Besides, the effective
sensitivity of a gravitational-wave detector would be improved if we
could know the coalescence time and/or sky location of a binary from
electromagnetic counterparts \cite{kochanek_piran1993}.

Among the candidates of electromagnetic counterparts, a short-hard
gamma-ray burst and its afterglow are vigorously studied both
theoretically and observationally (see Refs.~\cite{nakar2007,berger2014}
for reviews). While whether compact binary coalescences can really drive
short-hard gamma-ray bursts is still an open question, future
simultaneous detection with gravitational-wave chirp signals will prove
this hypothesis. Prompt emission is so bright that it can be easily
detected by gamma-ray satellites within the horizon distance of
gravitational-wave detectors. Accurate localization is possible if an
associated afterglow is observed at longer wavelengths. Short-hard
gamma-ray bursts do not, however, always serve as counterparts to
gravitational waves because of their presumably jetlike geometry. If the
typical jet opening angle is $\lesssim \ang{10}$ as suggested by
jet-break observations \cite{fong_etal2012,fong_etal2014}, the fraction
of gravitational-wave events accompanied by observable short-hard
gamma-ray bursts will be a few percent at best.

In recent years, electromagnetic counterparts have been getting a lot
more attention, and many isotropic emission models are studied. Most of
the proposed models require the ejection of unbound material from the
binary,\footnote{Precursor emission may not require mass ejection
\cite{ioka_taniguchi2000,mcwilliams_levin2011,tsang_rhpb2012,paschalidis_es2013},
and we do not consider them in this study.} where examples include a
macronova/kilonova powered by decay heat of unstable \textit{r}-process
elements
\cite{li_paczynski1998,kulkarni2005,metzger_mdqaktnpz2010,kisaka_it2015}
and nonthermal radiation from electrons accelerated at blast waves
between the ejecta and interstellar medium
\cite{nakar_piran2011,takami_ki2014}. Possible emission from the ejecta
will be isotropic if the ejecta has spherical geometry and/or a
subrelativistic velocity. Such a ``$4\pi$-counterpart'' is ideal for
followup observations, because it will accompany a majority of
gravitational-wave events unlike beamed radiation of short-hard
gamma-ray bursts.

Despite its 40-years-old history in theoretical astrophysics
\cite{lattimer_schramm1974,lattimer_schramm1976}, mass ejection from the
compact binary merger is a young research topic in numerical
relativity. Most of the previous black hole-neutron star binary
simulations in numerical relativity were performed aiming at deriving
gravitational waves in the late inspiral and merger phases and at
clarifying properties of remnant accretion disks formed after the tidal
disruption of neutron stars (see Ref.~\cite{shibata_taniguchi2011} and
references therein for earlier work). Mass ejection has not been studied
in detail compared to these topics in full general relativity
\cite{kyutoku_ost2011,foucart_ddkmopsst2013,lovelace_dfkpss2013,deaton_dfookmss2013,foucart_etal2014},
whereas a substantial effort to clarify mass ejection has been made in
simulations performed in Newtonian gravity or approximate general
relativity \cite{rosswog2005,rosswog_pn2013,just_bagj2015} (see also
Refs.~\cite{lee_kluzniak1999-2,lee2000,lee2001,faber_bst2006,faber_bstr2006,rantsiou_klr2008,ruffert_janka2010}). It
is pointed out that dynamical ejecta from binary neutron star mergers
become less massive and more isotropic in full general relativity
\cite{hotokezaka_kkosst2013} or the conformal flatness approximation
\cite{bauswein_gj2013} than in Newtonian gravity
\cite{rosswog_ltdbp1999,rosswog_dtp2000,rosswog_pn2013}. The difference
due to realism of the gravity should be most pronounced when a black
hole is involved as already suggested by existing work. Thus, it is
natural that numerical relativity is vital to study mass ejection from
the black hole-neutron star binary merger.

In this study, we perform simulations of black hole-neutron star binary
mergers using numerical-relativity code {\small SACRA}
\cite{yamamoto_st2008} and investigate dynamical mass ejection extending
our preceding work \cite{kyutoku_is2013}. In particular, we focus on
kinematical properties of dynamical ejecta such as the mass and
velocity. Compared to our previous simulations \cite{kyutoku_ost2011},
we adopt large computational domains to track long-term evolution of the
ejecta. Because the dynamical ejecta has a velocity comparable (a few
tens of percent) to the speed of light as shown in this study, the large
computational domains are essential for the reliable estimation of
ejecta properties. We also improve the treatment of artificial
atmosphere (inevitable in conservative hydrodynamic schemes) from our
previous work
\cite{shibata_kyt2009,yamamoto_st2008,shibata_kyt2009e,kyutoku_st2010,kyutoku_st2010e,kyutoku_ost2011}
and confirm that characteristic quantities of dynamical ejecta such as
the mass and velocity depend only weakly on the atmosphere. We do not,
however, study disk winds expected to be driven by unincorporated
physics.

This paper is organized as follows. Section \ref{sec:setup} describes
our models of black hole-neutron star binaries including neutron-star
equations of state. Our numerical methods are also described, and
diagnostics of simulations are presented with particular emphasis on the
ejecta defined as unbound material. Numerical results of the simulations
are presented in Sec.~\ref{sec:res}. After briefly reviewing the merger
dynamics in Sec.~\ref{sec:res_review}, dynamical mass ejection processes
are described in Sec.~\ref{sec:res_ej}. The dependence of characteristic
quantities on binary parameters is discussed in Sec.~\ref{sec:res_char},
and the material structure is investigated in
Sec.~\ref{sec:res_struc}. We also study fallback material, remnant black
hole-disk systems, and gravitational waves in Secs.~\ref{sec:res_fb},
\ref{sec:res_diskbh}, and \ref{sec:res_gw}, respectively. Possible
electromagnetic counterparts from black hole-neutron star binaries are
discussed based on the results of simulations in
Sec.~\ref{sec:cp}. Specifically, Sec.~\ref{sec:cp_mk} describes the
macronova/kilonova and Sec.~\ref{sec:cp_sync} describes synchrotron
radio emission from accelerated electrons. Section \ref{sec:summary} is
devoted to a summary. Numerical values of characteristic ejecta
quantities derived by simulations are summarized in Table
\ref{table:result}. Readers interested primarily in electromagnetic
counterparts should read Sec.~\ref{sec:cp}, of which important results
are described in Ref.~\cite{kyutoku_is2013}.

Notational conventions are summarized as follows. Throughout this paper,
we adopt geometrical units in which $G=c=1$, where $G$ and $c$ are the
gravitational constant and speed of light, respectively. Exceptionally,
$c$ is sometimes inserted for clarity when we discuss the velocity of
the ejecta or fluid element. Greek and Latin indices denote the
spacetime and space components, respectively. The black-hole mass,
neutron-star gravitational mass, and neutron-star circumferential radius
in isolation are denoted by $M_\mathrm{BH}$, $M_\mathrm{NS}$, and
$R_\mathrm{NS}$, respectively. The dimensionless spin parameter of the
black hole,\footnote{In our previous work
\cite{kyutoku_st2010,kyutoku_st2010e,kyutoku_ost2011}, this parameter is
denoted as $a$. We change the convention, because $a$ is sometimes
reserved for the specific spin angular momentum, $S_\mathrm{BH} /
M_\mathrm{BH} = \chi M_\mathrm{BH}$.} total mass of the system at an
infinite separation, mass ratio, and compactness of the neutron star are
defined as $\chi \equiv S_\mathrm{BH} / M_\mathrm{BH}^2$, $m_0 \equiv
M_\mathrm {BH} + M_\mathrm{NS}$, $Q \equiv M_\mathrm{BH} /
M_\mathrm{NS}$, and $\mathcal{C} \equiv M_\mathrm{NS} / R_\mathrm{NS}$,
respectively, where $S_\mathrm{BH}$ is the black-hole spin angular
momentum.

\section{Numerical method} \label{sec:setup}

\subsection{Zero-temperature equation of state} \label{sec:setup_eos}

\begin{table*}
 \caption{Parameters and key ingredients of the adopted equations of
 state, where $P_2$ (the pressure at $\rho =
 \SI{e14.7}{\gram\per\cubic\cm}$) is shown in units of
 \si{dyne.cm^{-2}}. $M_\mathrm{max}$ is the maximum gravitational mass
 of a spherical neutron star for a given equation of state. $R_{1.35}$,
 $\rho_{1.35}$, $M_{*,1.35}$, $\mathcal{C}_{1.35}$, $k_{1.35}$, and
 $\Lambda_{1.35}$ are the circumferential radius, central rest-mass
 density, baryon rest mass, compactness, quadrupolar tidal Love number,
 and dimensionless quadrupolar tidal deformability of a $1.35 M_\odot$
 neutron star, respectively.}
 \begin{tabular}{c|cccc|ccccccc} \hline
  ~Model~ & ~$\log_{10} ( P_2 )$~ & ~$\Gamma_1$~ & ~$\Gamma_2$~ &
  ~$\Gamma_3$~ & ~$M_\mathrm{max} [M_\odot]$~ & ~$R_{1.35}$
  (\si{\kilo\meter})~ & $\rho_{1.35}$ (\si{\gram\per\cubic\cm})~ &
  ~$M_{*,1.35} [M_\odot]$~ & ~$\mathcal{C}_{1.35}$~ & ~$k_{1.35}$~ &
  ~$\Lambda_{1.35}$~ \\
  \hline \hline
  APR4 & 34.269 & 2.830 & 3.445 & 3.348 & 2.20 & 11.1 & \num{8.9e14} &
				  1.50 & 0.180 & 0.0908 & 323 \\
  ALF2 & 34.616 & 4.070 & 2.411 & 1.890 & 1.99 & 12.4 & \num{6.4e14} &
				  1.49 & 0.161 & 0.120 & 734 \\
  H4 & 34.669 & 2.909 & 2.246 & 2.144 & 2.03 & 13.6 & \num{5.5e14} &
				  1.47 & 0.147 & 0.115 & 1110 \\
  MS1 & 34.858 & 3.224 & 3.033 & 1.325 & 2.77 & 14.4 & \num{4.2e14} &
				  1.46 & 0.138 & 0.132 & 1740 \\
  \hline
 \end{tabular}
 \label{table:eos}
\end{table*}

We model equations of state of zero-temperature neutron-star matter by
piecewise polytropes \cite{read_lof2009}. Neutron stars in compact
binaries right before the merger will be cold enough to be modeled by
zero-temperature equations of state (see, e.g.,
Ref.~\cite{yakovlev_pethick2004}). However, the realistic equation of
state of neutron-star matter is not known precisely yet. Therefore, it
is necessary to adopt various equations of state systematically to span
a plausible range of neutron-star properties. Piecewise polytropes are
suitable for this purpose, because those with one and three pieces for
crust and core regions, respectively, are known to be able to
approximate nuclear-theory-based equations of state accurately with a
small number of parameters \cite{read_lof2009}. Following
Ref.~\cite{read_lof2009}, we employ piecewise polytropes of the form
\begin{equation}
 P ( \rho ) = \kappa_i \rho^{\Gamma_i} \; ( \rho_i \le \rho \le
  \rho_{i+1} ) ,
\end{equation}
where $P$ is the pressure and $\rho$ is the rest-mass density, with $i
\in \{0,1,2,3\}$ in this study. It is always assumed that $\rho_0 = 0$
and $\rho_4 \to \infty$. We fix parameters for the lowest-density, crust
region to be
\begin{align}
 \Gamma_0 & = \num{1.35692395} , \label{eq:crustgam} \\
 \kappa_0 & = \SI{3.99873692e-8}{(g~cm^{-3})}^{1-\Gamma_0} .
\end{align}
We further set $\rho_2 = \SI{e14.7}{\gram\per\cubic\cm} \approx
\SI{5.0e14}{\gram\per\cubic\cm}$ and $\rho_3 =
\SI{e15.0}{\gram\per\cubic\cm}$ to reduce the number of free
parameters. Requiring the continuity of $P (\rho)$, each piecewise
polytrope is characterized by four parameters. We choose the free
parameters to be the pressure at $\rho_2$, denoted by $P_2 = P ( \rho_2
)$, and adiabatic indices for the core, $\{ \Gamma_1 , \Gamma_2 ,
\Gamma_3\}$.

Table \ref{table:eos} lists parameters of piecewise polytropes adopted
in this study as well as neutron-star properties computed using
them. The naming convention and parameters follow
Ref.~\cite{read_lof2009}. APR4 \cite{akmal_pr1998} is computed by a
variational method incorporating three-nucleon interactions and
relativistic boost corrections. This equation of state gives the
smallest radius of a $1.35 M_\odot$ neutron star, $R_{1.35} =
\SI{11.1}{\kilo\meter}$, and thus APR4 is the softest equation of state
among those adopted in this study. Accordingly, tidal disruption is less
pronounced for neutron stars modeled with APR4 than those modeled with
the other equations of state. ALF2 \cite{alford_bpr2005} is a hybrid
equation of state obtained combining a nucleonic, APR-type equation of
state at low density and a quark-matter equation of state with quantum
chromodynamics corrections at high density. H4
\cite{glendenning_moszkowski1991,lackey_no2006} is computed by
relativistic mean-field theory incorporating hyperons with the stiffest
possible parameters (at the time). MS1 \cite{muller_serot1996} is also
derived by relativistic mean-field theory for nucleonic matter and gives
$R_{1.35} = \SI{14.4}{\kilo\meter}$, which is the largest value in this
study. Thus, MS1 is an extreme example with which tidal disruption
occurs most violently.

In practice, a very-high-density regime is not relevant to black
hole-neutron star binary coalescences as far as canonical-mass neutron
stars with $M_\mathrm{NS} \approx 1.35 M_\odot$
\cite{ozel_pnv2012,lattimer2012} are concerned. The reason for this is
that the maximum rest-mass density of the system, i.e., the central
density of the neutron star and maximum density in the remnant accretion
disk, is a decreasing function of time except for subdominant
oscillations. The rest-mass density at the center of an isolated $1.35
M_\odot$ neutron star, $\rho_{1.35}$, never exceeds $\rho_3$ for the
equations of state adopted in this study (see Table \ref{table:eos}),
and hence $\Gamma_3$ never plays a role in black hole-neutron star
binary coalescences. For MS1, even $\Gamma_2$ is irrelevant, because
$\rho_{1.35}$ is lower than $\rho_2$.\footnote{This means that
two-piecewise polytropes adopted in
Refs.~\cite{kyutoku_st2010,kyutoku_st2010e,kyutoku_ost2011} can fully
replace four-piecewise polytropes adopted here for modeling such a stiff
equation of state in simulations of black hole-neutron star binary
coalescences.} This situation is in stark contrast to that of binary
neutron star coalescences, which depend crucially on the high-density
regime of the equations of state.

In this study, we regard quantities associated with $1.35 M_\odot$
neutron stars as characteristic quantities of the equation of state
rather than the maximum mass $M_\mathrm{max}$, which is sensitive to the
behavior of matter at high density. Table \ref{table:eos} shows the
baryon rest mass $M_*$, compactness $\mathcal{C}$, quadrupolar tidal
Love number $k$ \cite{hinderer2008,hinderer2008e}, and dimensionless
quadrupolar tidal deformability $\Lambda \equiv (2/3) k
\mathcal{C}^{-5}$ of a $1.35 M_\odot$ neutron star, in addition to
$R_{1.35}$, $\rho_{1,35}$, and $M_\mathrm{max}$. Note that all the
equations of state can support $\sim 2 M_\odot$ neutron stars and
satisfy constraints from observations of massive pulsars
\cite{demorest_prrh2010,antoniadis_etal2013}, and thus they are possible
candidates of the realistic equation of state.

\subsection{Initial condition} \label{sec:setup_init}

We adopt quasiequilibrium states of black hole-neutron star binaries as
initial data of our simulations in the same manner as
Refs.~\cite{kyutoku_st2010,kyutoku_st2010e,kyutoku_ost2011}. Here, we
briefly describe the computational method of quasiequilibrium states,
and the details are found in
Refs.~\cite{kyutoku_st2009,kyutoku_ost2011}. Numerical computations are
performed using the multidomain spectral method library {\small LORENE}
\cite{LORENE}.

We solve a subset of the Einstein equation and the hydrostatic
equilibrium equations assuming the existence of helical
symmetry. Hamiltonian and momentum constraints are solved by a mixture
of the conformal transverse-traceless decomposition \cite{york1979} and
extended conformal thin-sandwich formulation
\cite{york1999,pfeiffer_york2003} imposing the spatial conformal
flatness, maximal slicing, and preservation of them in time. The
singularity associated with the black hole is handled in the puncture
framework \cite{brandt_brugmann1997}, and thus we obtain initial data of
the induced metric $\gamma_{ij}$ and extrinsic curvature $K_{ij}$
everywhere on the initial hypersurface (except for the exact location of
the puncture, with which simulation grids are chosen not to
coincide). The neutron-star matter is modeled by a perfect fluid
expressed by an energy-momentum tensor of the form
\begin{equation}
 T_{\mu \nu} = \rho h u_\mu u_\nu + P g_{\mu \nu} ,
\end{equation}
where $h \equiv 1 + \varepsilon + ( P / \rho )$ is the specific
enthalpy, $\varepsilon$ is the specific internal energy, and $u^\mu$ is
the fluid four-velocity. We further assume that the fluid is in a
zero-temperature and irrotational state during the computation of the
initial data, and hydrostatic equilibrium configurations are obtained by
solving the continuity equation and local energy-momentum conservation
equation \cite{bonazzola_gm1997,asada1998,teukolsky1998,shibata1998}.

Parameters characterizing a black hole-neutron star binary are specified
in initial data computations (see
Refs.~\cite{kyutoku_st2009,kyutoku_ost2011} for the details). For
simplicity, we always choose $M_\mathrm{NS}$ to be a typical mass of
observed binary neutron stars, $M_\mathrm{NS} = 1.35 M_\odot$, in this
study \cite{ozel_pnv2012,lattimer2012}. With this choice, the black-hole
mass, $M_\mathrm{BH}$, is uniquely determined by the mass ratio, $Q$,
which we regard as an independent parameter instead of $M_\mathrm{BH}$
itself. We only consider cases in which the spin angular momentum of the
black hole is zero or aligned with the orbital angular momentum of the
binary,\footnote{We will report results of cases in which the black-hole
spin angular momentum is inclined with respect to the orbital angular
momentum in a subsequent paper \cite{kawaguchi_knost2015}.} and thus the
spin is fully characterized by its dimensionless magnitude, $\chi$. The
orbital angular velocity of a binary $\Omega$ is determined by a
force-balance condition at the center of the neutron star for a given
orbital separation. We use a dimensionless orbital angular velocity $m_0
\Omega$ to characterize the initial data rather than the orbital
separation.

\subsection{Dynamical simulation} \label{sec:setup_sim}

Our numerical simulations are performed using an
adaptive-mesh-refinement code, {\small SACRA}
\cite{yamamoto_st2008}. The Einstein evolution equations are solved in a
Baumgarte--Shapiro--Shibata--Nakamura formulation
\cite{shibata_nakamura1995,baumgarte_shapiro1998}. We evolve the
conformal-factor variable $W$, conformal metric $\tilde{\gamma}_{ij}$,
conformal connection function $\tilde{\Gamma}^i$, extrinsic curvature
trace $K$, and conformally weighted traceless part of the extrinsic
curvature $\tilde{A}_{ij}$ defined by
\begin{align}
 W & \equiv \gamma^{-1/6} \; , \; \tilde{\gamma}_{ij} \equiv
 \gamma^{-1/3} \gamma_{ij} \; , \; \tilde{\Gamma}^i \equiv - \partial_j
 \tilde{\gamma}^{ij} , \\
 K & \equiv K_{ij} \gamma^{ij} \; , \; \tilde{A}_{ij} \equiv
 \gamma^{-1/3} \left( K_{ij} - \frac{1}{3} K \gamma_{ij}\right) ,
\end{align}
where Cartesian coordinates are adopted. The lapse function $\alpha$ and
shift vector $\beta^i$ are evolved by a moving puncture gauge condition
\cite{campanelli_lmz2006,baker_cckm2006} of the form
\begin{align}
 \left( \partial_t - \beta^j \partial_j \right) \alpha & = - 2 \alpha K
 , \\
 \left( \partial_t - \beta^j \partial_j \right) \beta^i & = \frac{3}{4}
 B^i , \\
 \left( \partial_t - \beta^j \partial_j \right) B^i & = \left(
 \partial_t - \beta^j \partial_j \right) \tilde{\Gamma}^i - \eta_s B^i ,
\end{align}
where $B^i$ is an auxiliary vectorial variable and $\eta_s$ is a free
parameter. Initial data of the lapse function are given by $\alpha = W$,
and the shift vector is initialized as $\beta^i = 0$ with $B^i = 0$. We
adopt $\eta_s \approx 1 / m_0$ in this study.

Hydrodynamic evolution equations are solved by a high-resolution
shock-capturing central scheme \cite{kurganov_tadmor2000} with
third-order piecewise parabolic reconstruction
\cite{colella_woodward1984}. We evolve the conserved rest-mass density
$\rho_*$, conserved momentum density $\rho_* \hat{u}_i$, and conserved
energy density $\rho_* \hat{e}$ defined via
\begin{equation}
 \rho_* \equiv \rho \alpha \sqrt{\gamma} u^t \; , \; \hat{u}_i \equiv h
  u_i \; , \; \hat{e} \equiv h \alpha u^t - \frac{P}{\rho \alpha u^t} .
\end{equation}
Equations of state adopted in dynamical simulations comprise cold and
thermal parts. The former is taken to be piecewise polytropes described
in Sec.~\ref{sec:setup_eos}, and the latter is given by an
ideal-gas-like form
\begin{equation}
 P_\mathrm{th} = ( \Gamma_\mathrm{th} - 1 ) \rho \varepsilon_\mathrm{th}
  ,
\end{equation}
where the thermal-part specific internal energy is defined by
$\varepsilon_\mathrm{th} ( \rho , \varepsilon ) \equiv \varepsilon -
\varepsilon_\mathrm{cold} ( \rho )$ with $\varepsilon_\mathrm{cold} (
\rho )$ the cold-part specific internal energy computed by piecewise
polytropes. Total pressure is given by $P = P_\mathrm{cold} ( \rho ) +
P_\mathrm{th} ( \rho , \varepsilon )$, where $P_\mathrm{cold} ( \rho )$
is computed by piecewise polytropes. We choose a fiducial value of
$\Gamma_\mathrm{th}$ to be 1.8 following
Ref.~\cite{hotokezaka_kkosst2013} (see also Ref.~\cite{bauswein_jo2010})
and also adopt 1.6 and 2.0 for selected models (see Appendix
\ref{app:err_gamth}). Note that these values are larger than that
adopted in our previous work
\cite{kyutoku_st2010,kyutoku_st2010e,kyutoku_ost2011}, in which
$\Gamma_\mathrm{th}$ is always chosen to be $\Gamma_0$ [see
Eq.~\eqref{eq:crustgam}].

In our simulations, all the postmerger material is governed effectively
by the same sub-nuclear-density equation of state irrespective of the
adopted piecewise polytrope. Specifically, when the rest-mass density
falls below $\rho_1$, the equation of state is given by the sum of the
crust polytrope and thermal correction,
\begin{equation}
 P ( \rho < \rho_1 , \varepsilon ) = \kappa_0 \rho^{\Gamma_0} + (
  \Gamma_\mathrm{th} - 1 ) \rho \varepsilon_\mathrm{th} .
\end{equation}
The values of $\rho_1$ are computed as $( \kappa_0 \rho_2^{\Gamma_1} /
P_2 )^{1 / ( \Gamma_1 - \Gamma_0 )}$ for each piecewise polytrope (see
Sec.~\ref{sec:setup_eos}) and take
0.9--\SI{2e14}{\gram\per\cubic\cm}. The rest-mass density never exceeds
these values after tidal disruption of neutron stars.

An artificial atmosphere has to be set carefully to study mass ejection
accurately. According to Ref.~\cite{hotokezaka_kkosst2013}, we put an
atmospheric density floor of the form
\begin{equation}
 \rho_{*,\mathrm{at}} = f_\mathrm{at} \rho_{*,0} \; \mathrm{min} \left[
								  1 ,
								  \left(
								   \frac{R_\mathrm{crit}}{r}
								  \right)^{n_\mathrm{at}}
								 \right]
 , \label{eq:atmosphere}
\end{equation}
where $\rho_{*,0}$ is the maximum (conserved) rest-mass density of the
initial configuration (see Ref.~\cite{yamamoto_st2008} for our previous
treatment). We typically choose $f_\mathrm{at} = \num{e-12}$ and
$n_\mathrm{at} = 3$ and vary them for selected models (see Appendix
\ref{app:err_atm}). The critical radius $R_\mathrm{crit}$ is chosen to
be $L / 16$, where $L$ is the size of the computational domain on one
side (see below). The atmospheric velocity is set to be zero, and the
atmospheric pressure is given by zero-temperature equations of state.

The grid structure of {\small SACRA} is summarized as
follows. Computational domains are composed of nested equidistant
Cartesian grids, and each grid has $(2N+1, 2N+1, N+1)$ points in
$(x,y,z)$ directions. The equatorial symmetry is imposed on the $z=0$
plane. We adopt $N=60$ as a fiducial value, with which the neutron-star
radius is covered by $\gtrsim 50$ points in the finest grid. We also
perform simulations with $N=40$ and 48 for selected models to check the
convergence of ejecta properties (see Appendix \ref{app:err_resol}). The
outer boundary is a cuboid covering $(x,y,z) \in [-L:L] \times [-L:L]
\times [0:L]$, and outgoing-wave boundary conditions are imposed except
for the $z=0$ plane. As for the adaptive-mesh-refinement grid structure,
we prepare $l_c$ coarser nonmoving grids and $l_f$ finer moving
grids. Namely, we have $l_c + 2 l_f$ computational grids spanning $l_c +
l_f$ refinement levels, which we always choose to be 9 in this
study. The nonmoving grids are fixed around an approximate center of
mass throughout the simulation. One set of the moving grids follows the
black hole, and the other set follows the neutron star. Starting from
the coarsest level as $l=0$, the $l$th level has a grid spacing $\Delta
x_l \equiv L / ( 2^l N )$, and we specifically denote the grid spacing
at the finest level by $\Delta x \equiv L / ( 2^{l_c + l_f - 1} N
)$. Finally, time steps of all the moving grids are chosen by setting
the Courant--Friedrichs--Lewy factor to be 0.5, and those of the
nonmoving grids are chosen to agree with that of the $l_c$th level
(i.e., the coarsest moving grid). In other words, the
Courant--Friedrichs--Lewy factor is given by $0.5 / 2^{l_c-l}$ in the
nonmoving grids.

\subsection{Binary model and grid setting} \label{sec:setup_model}

\begin{table*}
 \caption{Key parameters of initial data and grid structures of
 simulations for models adopted in this study. The names of models
 represent the equation of state (EOS), mass ratio ($Q$), and
 dimensionless spin parameter of the black hole ($\chi$). $m_0
 \Omega_0$, $M_0$, and $J_0$ are the dimensionless initial orbital
 angular velocity, Arnowitt--Deser--Misner mass, and orbital angular
 momentum of the system, respectively. As for grid configurations, $l_c$
 and $l_f$ are the numbers of coarser nonmoving grids and of a half of
 finer moving grids, respectively. The grid spacing at the finest level
 for $N=60$ (fiducial resolution) is shown in physical units as well as
 a value normalized by $m_0 \approx 2(Q+1)$ km. The grid number assigned
 to the semimajor diameter of the neutron star is given by
 $R_\mathrm{diam} / \Delta x$ for the direction along the binary
 separation. The box size $L$ is shown in physical units as well as a
 value normalized by the initial gravitational wavelength $\lambda_0 =
 \Omega_0 / \pi$.}
 \begin{tabular}{c|ccc|ccc|ccccccc} \hline
  ~Model~ & ~EOS~ & ~$Q$~ & ~$\chi$~ & ~$m_0 \Omega_0$~ & ~$M_0
  [M_\odot]$~ & ~$J_0 [M_\odot^2]$~ & ~$l_c$~ & ~$l_f$~ & ~$\Delta x$
  (\si{\meter})~ & ~$\Delta x / m_0$~ & ~$R_\mathrm{diam} / \Delta x$~ &
  ~$L$ (\si{\km})~ & ~$L / \lambda_0$~ \\
  \hline \hline
  APR4-Q3a75 & APR4 & 3 & 0.75 & 0.036 & 5.35 & 18.74 & 5 & 4 & 162 &
					  0.0203 & 102 & 2486 & 3.6 \\
  ALF2-Q3a75 & ALF2 & 3 & 0.75 & 0.036 & 5.35 & 18.74 & 5 & 4 & 186 &
					  0.0233 & 102 & 2858 & 4.1 \\
  H4-Q3a75 & H4 & 3 & 0.75 & 0.036 & 5.35 & 18.74 & 5 & 4 & 209 & 0.0263
					  & 102 & 3215 & 4.6 \\
  MS1-Q3a75 & MS1 & 3 & 0.75 & 0.036 & 5.35 & 18.74 & 5 & 4 & 228 &
					  0.0286 & 101 & 3501 & 5.0 \\
  \hline
  APR4-Q3a5 & APR4 & 3 & 0.5 & 0.036 & 5.35 & 19.15 & 5 & 4 & 162 &
					  0.0203 & 102 & 2486 & 3.6 \\
  ALF2-Q3a5 & ALF2 & 3 & 0.5 & 0.036 & 5.35 & 19.15 & 5 & 4 & 186 &
					  0.0233 & 102 & 2858 & 4.1 \\
  H4-Q3a5 & H4 & 3 & 0.5 & 0.036 & 5.35 & 19.15 & 5 & 4 & 209 & 0.0263 &
					      102 & 3215 & 4.6 \\
  MS1-Q3a5 & MS1 & 3 & 0.5 & 0.036 & 5.35 & 19.15 & 5 & 4 & 228 & 0.0286
					  & 101 & 3501 & 5.0 \\
  \hline
  APR4-Q3a0 & APR4 & 3 & 0 & 0.036 & 5.35 & 19.98 & 5 & 4 & 162 & 0.0203
					  & 102 & 2486 & 3.6 \\
  ALF2-Q3a0 & ALF2 & 3 & 0 & 0.036 & 5.35 & 19.98 & 5 & 4 & 186 & 0.0233
					  & 102 & 2858 & 4.1 \\
  H4-Q3a0 & H4 & 3 & 0 & 0.036 & 5.35 & 19.98 & 5 & 4 & 209 & 0.0263 &
					      102 & 3215 & 4.6 \\
  MS1-Q3a0 & MS1 & 3 & 0 & 0.036 & 5.35 & 19.98 & 5 & 4 & 228 & 0.0286 &
					      101 & 3501 & 5.0 \\
  \hline
  APR4-Q5a75 & APR4 & 5 & 0.75 & 0.040 & 8.04 & 30.13 & 4 & 5 & 158 &
					  0.0132 & 102 & 2429 & 2.6 \\
  ALF2-Q5a75 & ALF2 & 5 & 0.75 & 0.040 & 8.04 & 30.13 & 4 & 5 & 181 &
					  0.0152 & 102 & 2786 & 3.0 \\
  H4-Q5a75 & H4 & 5 & 0.75 & 0.040 & 8.04 & 30.13 & 4 & 5 & 205 & 0.0171
					  & 101 & 3144 & 3.3 \\
  MS1-Q5a75 & MS1 & 5 & 0.75 & 0.040 & 8.04 & 30.13 & 4 & 5 & 219 &
					  0.0183 & 102 & 3358 & 3.6 \\
  \hline
  APR4-Q5a5 & APR4 & 5 & 0.5 & 0.040 & 8.04 & 30.99 & 4 & 5 & 158 &
					  0.0132 & 102 & 2429 & 2.6 \\
  ALF2-Q5a5 & ALF2 & 5 & 0.5 & 0.040 & 8.04 & 30.99 & 4 & 5 & 181 &
					  0.0152 & 102 & 2786 & 3.0 \\
  H4-Q5a5 & H4 & 5 & 0.5 & 0.040 & 8.04 & 30.99 & 4 & 5 & 205 & 0.0171 &
					      101 & 3144 & 3.3 \\
  MS1-Q5a5 & MS1 & 5 & 0.5 & 0.040 & 8.04 & 30.99 & 4 & 5 & 219 & 0.0183
					  & 102 & 3358 & 3.6 \\
  \hline
  APR4-Q7a75 & APR4 & 7 & 0.75 & 0.044 & 10.73 & 40.96 & 4 & 5 & 153 &
					  0.0096 & 103 & 2358 & 2.1 \\
  ALF2-Q7a75 & ALF2 & 7 & 0.75 & 0.044 & 10.73 & 40.96 & 4 & 5 & 179 &
					  0.0112 & 101 & 2751 & 2.4 \\
  H4-Q7a75 & H4 & 7 & 0.75 & 0.044 & 10.73 & 40.96 & 4 & 5 & 200 &
					  0.0125 & 102 & 3072 & 2.7 \\
  MS1-Q7a75 & MS1 & 7 & 0.75 & 0.044 & 10.73 & 40.96 & 4 & 5 & 215 &
					  0.0135 & 102 & 3301 & 2.9 \\
  \hline
  APR4-Q7a5 & APR4 & 7 & 0.5 & 0.044 & 10.73 & 42.35 & 4 & 5 & 154 &
					  0.0097 & 102 & 2372 & 2.1 \\
  ALF2-Q7a5 & ALF2 & 7 & 0.5 & 0.044 & 10.73 & 42.35 & 4 & 5 & 179 &
					  0.0112 & 102 & 2743 & 2.4 \\
  H4-Q7a5 & H4 & 7 & 0.5 & 0.044 & 10.74 & 42.35 & 4 & 5 & 201 & 0.0126
					  & 101 & 3086 & 2.7 \\
  MS1-Q7a5 & MS1 & 7 & 0.5 & 0.044 & 10.74 & 42.35 & 4 & 5 & 217 &
					  0.0136 & 101 & 3329 & 2.9 \\
  \hline
 \end{tabular}
 \label{table:model}
\end{table*}

Table \ref{table:model} lists black hole-neutron star binary models
considered in this study. We name each model after the equation of
state, mass ratio, and black-hole spin. For example, APR4-Q3a75 is a
binary modeled with the APR4 equation of state, $Q=3$, and
$\chi=0.75$. Recall that $M_\mathrm{NS} = 1.35 M_\odot$ for all the
models. Table \ref{table:model} also presents the dimensionless initial
orbital angular velocity $m_0 \Omega_0$, Arnowitt--Deser--Misner mass
$M_0$, and orbital angular momentum of the system $J_0$. Here, $J_0$ is
defined from an Arnowitt--Deser--Misner type integral by subtracting the
spin angular momentum associated with the puncture.

We take the mass ratio, $Q$, from $\{3,5,7\}$, and the dimensionless
spin parameter, $\chi$, from $\{0.75,0.5,0\}$, where the spins are
always prograde, i.e., parallel to the orbital angular
momentum. Currently, neither the typical mass nor the typical spin of
stellar-mass black holes is known from observations. Thus, we perform
simulations systematically adopting various values of them along with
equations of state to predict possible outcomes of binary mergers. Here,
$Q=3,5$, and 7 correspond to $M_\mathrm{BH} = 4.05 M_\odot$, $6.75
M_\odot$, and $9.45 M_\odot$, respectively. The low-mass black hole with
$\approx 4 M_\odot$ is consistent with an observation of a black hole-Be
star binary \cite{casares_nrrphs2014}, which could evolve into a black
hole-neutron star binary, whereas the existence of a mass gap around
3--$5 M_\odot$ is frequently debated
\cite{ozel_pnm2010,kreidberg_bfk2012}. The middle-mass, $\approx 7
M_\odot$, and massive, $\approx 10 M_\odot$, black holes are safely
expected to exist from observations of x-ray binaries
\cite{ozel_pnm2010,kreidberg_bfk2012}. The spin parameter is even less
constrained than the mass is \cite{mcclintock_ns2014}, and we simply
take various values within our computational capabilities (see
Ref.~\cite{lovelace_dfkpss2013} for simulations of a near-extremal black
hole-neutron star binary). We pay, however, less attention to high-mass
and low-spin black holes. This is because such black holes are not able
to disrupt companion neutron stars before they reach the innermost
stable circular orbit \cite{shibata_taniguchi2011}, and thus the merger
process is essentially the same as that of binary black holes
\cite{foucart_bdgkmmpss2013}. We also do not pay attention to retrograde
spins, i.e., antiparallel to the orbital angular momentum, irrespective
of the value of $Q$ due to the same reason.

Table \ref{table:model} also shows the adaptive-mesh-refinement grid
structure for each simulation. We always choose $(l_c , l_f) = (5,4)$
for $Q=3$ and $(4,5)$ for $Q=5$ and 7. In all the cases, the
hydrodynamic evolution equations are solved only within $L/2 \approx
\SI{1500}{\km}$ for one side. Because it turns out later that a typical
velocity of dynamical ejecta is 0.2--$0.3c$, the ejecta motion can be
safely tracked over $\sim \SI{10}{\ms}$. At the same time, the box size
is larger than the initial gravitational wavelength, and thus
outgoing-wave boundary conditions are appropriate there as far as the
gravitational wavelength is covered by $\gtrsim 10$ grid points.

\subsection{Diagnostics} \label{sec:setup_diag}

\subsubsection{Ejecta} \label{sec:setup_diag_ej}

We analyze global ejecta properties by integrals over unbound material
\cite{hotokezaka_kkosst2013}. We define the ejecta to be unbound
material identified by a criterion $u_t < -1$, which becomes correct for
a particle moving along its geodesics in a stationary spacetime. Because
we are handling a fluid in a dynamical spacetime, this criterion is only
approximate and becomes especially poor in the vicinity of remnant black
hole-disk systems. Our computational domains always extend to $\gtrsim$
\SI{1000}{\km}, where the gravitational potential in geometrical units
is $\lesssim 0.01$--0.02, and thus we expect that typical errors
associated with this approximate criterion are a few percent. Strictly
speaking, $h u_t$ rather than $u_t$ is a conserved quantity associated
with a fluid in a stationary spacetime. We check that the results depend
only weakly on the choice of criteria, because shock heating does not
play an important role in dynamical mass ejection from black
hole-neutron star binaries (see Sec.~\ref{sec:res_ej}). In consideration
of the fact that our current simulations do not incorporate any process
other than shocks responsible for heating and cooling such as neutrino
interaction, we decide to neglect thermal effects for the purpose of
classification. Because $h \ge 1$ by definition, our estimates should be
regarded as conservative. In addition, this allows us to compare our
results directly with those of existing studies in numerical relativity
(e.g., Refs.~\cite{hotokezaka_kkosst2013,foucart_etal2014}).

The rest mass outside the apparent horizon including both bound and
unbound portions is computed by the integral
\begin{equation}
 M_{r>r_\mathrm{AH}} \equiv \int_{r>r_\mathrm{AH}} \rho_* d^3 x ,
\end{equation}
where $r_\mathrm{AH}$ is the angle-dependent coordinate radius of the
apparent horizon. The ejecta mass is defined by an unbound portion of
the rest mass as
\begin{equation}
 M_\mathrm{ej} \equiv \int_{r>r_\mathrm{AH},u_t < -1} \rho_* d^3 x
  . \label{eq:unbound}
\end{equation}
We also define the bound mass by
\begin{align}
 M_\mathrm{bd} & \equiv \int_{r>r_\mathrm{AH},u_t \ge -1} \rho_* d^3 x
 \notag \\
 & = M_{r>r_\mathrm{AH}} - M_\mathrm{ej} , \label{eq:bound}
\end{align}
which may be composed of the remnant disk and fallback material. We do
not, however, rigorously distinguish these two components due to the
absence of reasonable criteria.

The kinetic energy of ejecta $T_\mathrm{ej}$ is defined following
Ref.~\cite{hotokezaka_kkosst2013}. First, the total energy of the ejecta
is defined by
\begin{equation}
 E_\mathrm{ej} \equiv \int_{r>r_\mathrm{AH},u_t < -1} \rho_* \hat{e} d^3
  x ,
\end{equation}
whereas the gravitational binding energy is not (and cannot be in
general relativity) appropriately subtracted. Next, the internal energy
of the ejecta is defined by
\begin{equation}
 U_\mathrm{ej} \equiv \int_{r>r_\mathrm{AH},u_t < -1} \rho_* \varepsilon
  d^3 x .
\end{equation}
Finally, the kinetic energy of the ejecta may be defined by subtracting
the rest mass and internal energy from the total energy as
\begin{equation}
 T_\mathrm{ej} \equiv E_\mathrm{ej} - M_\mathrm{ej} - U_\mathrm{ej} .
\end{equation}
Although the internal energy is likely to be converted to the kinetic
energy in the long run, we do not count $U_\mathrm{ej}$ as a part of
$T_\mathrm{ej}$ in this study. This does not affect the results, because
$U_\mathrm{ej}$ is smaller than $T_\mathrm{ej}$ by orders of
magnitude. Using the mass and kinetic energy of the ejecta, we may also
define their average velocity as
\begin{equation}
 v_\mathrm{ave} \equiv \sqrt{\frac{2 T_\mathrm{ej}}{M_\mathrm{ej}}}
  \label{eq:vave}
\end{equation}
using the Newtonian relation. It should be cautioned that the kinetic
energy and average velocity defined in this manner are not calculated
taking the gravitational binding energy associated with remnant black
hole-disk systems into account. This implies that these measures
overestimate asymptotic values when evaluated in the vicinity of black
hole-disk systems independently of the validity of $u_t < -1$ and that
they are reliable only for distant regions. For this reason, we
typically measure the quantities of the ejecta at \SI{10}{\ms} after the
onset of merger, when the dominant portion of the ejecta leaves the
central region but still resides in our computational domains.

We also compute the linear momentum of ejecta, which indicates the
degree of ejecta anisotropy. Components of the linear momentum of the
ejecta may be defined by
\begin{equation}
 P_{\mathrm{ej},i} \equiv \int_{r>r_\mathrm{AH},u_t < -1} \rho_*
  \hat{u}_i d^3 x ,
\end{equation}
where the $z$ component vanishes identically due to the equatorial
symmetry in this study. The magnitude of the linear momentum is given by
\begin{equation}
 P_\mathrm{ej} = \sqrt{\sum_i ( P_{\mathrm{ej},i}^2 )} \; ,
  \label{eq:ejlinmag}
\end{equation}
and the center-of-mass velocity of the ejecta may be defined by
\begin{equation}
 v_\mathrm{ej} \equiv \frac{P_\mathrm{ej}}{M_\mathrm{ej}} ,
\end{equation}
which we call the bulk velocity in this paper. When the system is
symmetric with respect to the equatorial plane, the bulk velocity
vanishes if (but not only if) the ejecta is axisymmetric. A relation
$v_\mathrm{ej} \le v_\mathrm{ave}$ always holds. If the ejecta is
modeled by an axisymmetric outflow truncated at an opening angle
$\varphi_\mathrm{ej}$, we have $v_\mathrm{ej} = v_\mathrm{ave} \sin (
\varphi_\mathrm{ej} / 2) / ( \varphi_\mathrm{ej} / 2)$. These measures
suffer from the gravitational binding energy in the vicinity of black
hole-disk systems as $T_\mathrm{ej}$ and $v_\mathrm{ave}$ do. Thus, they
should also be estimated at a distant region. The propagation direction
of the ejecta with respect to our coordinate system may be characterized
by an angle defined from the linear momentum,
\begin{equation}
 \Phi_\mathrm{ej} \equiv \arctan \left(
				  \frac{P_{\mathrm{ej},y}}{P_{\mathrm{ej},x}}
				 \right) . \label{eq:ejlinang}
\end{equation}

In addition to these integral quantities, the mass spectrum with respect
to the asymptotic velocity, or simply the velocity distribution of the
ejecta, is estimated. The asymptotic velocity of each fluid element is
defined from an asymptotic Lorentz factor $- u_t$ as
\begin{equation}
 v \equiv \sqrt{1 - \frac{1}{( - u_t )^2}} \; .
\end{equation}
Here, we use $- u_t$ instead of the Lorentz factor seen from the
Eulerian observer, $\alpha u^t$, because the latter predicts the lower
end of ejecta velocity to be the local escape velocity rather than
zero. To derive the velocity distribution, we only analyze unbound
material on the equatorial plane and rescale the total mass to
$M_\mathrm{ej}$ measured over the full region by
Eq.~\eqref{eq:unbound}. To compensate this geometrical restriction, the
mass of each fluid element is weighted by the distance from the
coordinate origin, $r$, when computing the total mass of unbound
material on the equatorial plane. This procedure is acceptable for black
hole-neutron star binary mergers, because material ejected dynamically
from neutron stars is concentrated around the equatorial plane (see
Sec.~\ref{sec:res_ej}).

\subsubsection{Fallback material} \label{sec:setup_diag_fb}

We estimate fallback rates of bound material based on Newtonian
relations \cite{rosswog2007}. The motion of the bound material is
assumed to follow a ballistic trajectory determined by the energy and
angular momentum of each fluid element. For this purpose, we only
analyze bound material on the equatorial plane and rescale the total
mass to $M_\mathrm{bd}$ measured over the full region by
Eq.~\eqref{eq:bound} in a similar manner to the computation of the
velocity distribution of the ejecta.

A fluid element on each grid point of the second-largest ($l=1$) domain,
which is the largest domain where the hydrodynamic evolution equations
are solved, is identified as an isolated test particle with the mass
$\rho ( \Delta x_1 )^3$ neglecting the spacetime curvature at a selected
time slice. The specific energy (excluding the rest mass) $\tilde{E}$
and specific angular momentum $\tilde{J}$ of the particle are estimated
to be
\begin{equation}
 \tilde{E} = - u_t - 1 \; , \; \tilde{J} = u_\varphi ,
\end{equation}
where we only consider bound material identified by $u_t \ge -1$ and
therefore $\tilde{E} \le 0$. We neglect $h-1$ in the same manner as the
classification of the bound and unbound material. The azimuthal
velocity, $u_\varphi$, is defined from Cartesian components by
transformation with respect to the coordinate origin, which does not
correspond exactly to the black-hole position nor center of mass (see
the discussions in Sec.~\ref{sec:res_fb}). Assuming the presence of a
central mass $M_c$, the semimajor axis and eccentricity of the orbit are
given by
\begin{equation}
 a_\mathrm{fb} = - \frac{M_c}{2 \tilde{E}} \; , \; e_\mathrm{fb} =
  \sqrt{1 + \frac{2 \tilde{E} \tilde{J}^2}{M_c^2}}
\end{equation}
in Newtonian gravity. Accordingly, the periapsis and apoapsis distances
are given by $r_p = a_\mathrm{fb} ( 1 - e_\mathrm{fb} )$ and $r_a =
a_\mathrm{fb} ( 1 + e_\mathrm{fb} )$, respectively.

We define the fallback time of each particle to be the duration to reach
the periapsis. The particle is assumed to obey the Newtonian equation of
motion,
\begin{equation}
 \frac{dr}{dt} = \frac{u_r}{|u_r|} \sqrt{2 \tilde{E} + \frac{2 M_c}{r} -
  \frac{\tilde{J}^2}{r^2}} \; ,
\end{equation}
regarding $u_r$ as the radial velocity. This equation can be integrated
analytically to give the fallback time for a particle at $r=r_i$ as
\begin{equation}
 t_\mathrm{fb} = \frac{P_\mathrm{fb}}{2} + \frac{u_r}{|u_r|} [ I ( r_a )
  - I ( r_i ) ] ,
\end{equation}
where $P_\mathrm{fb} \equiv 2 \pi \sqrt{a_\mathrm{fb}^3 / M_c}$ is the
orbital period and
\begin{align}
 I ( r ) & \equiv \frac{\sqrt{2 \tilde{E} r^2 + 2 M_c r -
 \tilde{J}^2}}{2 \tilde{E}} \notag \\
 & - \frac{M_c}{\sqrt{-8 \tilde{E}^3}} \arcsin \left( \frac{2 \tilde{E}
 r + M_c}{M_c e_\mathrm{fb}} \right) .
\end{align}
Specifically, $I ( r_a )$ is $P_\mathrm{fb}/4$. It would be useful to
recall that the orbital period is given by
\begin{equation}
 P_\mathrm{fb} = \SI{5.5}{\ms} \left(
				\frac{a_\mathrm{fb}}{\SI{100}{\km}}
			       \right)^{3/2} \left( \frac{M_c}{10
			       M_\odot} \right)^{-1/2} .
\end{equation}
For a particle with $e_\mathrm{fb} = 0$, which appears in the central
region, we simply set $t_\mathrm{fb} = P_\mathrm{fb} / 2$. Physically,
components with $e_\mathrm{fb} \approx 0$ should be regarded as the
accretion disk rather than fallback material, while we do not have
quantitative criteria to distinguish them. Such a particle does not
contribute in any way to the long-term fallback rate due to its short
orbital period. We apply the same remedy for a particle that happens to
satisfy $e_\mathrm{fb}^2 < 0$ and/or $r_i < r_p$ due to numerical
errors, approximate identification of the azimuthal velocity, or abuse
of Newtonian relations. In this study, $M_c$ is always approximated by
$m_0$ ignoring the energy loss due to gravitational waves and existence
of the mass outside the black hole, $M_{r>r_\mathrm{AH}}$. We checked
that the results depend only weakly on the precise value of $M_c$.

Finally, the fallback rate is computed by dividing the material into
small segments according to the fallback time as
\begin{equation}
 \dot{M}_\mathrm{fb} (t) \equiv \frac{\Delta M (t)}{\Delta t (t)} ,
\end{equation}
where $\Delta M (t)$ is the mass of the fluid elements satisfying $t \le
t_\mathrm{fb} < t + \Delta t$ and $\Delta t (t)$ is arbitrarily chosen
to be $\approx t/10$. When we evaluate $\Delta M (t)$, the mass of each
fluid element is weighted by $r$ in the same way as done in the
computation of the velocity distribution of the ejecta. It should be
cautioned that, however, $\dot{M}_\mathrm{fb}$ does not necessarily
correspond to the black-hole accretion rate nor electromagnetic
luminosity, because a part of the fallback material may be blown off
from the disk as a wind or envelope \cite{rossi_begelman2009}. We do not
discuss the fate of the fallback material in this study.

\subsubsection{Black hole} \label{sec:setup_diag_bh}

Properties of remnant black holes are estimated by integrals on apparent
horizons as in our previous work
\cite{shibata_kyt2009,shibata_kyt2009e,kyutoku_st2010,kyutoku_st2010e,kyutoku_ost2011}. Assuming
that the spacetime is approximately stationary, the black-hole mass is
estimated by
\begin{equation}
 M_\mathrm{BH,f} = \frac{C_e}{4 \pi} ,
\end{equation}
where $C_e$ is the equatorial circumferential radius of the apparent
horizon. The spin parameter of the remnant black hole $\chi_\mathrm{f}$
is estimated via the relation of Kerr black holes,
\begin{equation}
 \frac{C_p}{C_e} = \frac{\sqrt{2 \hat{r}_+}}{\pi} E \left(
						     \frac{\chi_\mathrm{f}^2}{2
						     \hat{r}_+} \right)
 ,
\end{equation}
where $C_p$ is the polar circumferential radius, $\hat{r}_+ = 1 +
\sqrt{1 - \chi_\mathrm{f}^2}$ is the normalized radius of the outer
horizon, and $E (z)$ is an elliptic integral defined by
\begin{equation}
 E (z) = \int_0^{\pi / 2} \sqrt{1 - z \sin^2 \theta} d \theta .
\end{equation}
Comparisons among different estimates of the spin parameter suggest that
the systematic error associated with this method is $\Delta
\chi_\mathrm{f} \lesssim 0.01$
\cite{shibata_kyt2009,shibata_kyt2009e,kyutoku_ost2011}, and we do not
repeat them here.

\subsubsection{Gravitational waves} \label{sec:setup_diag_gw}

Our method to compute gravitational waves and related quantities is
summarized as follows (see Appendix B of Ref.~\cite{kyutoku_st2014} for
the details). We extract the Weyl scalar $\Psi_4$ at $\approx 400
M_\odot$ from the coordinate origin by projecting onto spin-weighted
spherical harmonics with $\ell \in \{2,3,4\}$ and extrapolate them to
null infinity by a perturbative method \cite{lousto_nzc2010}. The
energy, linear momentum, and angular momentum carried by gravitational
waves are computed by integrating $\Psi_4$ in time
\cite{ruiz_ant2008}. The time integration for calculating them and for
deriving gravitational waveforms are performed by a fixed frequency
integration method \cite{reisswig_pollney2011}. Because we always impose
the equatorial symmetry, we only consider the $z$ component for the
radiated angular momentum and denote it as $\Delta J_\mathrm{GW}$. The
radiated linear momentum, which only has the $x$ and $y$ components, is
decomposed into the magnitude $\Delta P_\mathrm{GW}$ and angle
$\Phi_\mathrm{GW}$ in the same way as the ejecta [see
Eqs.~\eqref{eq:ejlinmag} and \eqref{eq:ejlinang}, respectively]. The
radiated energy is denoted as $\Delta E_\mathrm{GW}$.

\section{Result of simulations} \label{sec:res}

\begin{table*}
 \caption{Characteristic physical quantities of the material measured at
 \SI{10}{\ms} after the onset of merger for our fiducial, $N=60$
 runs. $M_{r>r_\mathrm{AH}}$ is the rest mass outside the apparent
 horizon. $M_\mathrm{bd}$ and $M_\mathrm{ej}$ are the bound and unbound
 masses, respectively, and the unbound material is identified as the
 ejecta. Note that $M_{r>r_\mathrm{AH}} = M_\mathrm{bd} +
 M_\mathrm{ej}$. $T_\mathrm{ej}$ and $P_\mathrm{ej}$ are the kinetic
 energy and linear momentum of the ejecta, respectively. $v_\mathrm{ave}
 \equiv \sqrt{2T_\mathrm{ej}/M_\mathrm{ej}}$ and $v_\mathrm{ej} \equiv
 P_\mathrm{ej} / M_\mathrm{ej}$ are the average and bulk velocities of
 the ejecta, respectively.}
 \begin{tabular}{c|ccccccc} \hline
  ~Model~ & ~$M_{r>r_\mathrm{AH}} [M_\odot]$~ & ~$M_\mathrm{bd}
  [M_\odot]$~ & ~$M_\mathrm{ej} [M_\odot]$~ & ~$T_\mathrm{ej}$
  (\si{erg})~ & ~$P_\mathrm{ej} [M_\odot]$~ & ~$v_\mathrm{ave}$~ &
  ~$v_\mathrm{ej}$~ \\
  \hline \hline
  APR4-Q3a75 & 0.19 & 0.18 & 0.01 & \num{5e50} & \num{2e-3} & 0.23 &
			      0.19 \\
  ALF2-Q3a75 & 0.27 & 0.23 & 0.05 & \num{3e51} & \num{9e-3} & 0.25 &
			      0.21 \\
  H4-Q3a75 & 0.33 & 0.29 & 0.05 & \num{2e51} & \num{9e-3} & 0.24 &
			      0.20 \\
  MS1-Q3a75 & 0.35 & 0.28 & 0.07 & \num{4e51} & 0.01 & 0.25 & 0.21 \\
  \hline
  APR4-Q3a5 & 0.08 & 0.08 & \num{2e-3} & \num{1e50} & \num{4e-4} & 0.21
			  & 0.17 \\
  ALF2-Q3a5 & 0.19 & 0.17 & 0.02 & \num{1e51} & \num{5e-3} & 0.24 &
			      0.20 \\
  H4-Q3a5 & 0.24 & 0.21 & 0.03 & \num{1e51} & \num{6e-3} & 0.23 &
			      0.20 \\
  MS1-Q3a5 & 0.26 & 0.21 & 0.05 & \num{3e51} & 0.01 & 0.24 & 0.21 \\
  \hline
  APR4-Q3a0 & \num{4e-4} & \num{4e-4} & \num{2e-5} & \num{6e47} &
		      \num{1e-6} & 0.20 & 0.08 \\
  ALF2-Q3a0 & 0.03 & 0.03 & \num{3e-3} & \num{1e50} & \num{3e-4} & 0.22
			  & 0.11 \\
  H4-Q3a0 & 0.10 & 0.10 & \num{6e-3} & \num{3e50} & \num{1e-3} & 0.22 &
			      0.18 \\
  MS1-Q3a0 & 0.16 & 0.14 & 0.02 & \num{8e50} & \num{3e-3} & 0.23 &
			      0.19 \\
  \hline
  APR4-Q5a75 & 0.07 & 0.06 & \num{8e-3} & \num{5e50} & \num{8e-4} & 0.25
			  & 0.10 \\
  ALF2-Q5a75 & 0.24 & 0.20 & 0.05 & \num{3e51} & 0.01 & 0.28 & 0.21 \\
  H4-Q5a75 & 0.32 & 0.27 & 0.05 & \num{3e51} & 0.01 & 0.27 & 0.22 \\
  MS1-Q5a75 & 0.36 & 0.28 & 0.08 & \num{6e51} & 0.02 & 0.28 & 0.23 \\
  \hline
  APR4-Q5a5 & \num{5e-4} & \num{4e-4} & \num{9e-5} & \num{4e48} &
		      \num{5e-6} & 0.23 & 0.05 \\
  ALF2-Q5a5 & 0.04 & 0.03 & 0.01 & \num{8e50} & \num{7e-4} & 0.27 &
			      0.06 \\
  H4-Q5a5 & 0.14 & 0.12 & 0.02 & \num{1e51} & \num{4e-3} & 0.26 &
			      0.19 \\
  MS1-Q5a5 & 0.23 & 0.18 & 0.05 & \num{3e51} & 0.01 & 0.27 & 0.21 \\
  \hline
  APR4-Q7a75 & \num{2e-3} & \num{2e-3} & \num{5e-4} & \num{4e49} &
		      \num{3e-5} & 0.27 & 0.06 \\
  ALF2-Q7a75 & 0.07 & 0.05 & 0.02 & \num{2e51} & \num{2e-3} & 0.29 &
			      0.07 \\
  H4-Q7a75 & 0.19 & 0.16 & 0.04 & \num{3e51} & \num{7e-3} & 0.29 &
			      0.19 \\
  MS1-Q7a75 & 0.30 & 0.23 & 0.07 & \num{5e51} & \num{1e-2} & 0.30 &
			      0.23 \\
  \hline
  APR4-Q7a5 & \num{1e-5} & \num{1e-5} & \num{3e-6} & \num{1e47} &
		      \num{1e-7} & 0.23 & 0.04 \\
  ALF2-Q7a5 & \num{5e-4} & \num{3e-4} & \num{2e-4} & \num{1e49} &
		      \num{9e-6} & 0.27 & 0.05 \\
  H4-Q7a5 & \num{6e-3} & \num{3e-3} & \num{3e-3} & \num{3e50} &
		      \num{2e-4} & 0.29 & 0.06 \\
  MS1-Q7a5 & 0.04 & 0.02 & 0.02 & \num{1e51} & \num{1e-3} & 0.30 &
			      0.07 \\
  \hline
 \end{tabular}
 \label{table:result}
\end{table*}

In this section, we present the results of our numerical
simulations. Numerical values of characteristic quantities are shown in
Table \ref{table:result}, to which we refer repeatedly throughout this
section. These values are estimated consistently at \SI{10}{\ms} after
the onset of merger.\footnote{We define the time of the onset of merger,
$t_\mathrm{merge}$, by the condition that a part of neutron-star matter
of mass $0.01 M_\odot$ falls into the apparent horizon in this and also
previous work \cite{kyutoku_st2010,kyutoku_st2010e,kyutoku_ost2011}.}

\subsection{Overview of merger dynamics} \label{sec:res_review}

\begin{figure*}
 \includegraphics[width=.95\linewidth]{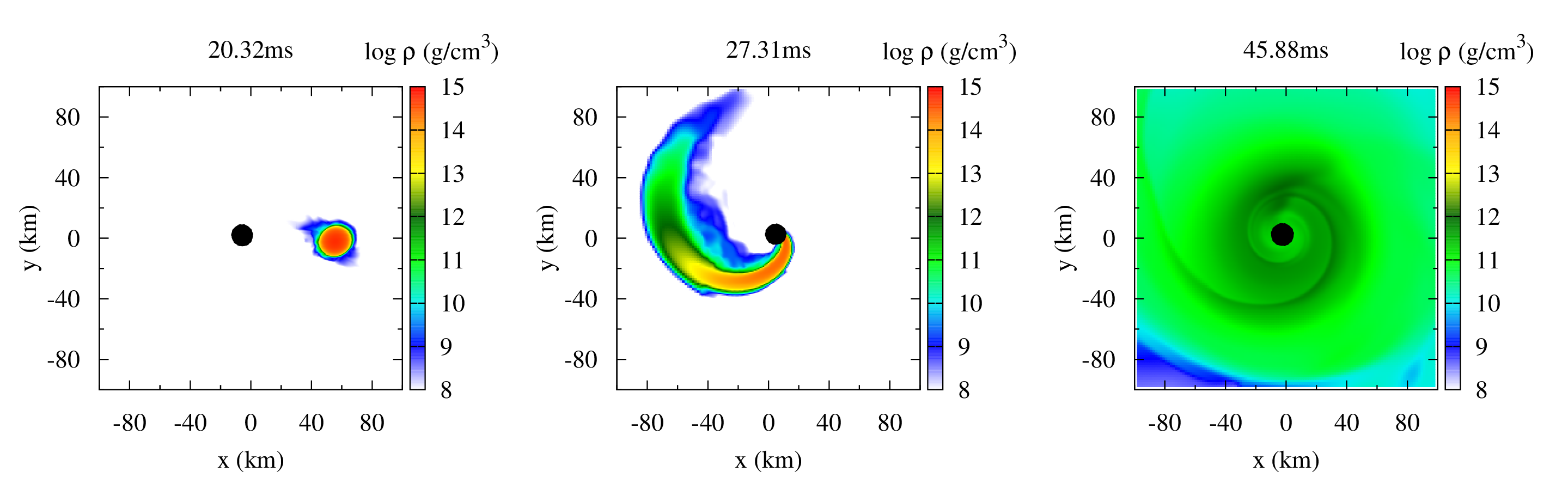} \caption{Rest-mass
 density profile on the equatorial plane in the central region at
 selected time slices for H4-Q5a75. Black filled circles show the
 interior of apparent horizons. The left, middle, and right panels
 correspond to a late inspiral phase, tidal tail formation, and
 quasistationary remnant accretion disk, respectively. The merger sets
 in at $t=$ \SI{26.49}{\ms} for this model.} \label{fig:snap_merger}
\end{figure*}

We begin with a brief review of the dynamics of black hole-neutron star
binary mergers (see Ref.~\cite{shibata_taniguchi2011} for details).
Black hole-neutron star binaries evolve as a result of gravitational
radiation reaction and eventually merge. Our initial conditions are
chosen to evolve for $\sim$ 3.5 to 7.5 orbits before the merger, where
the exact numbers depend on model parameters. Eccentricities are
estimated to be $e \sim 0.01$--0.02 for all the models using methods
described in Ref.~\cite{kyutoku_st2014}, and they introduce
uncertainties of the same order in the ejecta properties (see
App.~\ref{app:err_sep} for the estimate).

The fate of the system after the merger is determined primarily by
competition between the orbital separation at which tidal disruption
occurs (hereafter, the tidal disruption radius), $r_\mathrm{td}$, and
the radius of the innermost stable circular orbit, $r_\mathrm{ISCO}$. If
$r_\mathrm{ISCO}$ is larger than $r_\mathrm{td}$, no appreciable tidal
disruption occurs, and the neutron star is simply swallowed by the black
hole. In this case, the remnant disk, fallback material, and ejecta are
all negligible for our astrophysical interest. Although we do not pay
particular attention to such cases in this study, models like APR4-Q3a0
and ALF2-Q7a5 fall into this category (see the next paragraph). By
contrast, if $r_\mathrm{td}$ is larger than $r_\mathrm{ISCO}$, part of
the disrupted material spreads around the black hole in the form of a
tidal tail, while more than a half is still swallowed. Figure
\ref{fig:snap_merger} shows rest-mass density profiles on the equatorial
plane in the central region at selected time slices for H4-Q5a75 as an
example of this category. Material that remains outside the apparent
horizon can be divided into bound and unbound material, and the former
always dominates the latter for the models considered in this
study.\footnote{Hierarchy among the swallowed mass, bound mass, and
unbound mass could change for extreme binary parameters
\cite{lovelace_dfkpss2013}.} The bound material may be further divided
into disk and fallback components. The unbound component is generated
primarily by tidal torque exerted on the elongated neutron star during
tidal disruption, and details of the dynamical mass ejection process are
described separately in Sec.~\ref{sec:res_ej}.

Appreciable tidal disruption occurs when (i) the neutron-star equation
of state is stiff and the compactness is small, (ii) the mass ratio is
small, and/or (iii) the black-hole spin is large (for a prograde
orbit). These three conditions are reflected in our naming convention of
the models. Note that, if we presume $M_\mathrm{NS}$ to be fixed,
condition i can be rephrased as ``the neutron-star radius is large'' and
condition ii as ``the black-hole mass is small.'' On one hand,
$r_\mathrm{td}$ is expected to scale in the same way as the
mass-shedding radius $r_\mathrm{ms}$, which is determined by the
condition that the black-hole tidal force becomes equal to the
neutron-star self-gravity at the stellar surface (see, e.g.,
Ref.~\cite{shibata_taniguchi2011}),
\begin{equation}
 r_\mathrm{td} \propto r_\mathrm{ms} \sim Q^{1/3} R_\mathrm{NS} ,
  \label{eq:msradius}
\end{equation}
and the dependence on the black-hole spin is not very strong
\cite{wiggins_lai2000,ferrari_gp2009}. On the other hand,
$r_\mathrm{ISCO}$ is written as $\hat{r}_\mathrm{ISCO} ( \chi )
M_\mathrm{BH}$, where $\hat{r}_\mathrm{ISCO} ( \chi )$ is a decreasing
function of the dimensionless spin parameter, $\chi$
\cite{bardeen_pt1972}. Recalling $R_\mathrm{NS} / M_\mathrm{BH} = 1 / (
\mathcal{C} Q )$, we expect the ratio to satisfy
\begin{equation}
 \frac{r_\mathrm{td}}{r_\mathrm{ISCO}} \propto \frac{1}{\mathcal{C}
  Q^{2/3} \hat{r}_\mathrm{ISCO} ( \chi )} ,
\end{equation}
and a large value of this ratio should signal appreciable tidal
disruption. This expectation has been verified by previous studies of
disk formation and gravitational-wave emission
\cite{shibata_taniguchi2011}, and Table \ref{table:result} indicates
that dynamical mass ejection also becomes efficient when these three
parameters ($\mathcal{C}$, $Q$, and $\chi$) are advantageous for tidal
disruption. The dependence of the ejecta properties on these parameters
is discussed in more detail in Sec.~\ref{sec:res_char}.

\subsection{Mass ejection process and morphology} \label{sec:res_ej}

We first explain mechanisms of dynamical mass ejection and general
properties of the ejecta by closely investigating APR4-Q3a75 in
Sec.~\ref{sec:res_ej_case}. Mass ejection mechanisms and qualitative
trends are the same for all the black hole-neutron star binary models
simulated in this study, whereas differences in (semi)quantitative
properties are found. We next discuss differences in the ejecta geometry
among models in Sec.~\ref{sec:res_ej_var}. Characteristic quantities and
their differences are described in Sec.~\ref{sec:res_char}.

\subsubsection{Case study: APR4-Q3a75} \label{sec:res_ej_case}

\begin{figure*}
 \includegraphics[width=.95\linewidth]{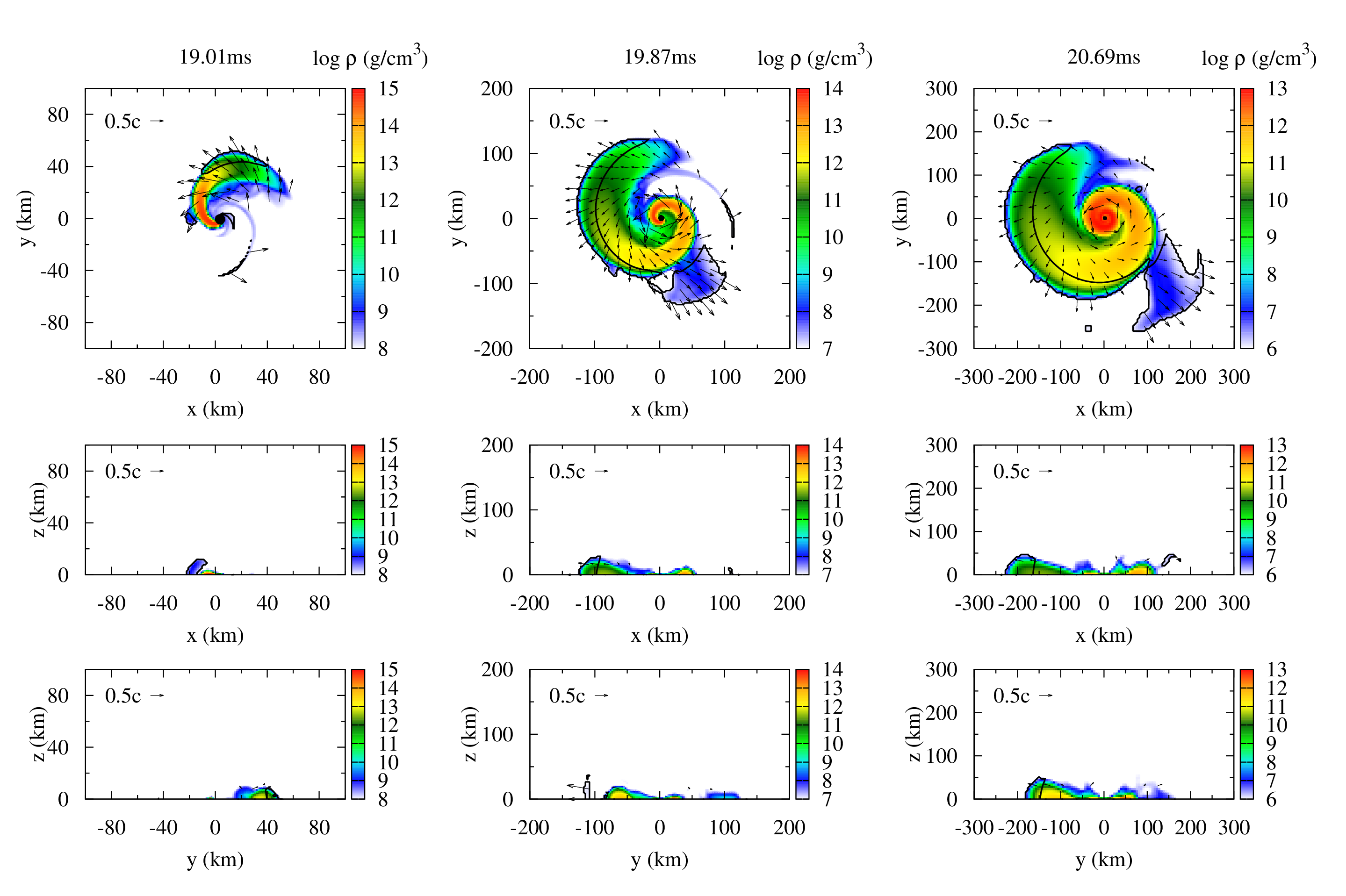} \caption{Rest-mass
 density profile for APR4-Q3a75 during the tidal disruption and
 dynamical mass ejection. Note the different spatial and density scales
 among the columns. Black filled circles show the interior of the
 apparent horizons. Black arrows show the spatial component of the
 covariant four-velocity, $u_i$. Unbound components satisfying $u_t <
 -1$ are marked by black curves, and we checked that contours marking $h
 u_t < -1$ nearly overlap with them. The top, middle, and bottom rows
 are the $xy$, $xz$, and $yz$ planes, respectively. The merger sets in
 at $t=$ \SI{18.48}{\ms} for this model. This figure should be compared
 with Figs.~3--5 of Ref.~\cite{hotokezaka_kkosst2013} where binary
 neutron stars are studied.} \label{fig:snap_eject}
\end{figure*}

Figure \ref{fig:snap_eject} depicts the typical process of dynamical
mass ejection at tidal disruption for model APR4-Q3a75. Once tidal
disruption sets in, the neutron star is drastically elongated and forms
a tidal tail. While the high-density innermost part is immediately
swallowed by the black hole, the outer part spreads to a distant orbit
and lags behind. Thus, the tidal tail exhibits a trailing one-armed
spiral structure, and the black hole exerts tidal torque on the tail,
increasing its orbital angular momentum. The outer part of the tail
moves further outward due to the gain of the angular momentum, and the
outermost part acquires enough kinetic energy to become unbound from the
system, as marked by black curves in Fig.~\ref{fig:snap_eject}. In the
course of this process, the pressure gradient in the tidal tail should
also boost the outer part. This angular momentum transport proceeds in
an unstable manner until the tidal tail winds around the black hole and
collides with itself to form a nearly axisymmetric black hole-disk
system. This mechanism generates most of the dynamical ejecta as well as
bound material which eventually falls back to the black hole-disk
system.

Although a small amount of unbound material appears to be ejected toward
$\varphi \sim \ang{-45}$ with a large velocity in
Fig.~\ref{fig:snap_eject}, where $\varphi$ is the azimuthal angle in
spherical coordinates, this appears to be an artifact created by the
artificial atmosphere and finite grid resolutions as we discuss in
Appendix \ref{app:err_hv}. This observation is consistent with
Ref.~\cite{foucart_etal2014}. The mass, energy, and linear momentum of
this component is negligible compared with the main component discussed
in the previous paragraph, and thus the values shown in Table
\ref{table:result} are not affected.

\begin{figure}
 \includegraphics[width=.95\linewidth]{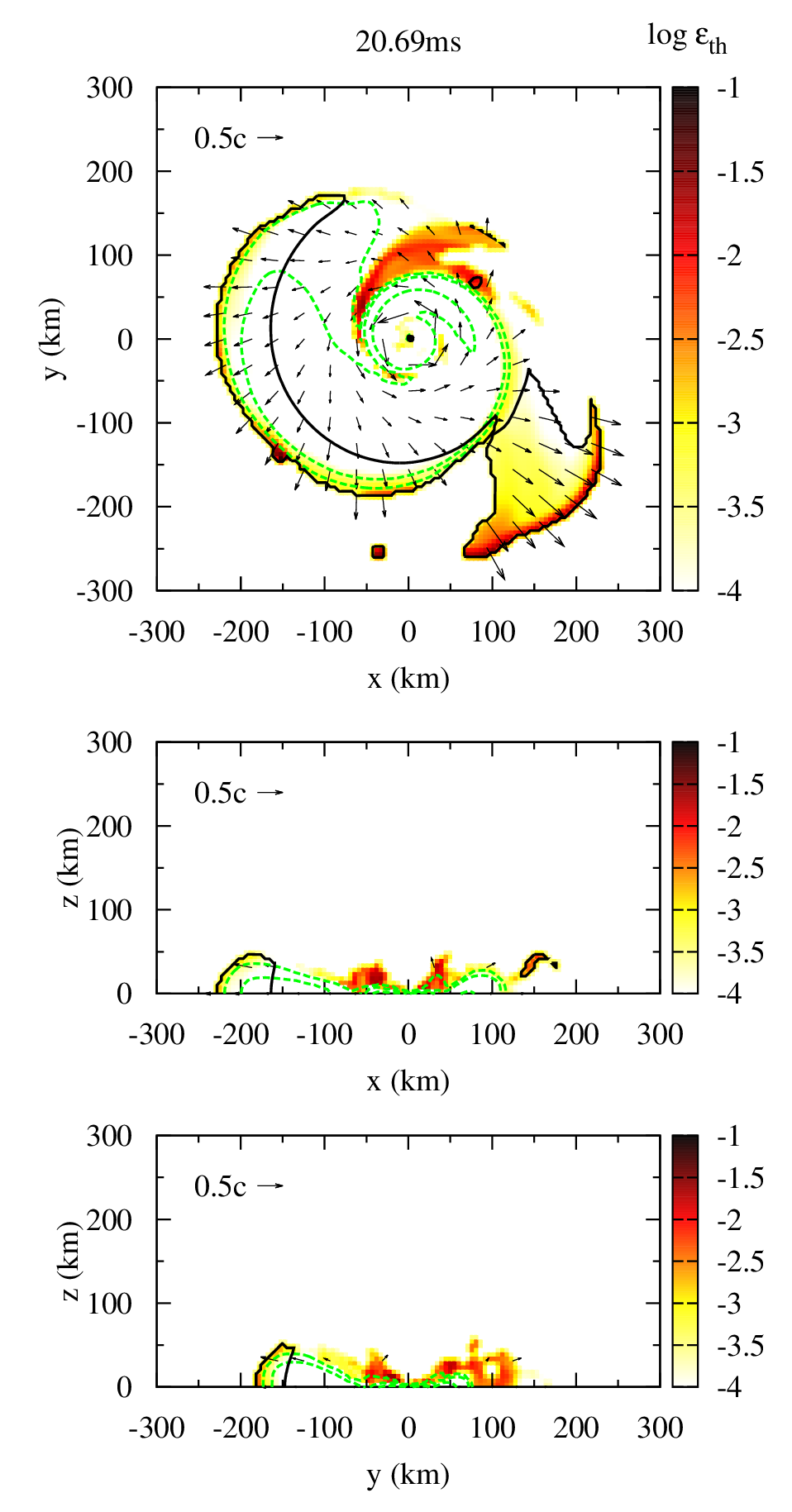} \caption{Profile of the
 thermal part of specific internal energy, $\varepsilon_\mathrm{th}$,
 for APR4-Q3a75. The time slice is taken to be the same as the right
 column of Fig.~\ref{fig:snap_eject}. The black filled circle, arrows,
 and curves have the same meanings as those in
 Fig.~\ref{fig:snap_eject}. Green dashed curves show contours of $\rho =
 \num{e8}$, \num{e10}, and \SI{e12}{\gram\per\cubic\cm}. This figure
 should be compared with Figs.~7 and 8 of
 Ref.~\cite{hotokezaka_kkosst2013} where binary neutron stars are
 studied, taking different spatial scales into account.}
 \label{fig:snap_thermal}
\end{figure}

Dynamical mass ejection from black hole-neutron star binaries is
anisotropic \cite{kyutoku_is2013}. Figure \ref{fig:snap_eject} shows
that the ejected material takes a crescentlike shape on the equatorial
plane during its early evolution for APR4-Q3a75. Although the relative
size of the central region occupied by bound material will become
negligible as the rear velocity approaches zero (see below), the ejecta
never sweep out the whole equatorial plane. Furthermore, it is
concentrated around the equatorial plane and does not extend above the
central black hole, because this mass ejection is driven by the tidal
torque, which works most efficiently in the equatorial plane. This
situation should be contrasted with dynamical mass ejection from binary
neutron stars, in which quasiradial oscillations of remnant massive
neutron stars eject an appreciable amount of material toward polar
regions via shock interaction \cite{hotokezaka_kkosst2013}. To elucidate
the difference, we show the thermal part of specific internal energy,
$\varepsilon_\mathrm{th}$, in Fig.~\ref{fig:snap_thermal}. As shocks do
not play a role, the tidal tail including the ejecta is not heated
significantly except for the self-colliding region of the tidal
tail. The self-colliding shock interaction eventually thermalizes and
circularizes material in the central region, and a hot accretion disk is
formed. We will discuss properties of the accretion disk later (see also
Sec.~\ref{sec:res_diskbh}). Apparent heating at the outermost part of
the tidal tail is caused by the artificial atmosphere and is thus
spurious.

Figures \ref{fig:snap_eject} and \ref{fig:snap_thermal} suggest that the
dynamical ejecta originates from the outer core and crust of the neutron
star retaining its very low electron fraction (the number of electrons
per baryon), $Y_e \lesssim 0.1$, at zero temperature
\cite{akmal_pr1998}.\footnote{Identifying the origin of postmerger
material is much more difficult in mesh-based simulations than in
smoothed-particle-hydrodynamics simulations. Rigorous confirmation would
require postprocess calculations using Lagrangian tracer particles.}
Because $M_\mathrm{ej}$ for APR4-Q3a75 is comparable to the typical mass
of neutron-star crusts, $0.01 M_\odot$ (see, e.g.,
Ref.~\cite{chamel_haensel2008}), the ejecta stripped from the outermost
part of the tidal tail in a highly nonspherical manner stems not only
from the crust but also from the core (see also Fig.~3 of
Ref.~\cite{just_bagj2015}). In fact, $M_\mathrm{ej}$ for other binary
models can easily exceed the typical crust mass. Nevertheless, the
ejecta would not come from the inner core, because the densest part of
the neutron star is swallowed by the black hole and bound material
separates the black hole and ejecta. Thus, the dynamical ejecta should
come mainly from the outer core and partly from the crust. The absence
of shocks suggests that the low electron fraction of the outer core is
not modified very much during dynamical mass ejection, and this is
consistent with results obtained by previous
smoothed-particle-hydrodynamics simulations
\cite{rosswog2005,rosswog_pn2013,just_bagj2015}. Such ejecta are
expected to be a promising site of \textit{r}-process nucleosynthesis
producing predominantly second- and third-peak elements via fission
cycling, while the production of first-peak elements may not accompany
\cite{wanajo_snkks2014,bauswein_ajg2014}. It has to be cautioned that
this estimation is speculative to some extent, because our simulations
are performed without taking the electron fraction into account. We plan
to revisit this topic with more sophisticated equations of state and
neutrino transport schemes \cite{sekiguchi_kks2015}.

\begin{figure*}
 \includegraphics[width=.95\linewidth]{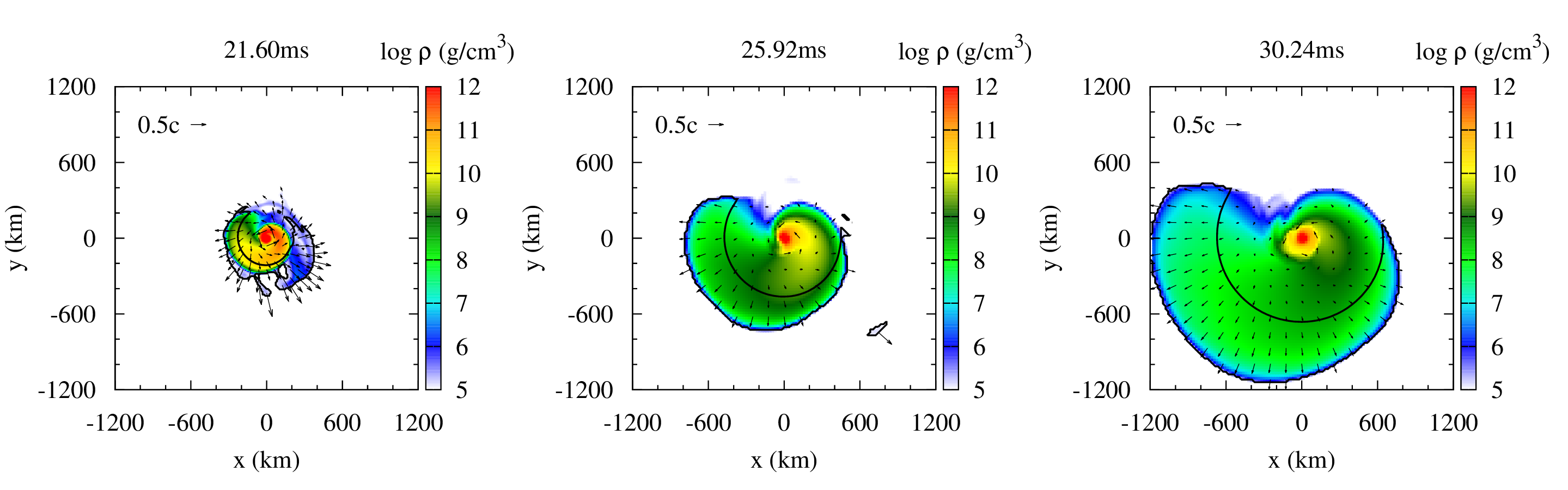} \caption{The same as
 Fig.~\ref{fig:snap_eject}, but on the equatorial plane at late times in
 the distant region. The black curves and arrows have the same meanings
 as those in Fig.~\ref{fig:snap_eject}. In this model, the ejecta linear
 momentum points toward $\Phi_\mathrm{ej} \approx \ang{-100}$, i.e.,
 close to the $-y$ direction [see Eq.~\eqref{eq:ejlinang} for the
 definition].} \label{fig:snap_distant}
\end{figure*}

Figure \ref{fig:snap_distant} shows the long-term evolution of the
dynamical ejecta in the distant region. This figure shows that the outer
edge of the ejecta expands in a nearly homogeneous manner after the
angular momentum transport by the tidal torque ceases. The azimuthal
component of velocity decreases approximately as $r^{-1}$ due to angular
momentum conservation and soon becomes negligible compared to the radial
component as shown in Fig.~\ref{fig:snap_distant}. This implies that the
kinetic energy of the ejecta is dominated by the radial velocity, and
thus the average velocity, $v_\mathrm{ave}$, estimated from the kinetic
energy approximately equals the typical radial velocity. Opening angles
of ejecta in the equatorial and also meridional (not shown in
Fig.~\ref{fig:snap_distant}) planes are approximately conserved, because
the direction of velocity does not change appreciably once hydrodynamic
interaction becomes negligible. Note that energy injection by the
\textit{r}-process heating will moderately change the ejecta geometry
\cite{rosswog_katp2014}.

Figure \ref{fig:snap_distant} also shows that the radial thickness of
the dynamical ejecta increases in the long-term evolution of the ejecta,
because the ejecta head is faster than the rear. Specifically, the head
will maintain a velocity on the order of the escape velocity of the
neutron star, while the rear velocity will approach zero (separation of
bound and unbound components) as the material climbs up the
gravitational potential well. The radius of the central bound region
will become negligible compared to the radial thickness of the dynamical
ejecta for exactly the same reason.

\begin{figure*}
 \includegraphics[width=.95\linewidth]{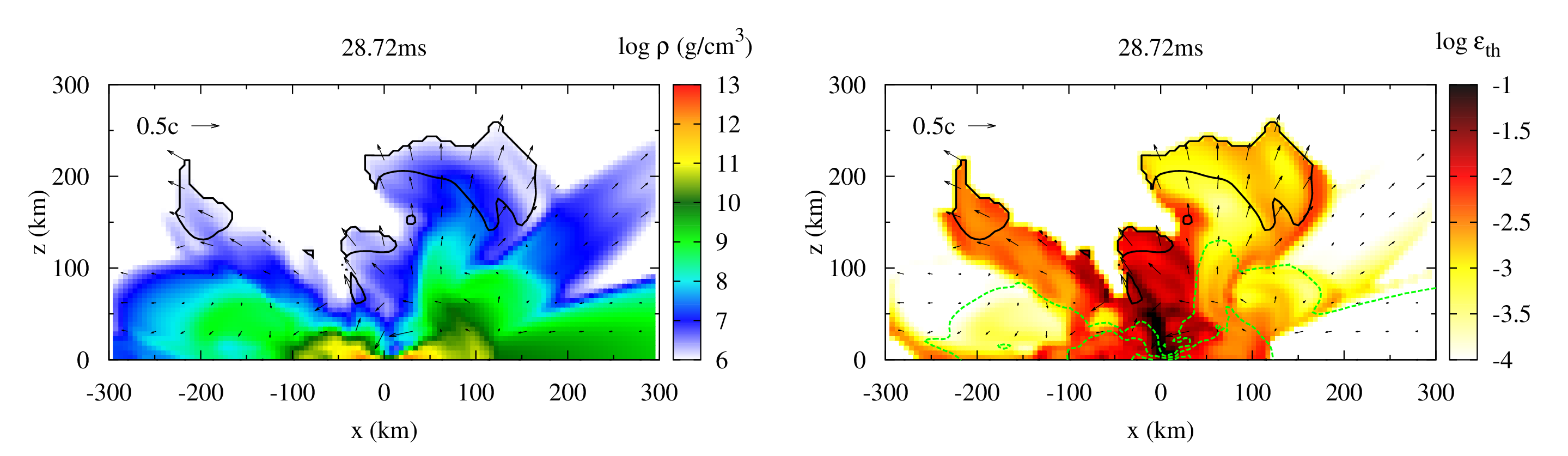} \caption{Profile of
 rest-mass density (left) and thermal part of the specific internal
 energy (right) on the $xz$ plane for APR4-Q3a75 after the disk
 formation. The black arrows and curves have the same meanings as those
 in Fig.~\ref{fig:snap_eject}. Green dashed curves show contours of
 $\rho = \num{e8}$, \num{e10}, and \SI{e12}{\gram\per\cubic\cm}.}
 \label{fig:snap_wind}
\end{figure*}

After disk formation, unbound material is newly generated from the disk
region due to shock heating. Figure \ref{fig:snap_wind} shows the shock
heating-driven disk outflow on the $xz$, meridional plane. When the
tidal tail collides with itself, shock interaction increases
$\varepsilon_\mathrm{th}$ near the contact surface. Because the
rest-mass density is not high in the relevant region, thermal pressure
dominates the cold-part pressure.\footnote{This should correspond to the
dominance of gas and radiation pressure over electron degeneracy
pressure.} The heated material expands, and some material is puffed up
off the equatorial plane. In addition, shock interaction circularizes
incoming tail material, and thus the disk region extends radially. Cold
fallback material eventually accumulates and circulates around the outer
edge of the hot disk material, as is visible from the right panel of
Fig.~\ref{fig:snap_wind} at $x \approx \SI{120}{\km}$. When the
accumulated cold material becomes very massive, shocks develop between
the cold and hot material. Shock heating occurs continually at the outer
edge of the disk due to this interaction, and material is also puffed up
there. Material off the equatorial plane exhibits (seemingly) random
motion, and a part of it collides with another part. Finally, some of
the material is ejected from the system as hot blobs, and the rest
eventually falls back to the disk surface. In contrast to dynamical mass
ejection due to the tidal torque, this mechanism ejects material mainly
toward nonequatorial (vertical) directions. As is evident from the left
panel of Fig.~\ref{fig:snap_wind}, however, the mass of the ejecta
generated by this heating is much smaller than that by the tidal
torque. The situation will change if magnetic fields
\cite{kiuchi_skstw2015}, neutrino heating
\cite{just_bagj2015,sekiguchi_kkst}, and/or nuclear interactions
\cite{lee_rl2009,fernandez_metzger2013} are taken into account, whereas
a significant fraction of the disk material has to be ejected to
dominate over dynamical mass ejection.

\subsubsection{Variety of ejecta morphology} \label{sec:res_ej_var}

The ejecta geometry may be characterized by an opening angle in the
equatorial plane, $\varphi_\mathrm{ej}$, and that in the meridional
plane, $\theta_\mathrm{ej}$, where the latter is defined to refer only
to material with $z \ge 0$, taking the equatorial symmetry into
account. In the nearly spherical mass ejection expected for supernovae
and binary neutron star mergers \cite{hotokezaka_kkosst2013},
$\varphi_\mathrm{ej}$ and $\theta_\mathrm{ej}$ should be regarded as $2
\pi$ and $\pi / 2$, respectively. We give estimates based on analytic
arguments of the opening angles for black hole-neutron star binaries in
Appendix \ref{app:ejang} to compare with numerical results.

\begin{figure*}
 \includegraphics[width=.95\linewidth]{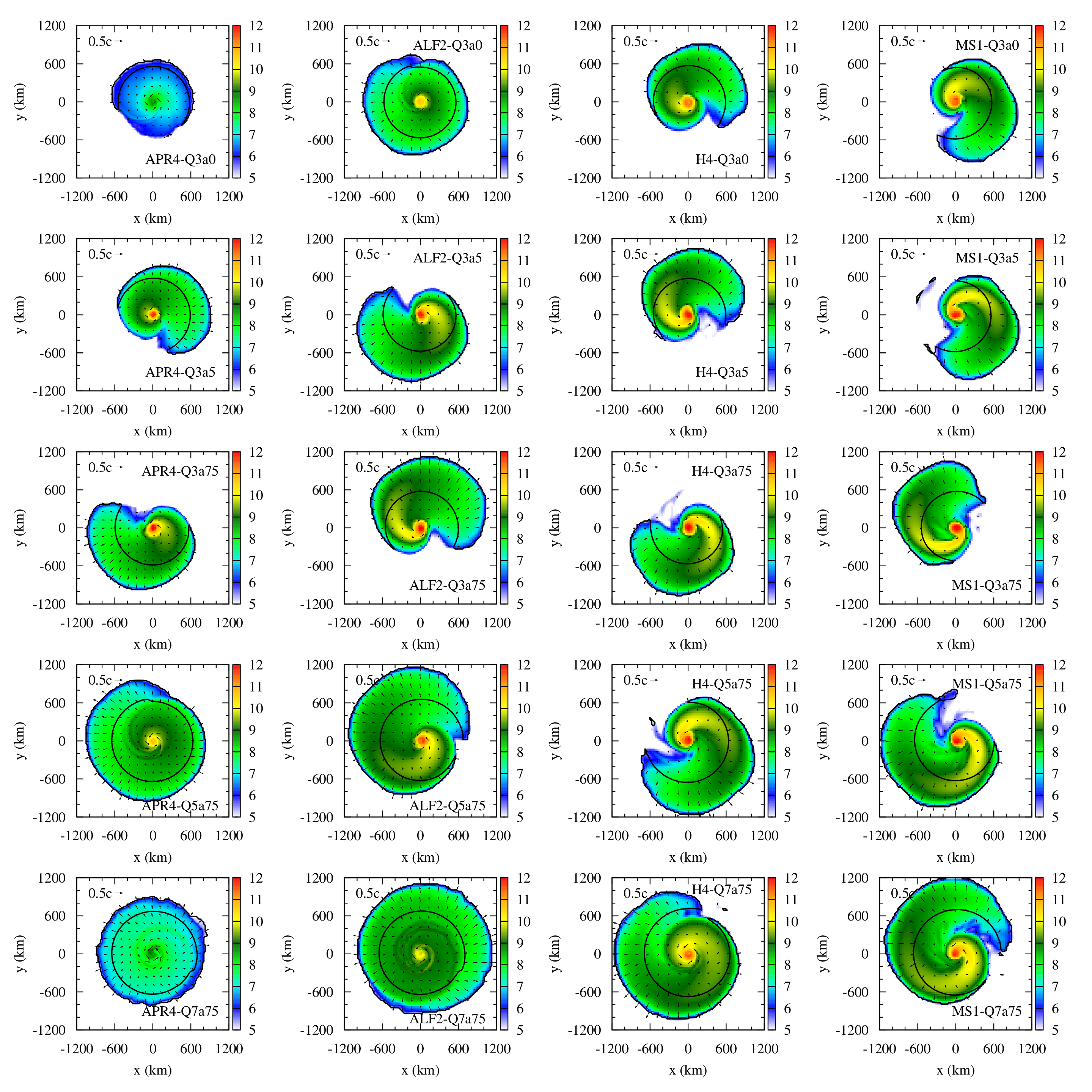} \caption{Rest-mass
 density profile on the equatorial plane for various models at $\approx
 \SI{10}{\ms}$ after the onset of merger. Unbound components satisfying
 $u_t < -1$ are marked by black curves. From left to right, the equation
 of state is APR4, ALF2, H4, and MS1. From top to bottom, $( Q , \chi )$
 is $(3,0)$, $(3,0.5)$, $(3,0.75)$, $(5,0.75)$, and $(7,0.75)$. The left
 panel on the third row is APR4-Q3a75 described in detail in
 Sec.~\ref{sec:res_ej_case}. Traces of rear-end collisions are found as
 bumps on inner black closed curves at $\varphi \approx \pi / 2$ for
 ALF2-Q3a0 and APR4-Q5a75 and at $\varphi \approx 0$ for APR4Q7a75 and
 ALF2Q7a75.} \label{fig:snap_eqvar}
\end{figure*}

Figure \ref{fig:snap_eqvar} shows the morphology of the dynamical ejecta
on the equatorial plane for various models. This figure implies that a
softer equation of state, a larger mass ratio, and a smaller black-hole
spin lead to a larger value of $\varphi_\mathrm{ej}$ when other
parameters are fixed.\footnote{APR4-Q3a0 might seem to have a smaller
$\varphi_\mathrm{ej}$ than ALF2-Q3a0, but this simply reflects the fact
that the ejecta of APR-Q3a0 is extremely tiny.} In particular for the
case in which mass ejection is not very substantial, an unbound portion
revolves more than one orbit ($\varphi_\mathrm{ej} > 2 \pi$) taking a
spiral shape at generation, and rear-end collisions occur in overlapping
directions to form a ring shape. Traces of the rear-end collisions are
observed as bumps on boundaries between bound and unbound material
(black closed curves) in Fig.~\ref{fig:snap_eqvar}. In these cases, the
bulk velocity, $v_\mathrm{ej}$, is lower than $0.1 c$ and is less than
half of the average velocity, $v_\mathrm{ave}$ (see Table
\ref{table:result}). The reason for this is that the ejecta linear
momentum, $P_\mathrm{ej}$, is very small for nearly axisymmetric mass
ejection.

This catalog suggests that $\varphi_\mathrm{ej}$ tends to become large
when tidal disruption occurs only weakly. This tendency does not agree
with the estimate obtained by time-scale arguments in Appendix
\ref{app:ejang}. A possible explanation of this tendency is the
periastron advance in general relativity, which is pronounced when tidal
disruption occurs very close to the innermost stable circular orbit
\cite{laguna_mwzd1993}. As an extreme example, orbital parameters of a
test particle can be finely tuned so that it experiences an arbitrarily
large number of revolutions traveling near marginally stable orbits
\cite{cutler_kp1994,glampedakis_kennefick2002}. Although the ejecta
material cannot be finely tuned due to its finite spatial extent and
does not experience infinitely many revolutions, i.e.,
$\varphi_\mathrm{ej}$ will not diverge, the dynamical ejecta should be
able to have a large value of $\varphi_\mathrm{ej}$ if the mass ejection
takes place near the innermost stable circular orbit. Indeed, tidal
disruption should have occurred very close to the innermost stable
circular orbit, i.e., $r_\mathrm{td} \approx r_\mathrm{ISCO}$, when the
ejecta mass is small but nonnegligible. This is consistent with the
tendency observed in Fig.~\ref{fig:snap_eqvar}.

From the observational viewpoint, dynamical ejecta with a large opening
angle, $\varphi_\mathrm{ej} \gtrsim 2 \pi$, may not be very important,
because a large opening angle is attained by the ejecta with a small
mass, for which electromagnetic radiation is expected to be weak. Strong
electromagnetic radiation should accompany substantial mass ejection,
say $M_\mathrm{ej} \gtrsim 0.01 M_\odot$, where $\varphi_\mathrm{ej}$
takes a value close to $\pi$ in most cases. However, substantial but
nearly axisymmetric dynamical mass ejection such as that for ALF2-Q7a75
is not completely excluded.

\begin{figure*}
 \includegraphics[width=.95\linewidth]{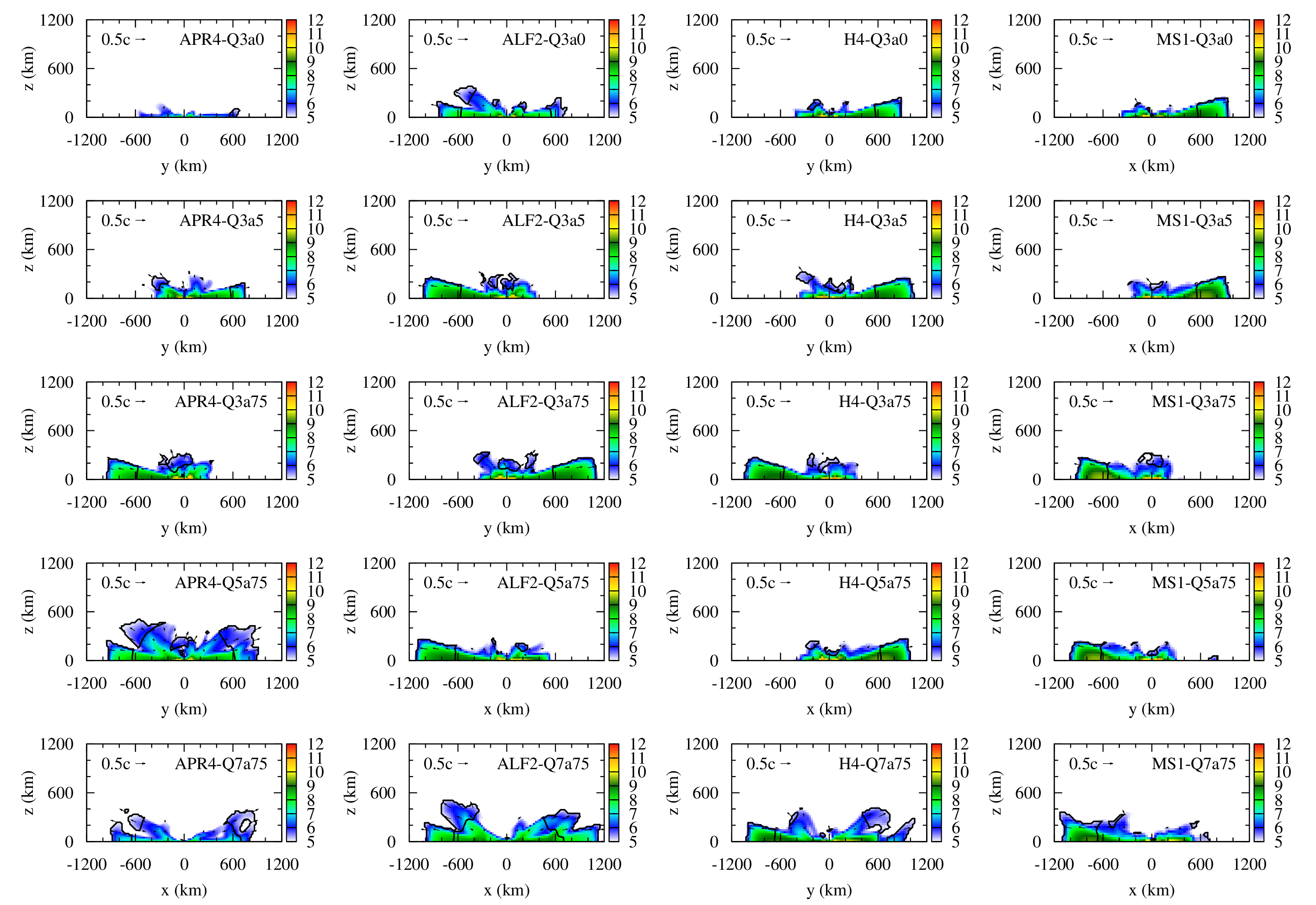} \caption{The same as
 Fig.~\ref{fig:snap_eqvar} but on a meridional, $xz$ or $yz$ plane,
 chosen to be the one which is closer to the direction of
 $\Phi_\mathrm{ej}$. Unbound components satisfying $u_t < -1$ are marked
 by black curves. From left to right, the equation of state is APR4,
 ALF2, H4, and MS1. From top to bottom, $( Q , \chi )$ is $(3,0)$,
 $(3,0.5)$, $(3,0.75)$, $(5,0.75)$, and $(7,0.75)$. The left panel on
 the third row is APR4-Q3a75 described in detail in
 Sec.~\ref{sec:res_ej_case}.} \label{fig:snap_mervar}
\end{figure*}

The opening angle in the meridional plane does not differ very much
among models as far as substantial mass ejection occurs. Figure
\ref{fig:snap_mervar} shows the morphology of the dynamical ejecta on
the meridional plane for various models. This figure shows that the
opening angle $\theta_\mathrm{ej}$ takes values between 1/5 and 1/3 (or
\ang{10} and \ang{20}) for cases with $M_\mathrm{ej} \gtrsim 0.01
M_\odot$. The variation of $\theta_\mathrm{ej}$ up to a factor of
$\lesssim 2$ is observed among models with substantial mass ejection,
but the ejecta driven by the tidal torque never extend to, say,
$\theta_\mathrm{ej} > \ang{30}$. At the same time, $\theta_\mathrm{ej}$
is very small when mass ejection is not efficient. Hence, sphericity is
never achieved even approximately for cases considered in this
study. This figure also suggests that $\theta_\mathrm{ej}$ tends to
become small when $Q$ is large. This is consistent with the analytic
expectation presented in Appendix \ref{app:ejang}.

\subsection{Characteristic quantities of ejecta} \label{sec:res_char}

Here we discuss characteristic quantities of dynamical ejecta such as
the mass and velocity, focusing on their dependence on binary
parameters. As described in the beginning of this section, we measure
ejecta quantities at \SI{10}{\ms} after the onset of merger. To check
that estimation at that time gives acceptable results, we first
investigate time evolution of the ejecta quantities in
Sec.~\ref{sec:res_char_evol}. Next, we discuss the dependence in
Sec.~\ref{sec:res_char_dep}.

\subsubsection{Time evolution} \label{sec:res_char_evol}

\begin{figure}
 \begin{tabular}{c}
  \includegraphics[width=.95\linewidth]{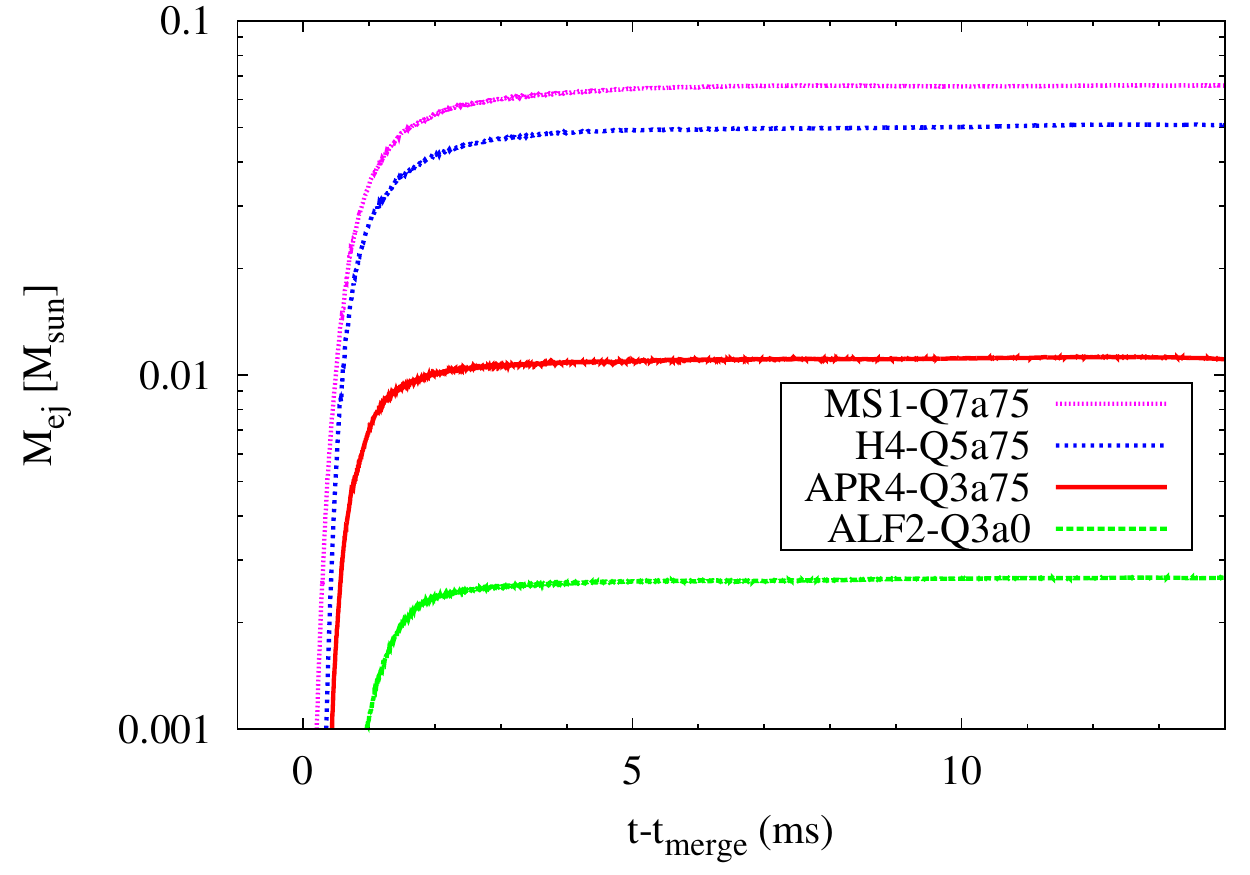} \\
  \includegraphics[width=.95\linewidth]{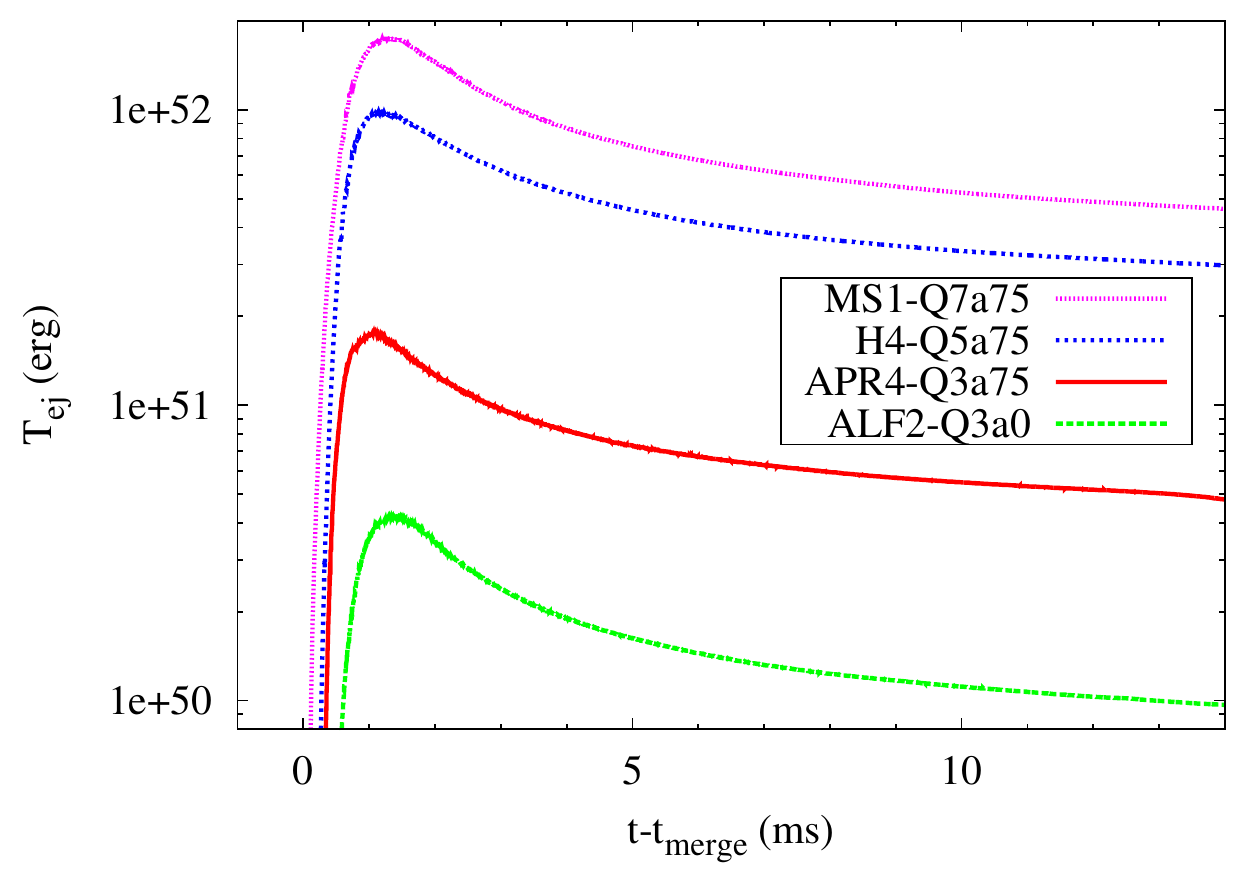} \\
  \includegraphics[width=.95\linewidth]{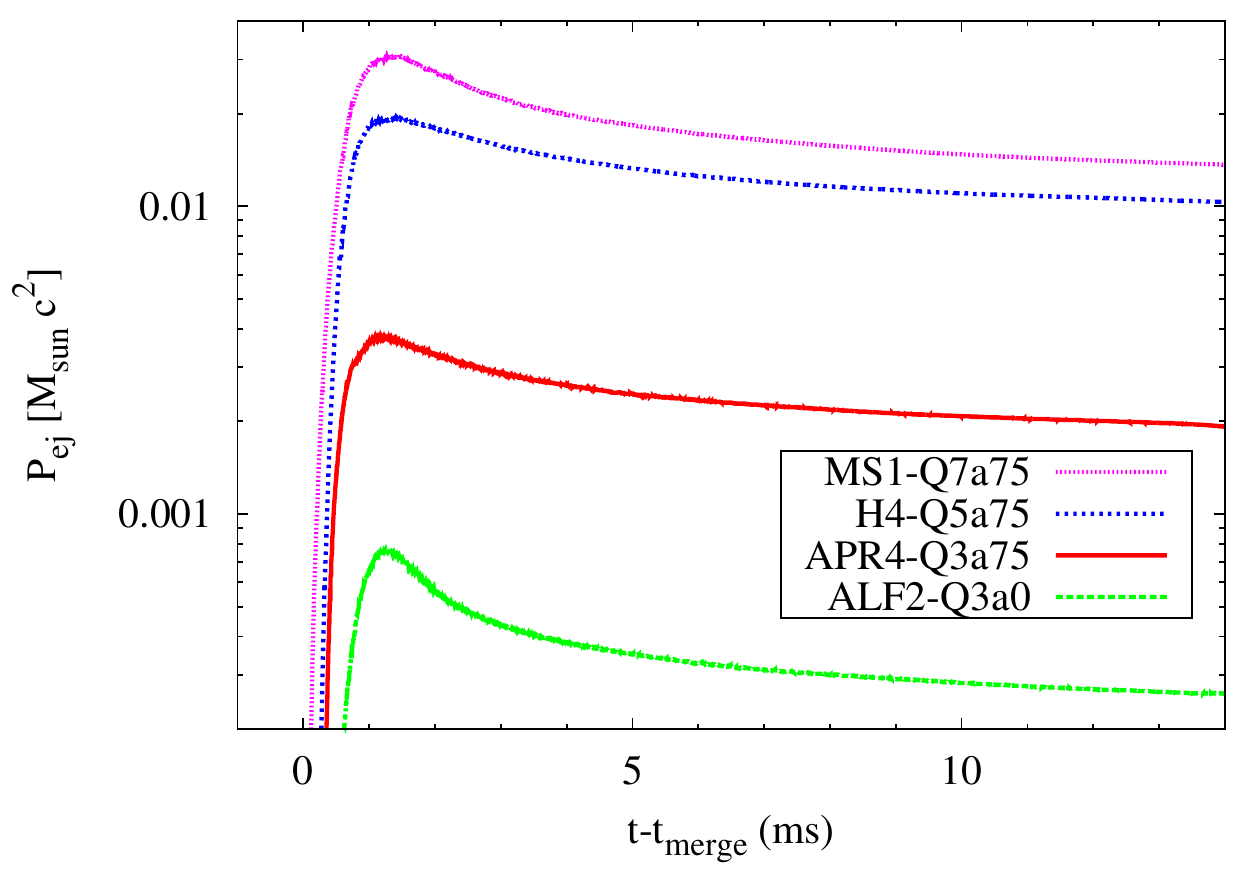}
 \end{tabular}
 \caption{Time evolution of the mass (top), kinetic energy (middle), and
 linear momentum (bottom) of the ejecta for selected models.}
 \label{fig:evol}
\end{figure}

Figure \ref{fig:evol} shows the time evolution of $M_\mathrm{ej}$,
$T_\mathrm{ej}$, and $P_\mathrm{ej}$ for selected models. All these
values suddenly increase right after the onset of merger. The time
evolution indicates that most of dynamical mass ejection progresses over
$\approx \SI{2}{\ms}$ and that the evolution relaxes afterward
irrespective of the models.

The ejecta mass settles to a quasistationary value within $\sim
\SI{5}{\ms}$. This confirms the observation in
Sec.~\ref{sec:res_ej_case} that mass ejection due to disk activity does
not contribute significantly to the total mass of the ejecta in our
simulations. Therefore, the measurement of $M_\mathrm{ej}$ at
\SI{10}{\ms} after the onset of merger is safely justified.

The kinetic energy and linear momentum peak at 1--\SI{2}{\ms} after the
onset of merger and decrease afterward. The reason of this decrease is
that the ejecta lose energy in climbing up the gravitational potential
well of the central black hole-disk system. The Newtonian formulas
indicate that $T_\mathrm{ej}$ measured at \SI{10}{\ms} after the onset
of merger overestimates its final value by $( m_0 M_\mathrm{ej} / r ) /
T_\mathrm{ej} \sim 30\%$--40\% for models shown in Fig.~\ref{fig:evol},
and this is consistent with the later evolution. This will result in
$\sim 15\%$--20\% overestimation of the ejecta velocity, and thus this
error has to be kept in mind in the following discussions, along with
those described in Appendix \ref{app:err}. If we measure these values at
$\lesssim$ \SI{5}{\ms} after the onset of merger and use them as proxies
for their final values, final ejecta velocities can be overestimated
nearly by 100\%. Hence, a large computational domain is a prerequisite
for an accurate study of mass ejection.\footnote{The amount of error
depends on estimation methods. For example, the kinetic energy can also
be defined by $\int \rho_* ( - u_t - 1 ) d^3 x$ (F.~Foucart, private
communication).}

\subsubsection{Dependence on binary parameters} \label{sec:res_char_dep}

\begin{figure}
 \includegraphics[width=.95\linewidth]{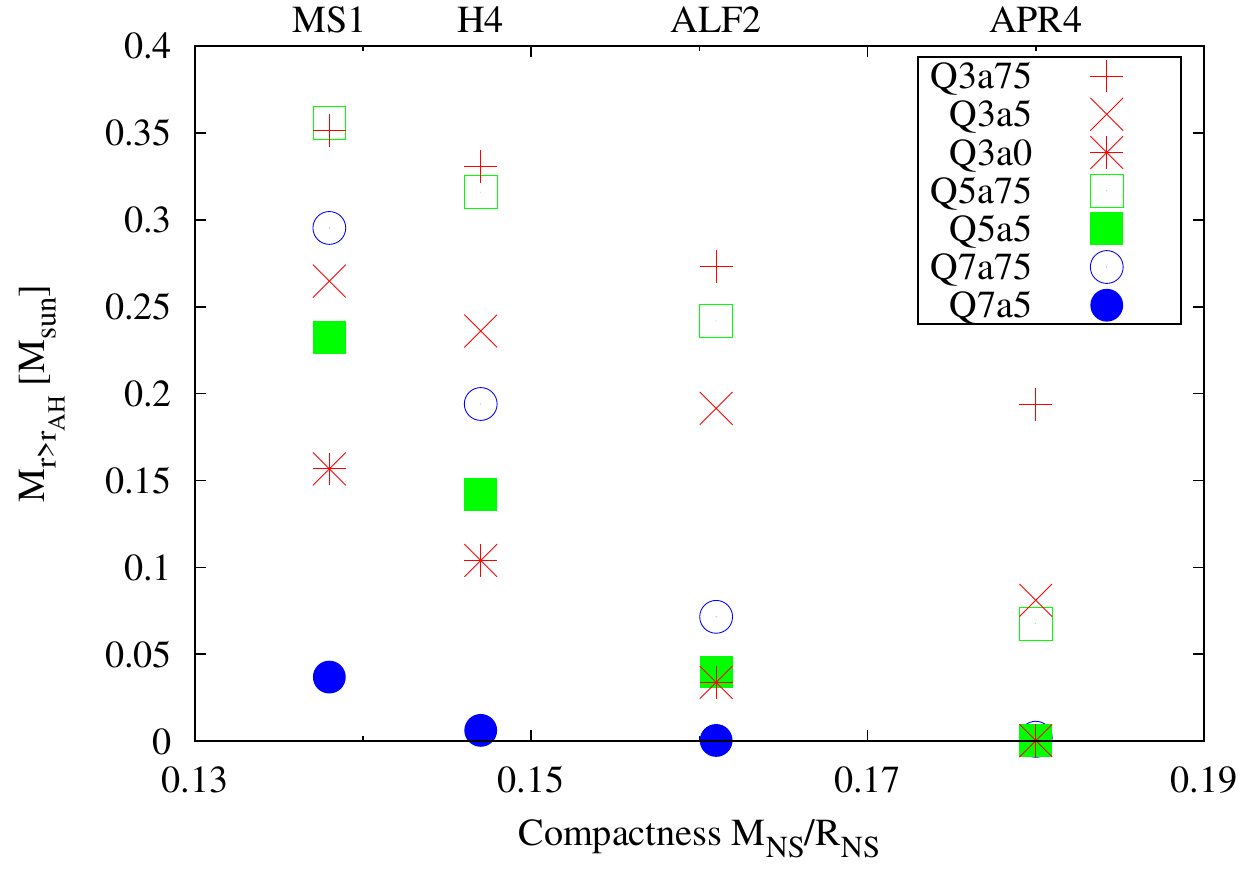} \caption{Mass remaining
 outside the apparent horizon composed of bound and unbound material,
 $M_{r>r_\mathrm{AH}} = M_\mathrm{bd} + M_\mathrm{ej}$, as functions of
 the compactness, $\mathcal{C}$.} \label{fig:depmtot}
\end{figure}

We start by looking at the total mass remaining outside the apparent
horizon, $M_{r>r_\mathrm{AH}} = M_\mathrm{bd} + M_\mathrm{ej}$, to check
consistency with previous work. Figure \ref{fig:depmtot} plots
$M_{r>r_\mathrm{AH}}$ measured at \SI{10}{\ms} after the onset of merger
(presented in Table \ref{table:result}) as a function of the
compactness, $\mathcal{C}$. This figure supports the discussion in
Sec.~\ref{sec:res_review}. That is, a small neutron-star compactness,
small mass ratio, and large black-hole spin increase the strength of the
tidal disruption resulting in the increase of $M_{r>r_\mathrm{AH}}$. Our
present simulations reproduce quantitatively the results of our previous
simulations \cite{kyutoku_st2010,kyutoku_st2010e,kyutoku_ost2011}, as
well as those by other authors (see Ref.~\cite{foucart2012} for a
compilation). The dependence of $M_{r>r_\mathrm{AH}}$ on $\mathcal{C}$
is approximately linear within the range studied here, until it levels
off at $\lesssim 0.01 M_\odot$. This suggests that the effect of
neutron-star properties on $M_{r>r_\mathrm{AH}}$ is reasonably captured
by the compactness, $\mathcal{C}$.

\begin{figure}
 \includegraphics[width=.95\linewidth]{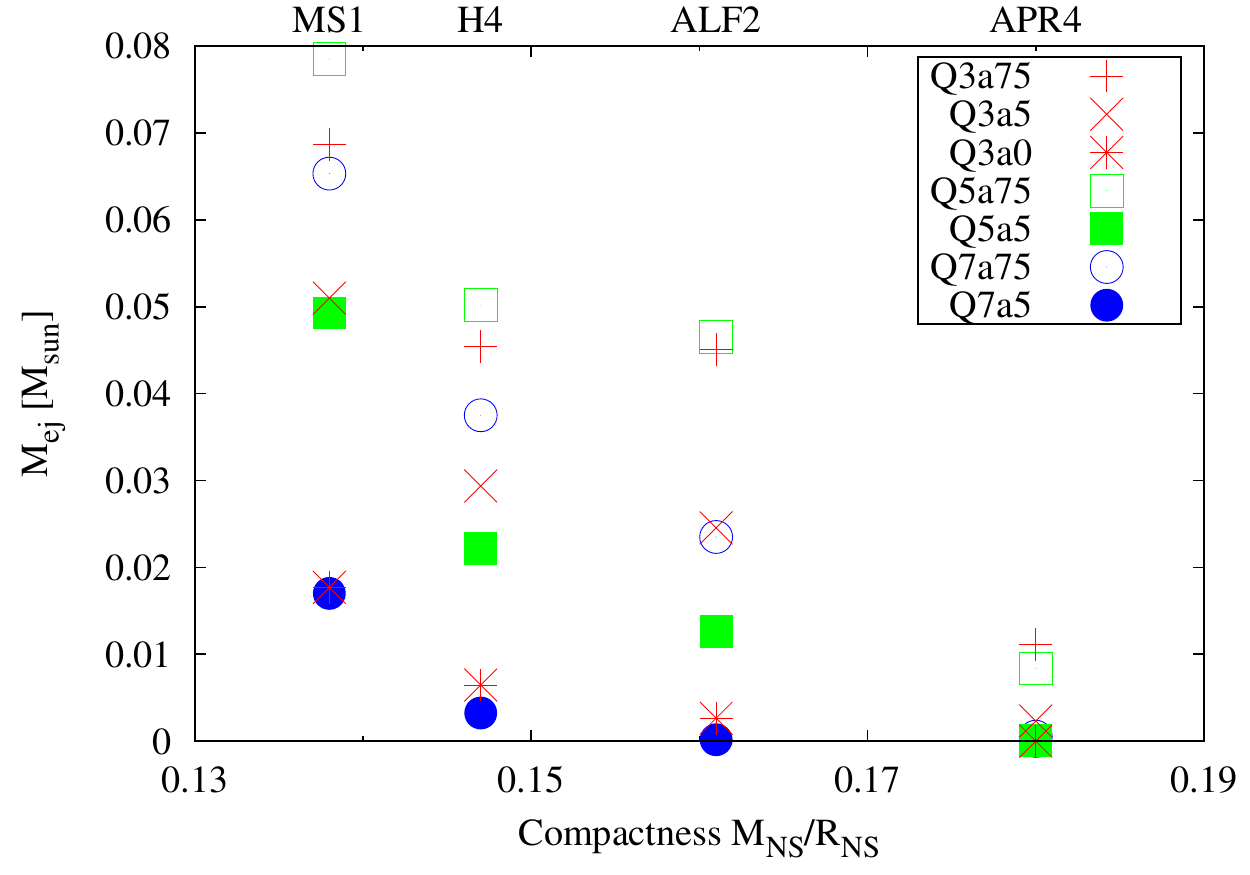} \caption{Ejecta mass,
 $M_\mathrm{ej}$, as a function of the compactness, $\mathcal{C}$.}
 \label{fig:depmej}
\end{figure}

The ejecta mass, $M_\mathrm{ej}$, is correlated with the strength of the
tidal disruption as $M_{r>r_\mathrm{AH}}$ is, but the dependence of
$M_\mathrm{ej}$ on binary parameters is complicated. Figure
\ref{fig:depmej} shows $M_\mathrm{ej}$ as a function of
$\mathcal{C}$. Plots of $T_\mathrm{ej}$ and $P_\mathrm{ej}$ exhibit
similar behavior. For fixed values of $Q$ and $\chi$, $M_\mathrm{ej}$
increases as $\mathcal{C}$ decreases. This is qualitatively the same as
$M_{r>r_\mathrm{AH}}$ and supports the expectation that strong tidal
disruption is accompanied by efficient mass ejection. However, the
correlation is weaker between $M_\mathrm{ej}$ and $\mathcal{C}$ than
between $M_{r>r_\mathrm{AH}}$ and $\mathcal{C}$ for fixed values of $Q$
and $\chi$. This suggests that the boundary separating bound and unbound
material, $u_t = -1$, is not determined solely by the compactness but is
also sensitive to the stellar structure. This observation is consistent
with Ref.~\cite{foucart_etal2014}, which found a similar fact by
comparing their results with some of our results reported in
Ref.~\cite{hotokezaka_ktkssw2013}. It is reasonable that detailed
properties of the equation of state could play an important role during
dynamical mass ejection via effects such as the pressure gradient and/or
central condensation.

\begin{figure}
 \includegraphics[width=.95\linewidth]{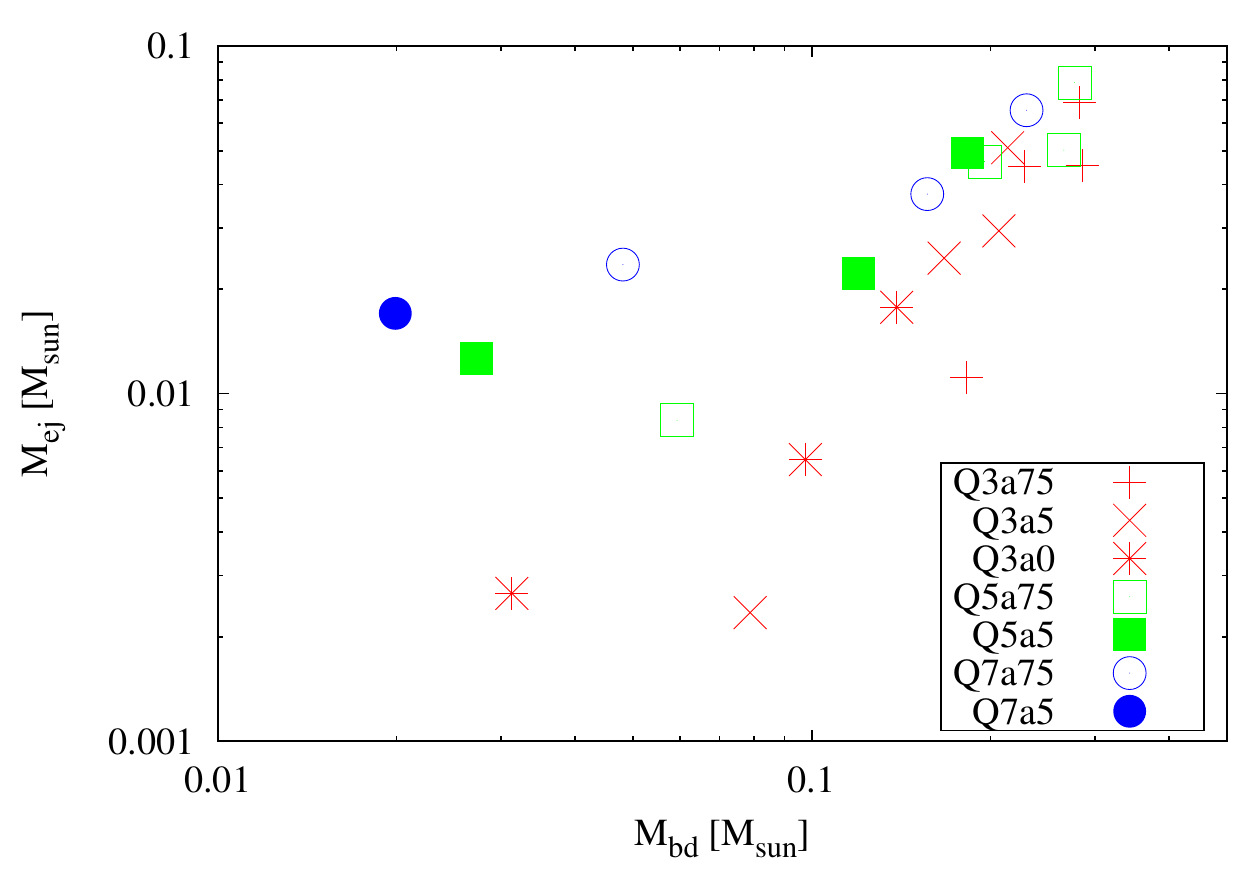} \caption{Correlation
 between the ejecta mass, $M_\mathrm{ej}$, and bound mass,
 $M_\mathrm{bd}$. We restrict the region to $M_\mathrm{ej} \ge 0.001
 M_\odot$ and $M_\mathrm{bd} \ge 0.01 M_\odot$, where the results are
 astrophysically interesting.} \label{fig:bdvsej}
\end{figure}

The ejecta mass, $M_\mathrm{ej}$, does not depend monotonically on the
mass ratio, $Q$, for fixed values of $\mathcal{C}$ and $\chi$ (see
Fig.~\ref{fig:depmej}). The reason for this is that the ejecta tends to
comprise a large fraction of material remaining outside the apparent
horizon for a large mass ratio, especially when tidal disruption is weak
and $M_{r>r_\mathrm{AH}}$ is not very large. Figure \ref{fig:bdvsej}
shows the correlation between the ejecta mass, $M_\mathrm{ej}$, and
bound mass, $M_\mathrm{bd}$. This figure indicates that $M_\mathrm{ej}$
does not decrease very rapidly with the decrease of $M_\mathrm{bd}$ (and
equivalently $M_{r>r_\mathrm{AH}}$) for a large value of
$Q$. Specifically, $M_\mathrm{ej} \ge 0.01 M_\odot$ can be achieved when
$M_\mathrm{bd} \gtrsim 0.01 M_\odot$ for $Q=7$, while it is possible
only when $M_\mathrm{bd} \gtrsim 0.1 M_\odot$ for $Q=3$. The fact that
mass ejection can be substantial even if tidal disruption is not very
strong for a large value of $Q$ is encouraging for electromagnetic
counterpart searches, because astrophysical black holes are expected to
prefer large mass ratios \cite{ozel_pnm2010,kreidberg_bfk2012}.

\begin{figure}
 \includegraphics[width=.95\linewidth]{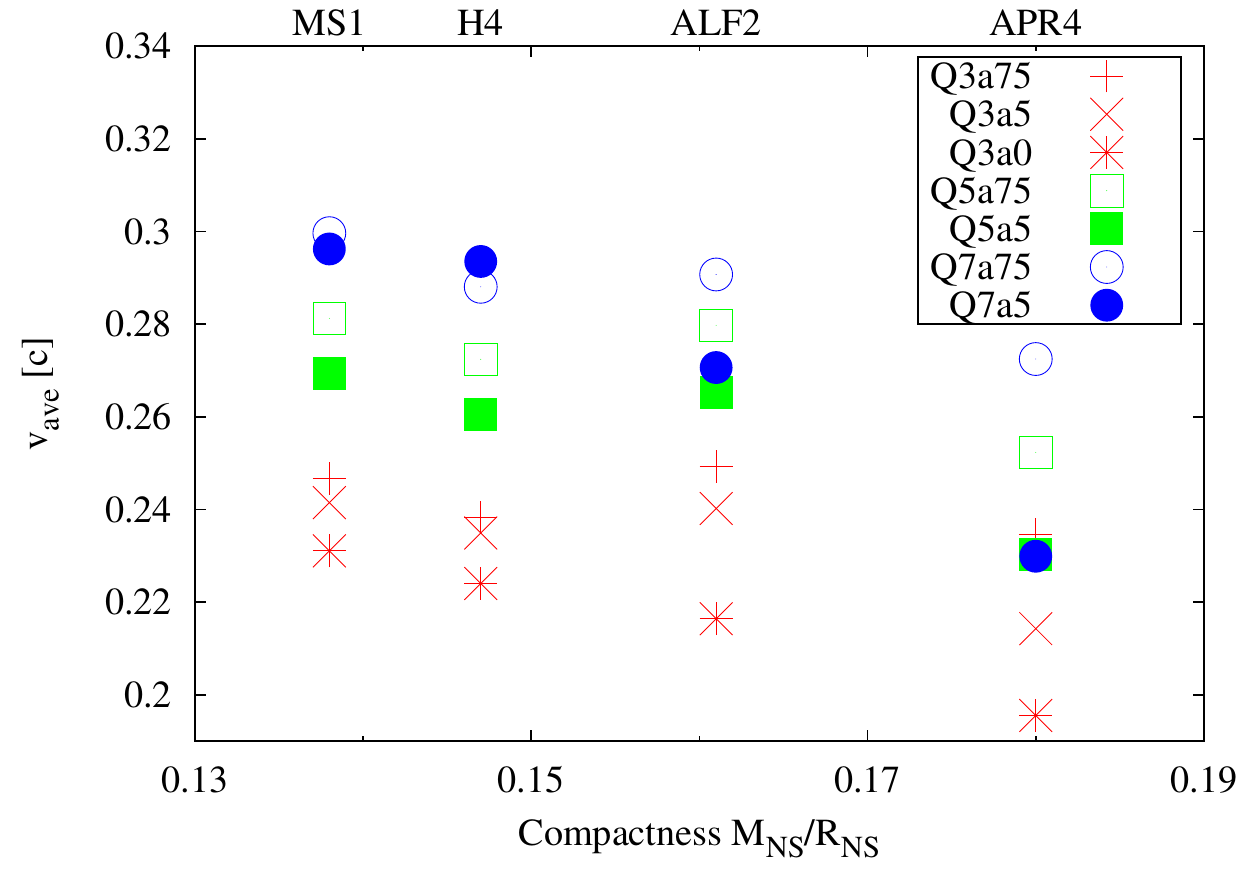} \caption{Average
 velocity of the ejecta, $v_\mathrm{ave}$, defined from the kinetic
 energy, $T_\mathrm{ej}$, as a function of the compactness,
 $\mathcal{C}$.} \label{fig:depvave}
\end{figure}

The increase of $M_\mathrm{ej} / M_\mathrm{bd}$ with the mass ratio,
$Q$, implies that material remaining outside the apparent horizon tends
to become more energetic when $Q$ is larger. This speculation is
supported by the fact that the average velocity of the ejecta,
$v_\mathrm{ave}$, is larger for a larger mass ratio. Figure
\ref{fig:depvave} shows $v_\mathrm{ave}$ as a function of
$\mathcal{C}$. A typical value of $v_\mathrm{ave}$ is 0.22--$0.25c$ for
$Q=3$, and this rises to 0.25--$0.28c$ for $Q=5$ and 0.28--$0.3c$ for
$Q=7$. This can be ascribed to the higher energy of material remaining
outside the apparent horizon for a larger value of $Q$. The effect of
the mass ratio on ejecta velocities via a gravitational potential is
pointed out in the context of the tidal disruption of a main sequence
star during a nearly parabolic encounter with a supermassive black hole,
where a half of the star is expected to become unbound
\cite{lacy_th1982}. Although the qualitative trend is the same,
dynamical processes should play a crucial role in realizing this
dependence in the inspiral of black hole-neutron star binaries, because
all the neutron-star material is bound to the system at the onset of
tidal disruption. Note that the systematic error in $v_\mathrm{ave}$
associated with the residual gravitational binding described in
Sec.~\ref{sec:res_char_evol} is not likely to modify this tendency
qualitatively, because all the values of $v_\mathrm{ave}$ are
systematically overestimated.

The dependence of the ejecta mass, $M_\mathrm{ej}$, on the black-hole
spin, $\chi$, is simpler than that on $\mathcal{C}$ and $Q$ (see
Fig.~\ref{fig:depmej}). Namely, a large black-hole spin increases the
amount of ejecta for fixed values of $\mathcal{C}$ and $Q$. We do not
find significant dependence of $M_\mathrm{ej} / M_\mathrm{bd}$ on
$\chi$. The average velocity, $v_\mathrm{ave}$, tends to increase as
$\chi$ increases.

The ejecta mass, $M_\mathrm{ej}$, is correlated with the mass remaining
outside the apparent horizon, $M_{r>r_\mathrm{AH}}$, as indicated in
Fig.~\ref{fig:bdvsej}. Quantitatively, we obtain
\begin{equation}
 \frac{M_\mathrm{ej}}{M_\odot} = ( 0.27 \pm 0.07 ) \left(
						    \frac{M_{r>r_\mathrm{AH}}}{M_\odot}
						   \right)^{1.3 \pm 0.2}
 ,
 \label{eq:ejfitall}
\end{equation}
by fitting all the data shown in Table \ref{table:result} with equal
weights, where the range indicates the 1-$\sigma$ asymptotic standard
error. If we fit the data of models with different values of $Q$
separately, relations become
\begin{equation}
 \frac{M_\mathrm{ej}}{M_\odot} =
  \begin{cases}
   (0.41 \pm 0.14) \left( M_{r>r_\mathrm{AH}} / M_\odot \right)^{1.8 \pm
   0.3} & (Q=3) \\
   (0.23 \pm 0.06) \left( M_{r>r_\mathrm{AH}} / M_\odot \right)^{1.1 \pm
   0.2} & (Q=5) \\
   (0.15 \pm 0.02) \left( M_{r>r_\mathrm{AH}} / M_\odot \right)^{0.73
   \pm 0.09} & (Q=7)
  \end{cases}
  . \label{eq:ejfit}
\end{equation}
It is evident that the power-law index is smaller for a larger value of
$Q$, and thus the separate fitting may be more appropriate. These
relations give us an approximate estimate of $M_\mathrm{ej}$ combined
with a fitting formula for $M_{r>r_\mathrm{AH}}$ provided in
Ref.~\cite{foucart2012}. Sources of the error come from both simulations
and fitting procedures, and only the latter is taken into account in
Eqs.~\eqref{eq:ejfitall} and \eqref{eq:ejfit}.

\subsection{Ejecta and envelope structure} \label{sec:res_struc}

First in Sec.~\ref{sec:res_struc_eq}, we investigate matter profiles on
the equatorial plane, where most of the material resides. It includes
disk, fallback, and ejecta components. Next, material distribution along
the $z$ axis is investigated in Sec.~\ref{sec:res_struc_z}. It will be
important for gamma-ray bursts, because a hypothetical jet (or fireball)
can achieve an ultrarelativistic velocity only if the baryon load is not
very high \cite{meszaros_rees2000}. Finally, we investigate the velocity
distribution of dynamical ejecta in Sec.~\ref{sec:res_struc_v}, which is
required to predict electromagnetic radiation quantitatively
\cite{piran_nr2013,kisaka_it2015}. Detailed structures of material
obtained from our simulations are not expected to be very realistic,
because the equation of state in a relevant regime is composed of a
single zero-temperature polytrope and ideal-gas-like thermal
correction. We still believe that our results capture qualitative
properties of the material structure, particularly for ejecta in distant
regions where hydrodynamic interaction does not play an important role.

\subsubsection{Equatorial plane} \label{sec:res_struc_eq}

\begin{figure*}
 \begin{tabular}{cc}
  \includegraphics[width=.45\linewidth]{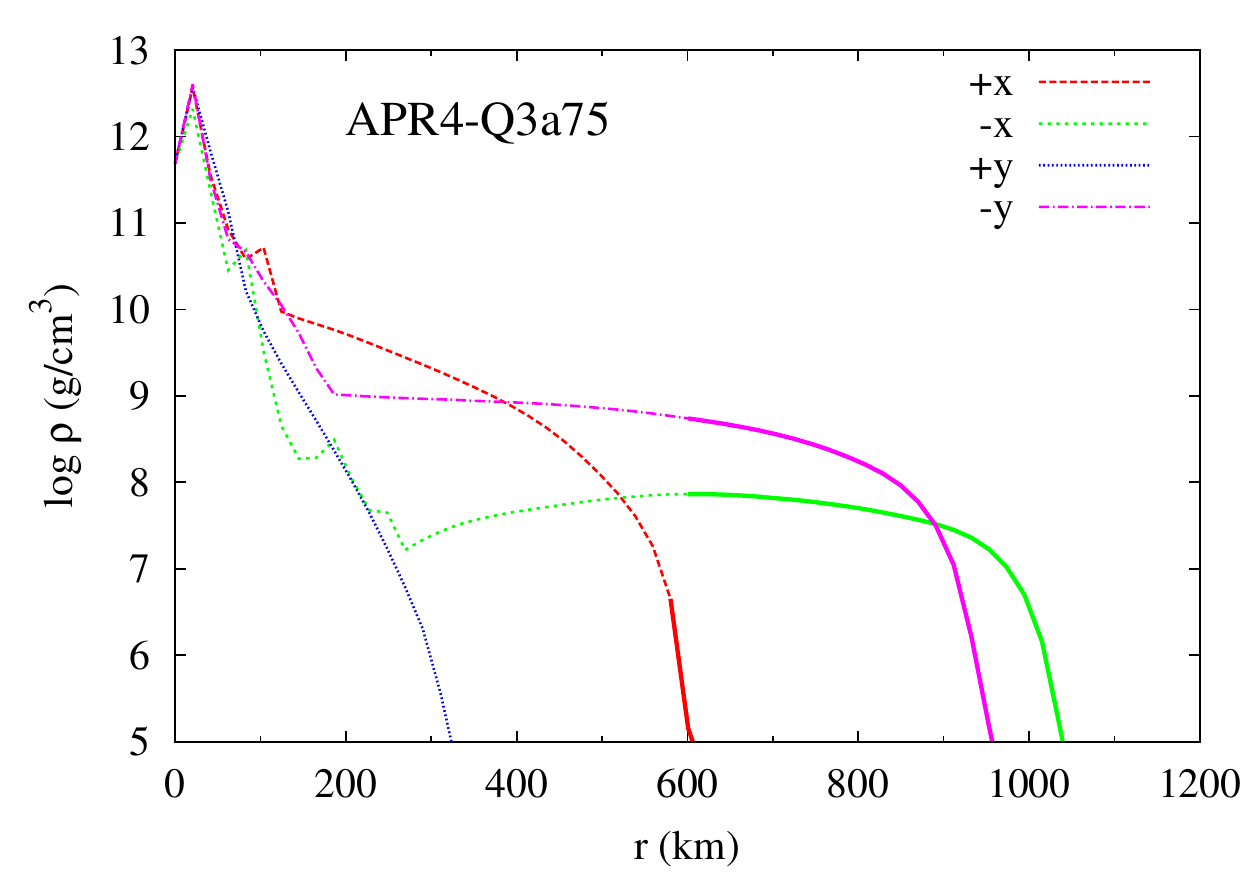} &
  \includegraphics[width=.45\linewidth]{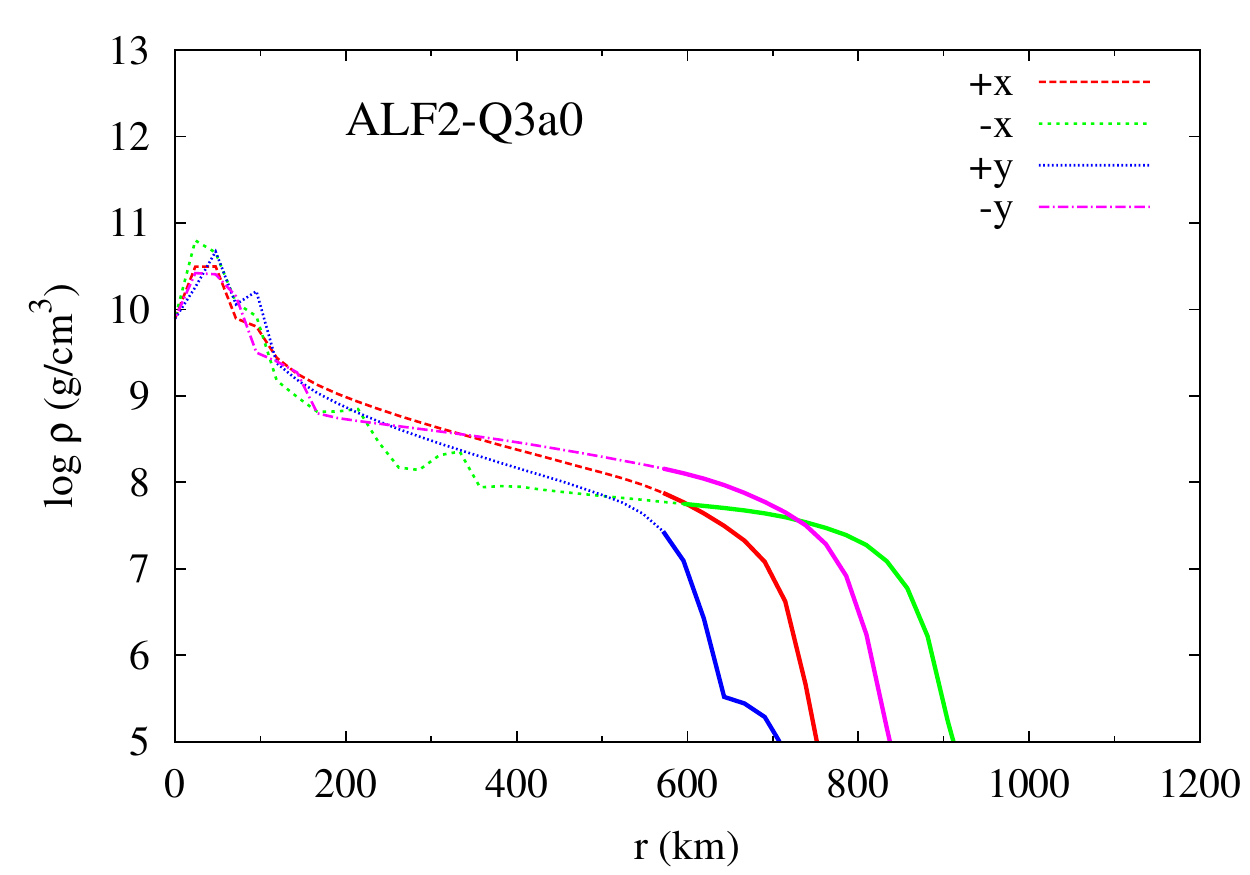} \\
  \includegraphics[width=.45\linewidth]{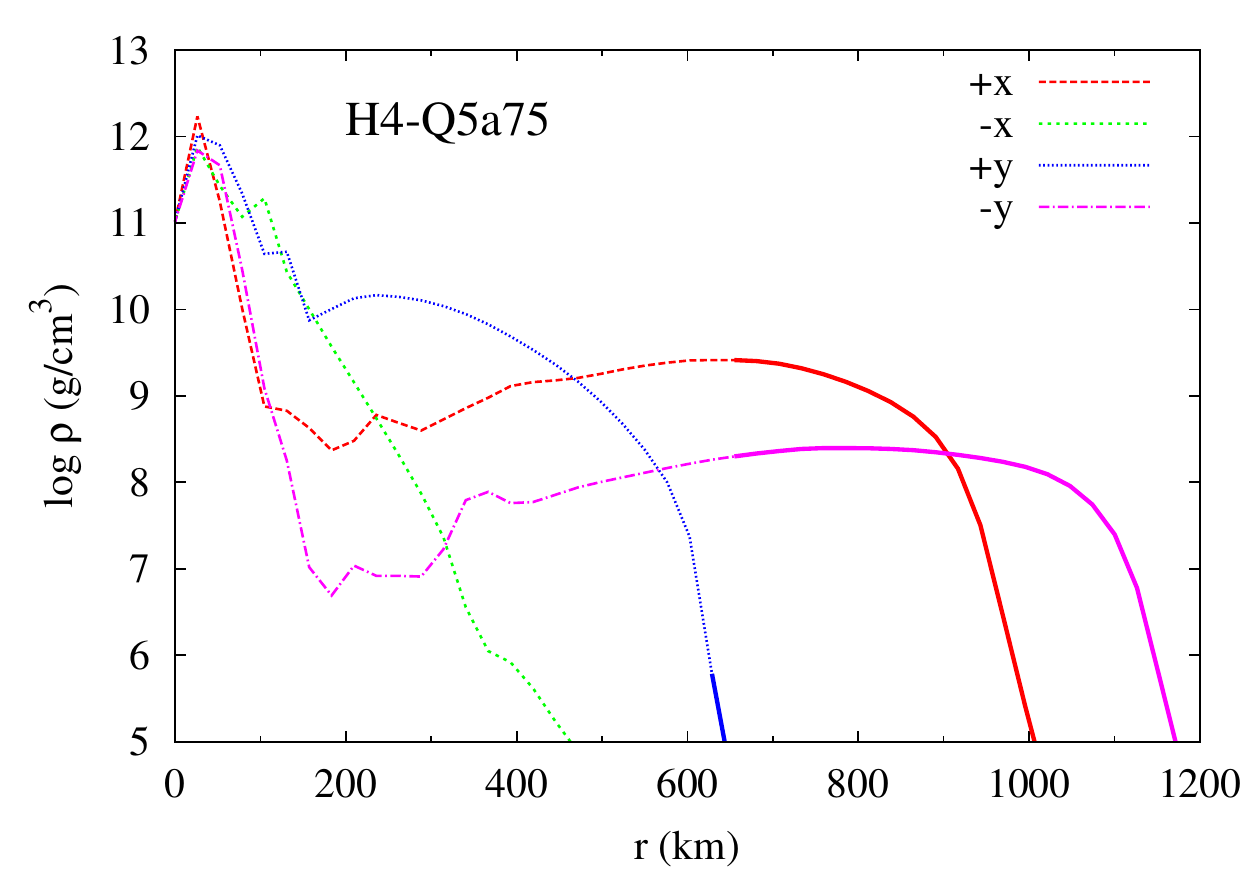} &
  \includegraphics[width=.45\linewidth]{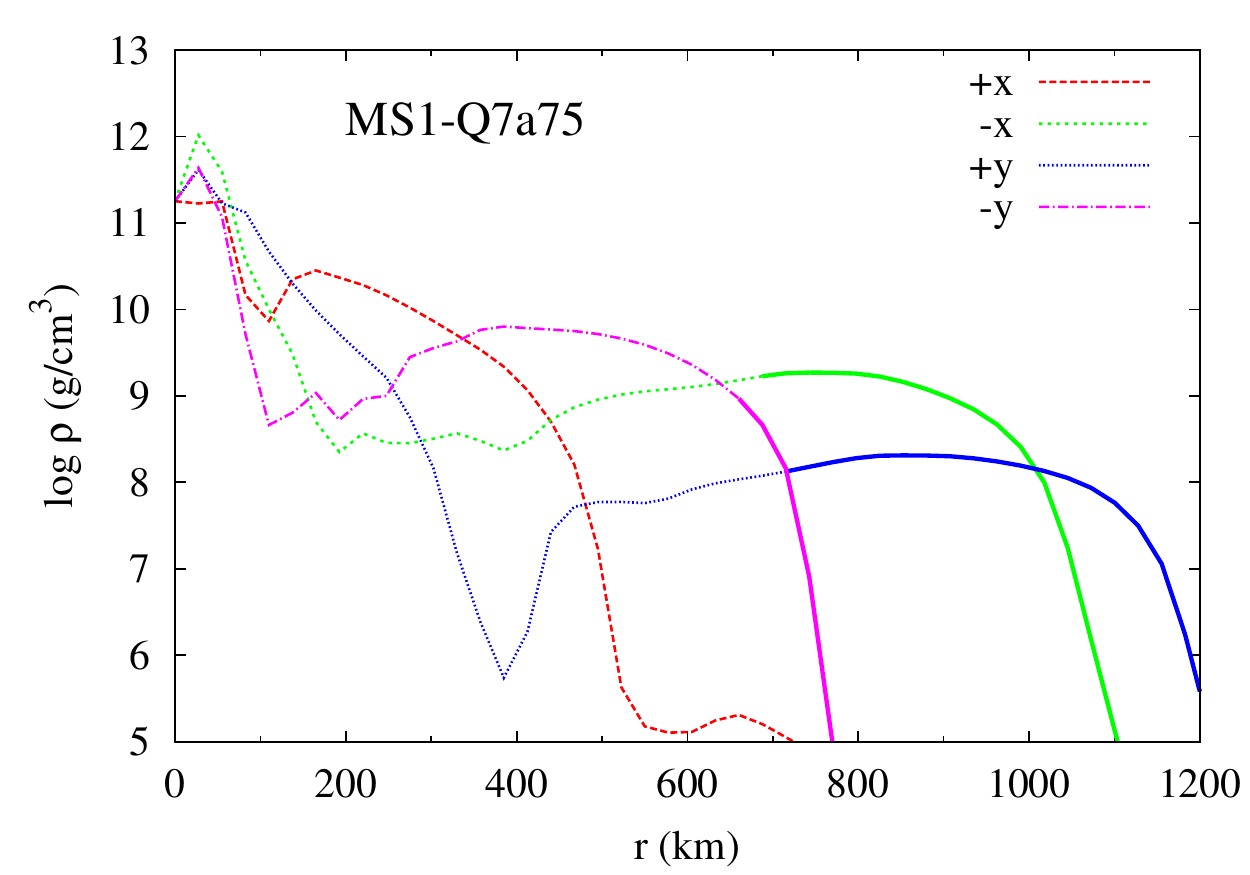}
 \end{tabular}
 \caption{Rest-mass density in the logarithmic scale along the $x$ and
 $y$ axes at \SI{10}{\ms} after the onset of merger for the selected
 models. Positive and negative directions are plotted separately. Solid
 and dashed portions of each curve denote unbound and bound material,
 respectively. Corresponding plots in Fig.~\ref{fig:snap_eqvar} are the
 third row and first column for the top left (APR4-Q3a75), first row and
 second column for the top right (ALF2-Q3a0), fourth row and third
 column for the bottom left (H4-Q5a75), and fifth row and fourth column
 for the bottom right (MS1-Q7a75).} \label{fig:profeqrho}
\end{figure*}

Figure \ref{fig:profeqrho} shows density profiles along the $x$ and $y$
axes at \SI{10}{\ms} after the onset of merger for selected
models. Corresponding snapshots are given in Fig.~\ref{fig:snap_eqvar}.
The material at $r \lesssim$ \SI{100}{\km} is in an approximately
axisymmetric state for all the models. This implies that accretion disks
are formed in the central regions at this time. For a given value of
$M_{r>r_\mathrm{AH}}$, the rest-mass density in the disk region is
higher when $Q$ is smaller. The reason for this is that characteristic
length scales are proportional to the total mass of the system, and thus
to $Q+1$. Accordingly, characteristic rest-mass density should be
proportional to $(Q+1)^{-2}$ for a given value of
$M_{r>r_\mathrm{AH}}$. This tendency was already reported in
Ref.~\cite{kyutoku_ost2011}.

Density profiles outside the disk region depend significantly on the
azimuthal angle. On one hand, the rest-mass density steeply decreases
along directions with no ejecta. In Fig.~\ref{fig:profeqrho}, the $+y$
direction of APR4-Q3a75 and $-x$ direction of H4-Q5a75 fall into this
category. The $+x$ direction of MS1-Q7a75 also corresponds to this case,
but a high-density region is still observed up to $\approx
\SI{500}{\km}$, because the tidal tail has not fallen back and collided
with itself yet in this direction. On the other hand, approximately
constant density plateaus extend up to $\sim \SI{1000}{\km}$ along
directions that the ejecta sweep. For example, the $-x$ and $-y$
directions of APR4-Q3a75 exhibit sudden changes of the structure at
$\approx \SI{200}{\km}$ from a steep decline to plateaus. Similar
situations are also found in the $+x$ and $-y$ directions of H4-Q5a75
and $-x$ and $+y$ directions of MS1-Q7a75, except for pronounced
low-density regions between disk regions and plateaus. These gaps are
more prominent for systems with a larger neutron-star radius at a fixed
time (i.e., \SI{10}{\ms}) from the onset of merger and eventually
disappear as tidal tails fall back. When material spreads in a nearly
axisymmetric manner with $\varphi_\mathrm{ej} \gtrsim 2 \pi$,
plateaulike profiles are observed in all the directions like
ALF2-Q3a0. In any case, the plateaus change to rapidly decaying profiles
at their outer edges.

The ejecta as an unbound portion is smoothly connected to a bound
portion in the plateau regions. When the ejecta mass is large, the
ejecta tends to occupy a large fraction of plateau material,
particularly along a direction with the fastest expansion. The
highest-density direction always disagrees with the fastest-expanding
direction, in which the rest-mass density is typically lower by an order
of magnitude at a given radius than the highest. For example, the
rest-mass density of the ejecta is the highest in the $-y$ direction for
APR4-Q3a75, whereas the fastest direction is the $-x$ direction. This is
because low-density material is ejected from the outer part of neutron
stars prior to the high-density material from the inner part during mass
ejection driven by the tidal torque. The ejecta of ALF2-Q3a0 is more
axisymmetric than those of the other models, and a bump at $\approx
\SI{650}{\km}$ in the $+y$ direction reflects the rear-end collision of
the tidal tail with $\varphi_\mathrm{ej} > 2 \pi$. Note that the spatial
distribution of the dynamical ejecta is different from that for binary
neutron star mergers, where a moderately steep power law with the index
$\approx -3.5$ is observed \cite{nagakura_hssi2014}.

\begin{figure}
 \begin{tabular}{c}
  \includegraphics[width=.95\linewidth]{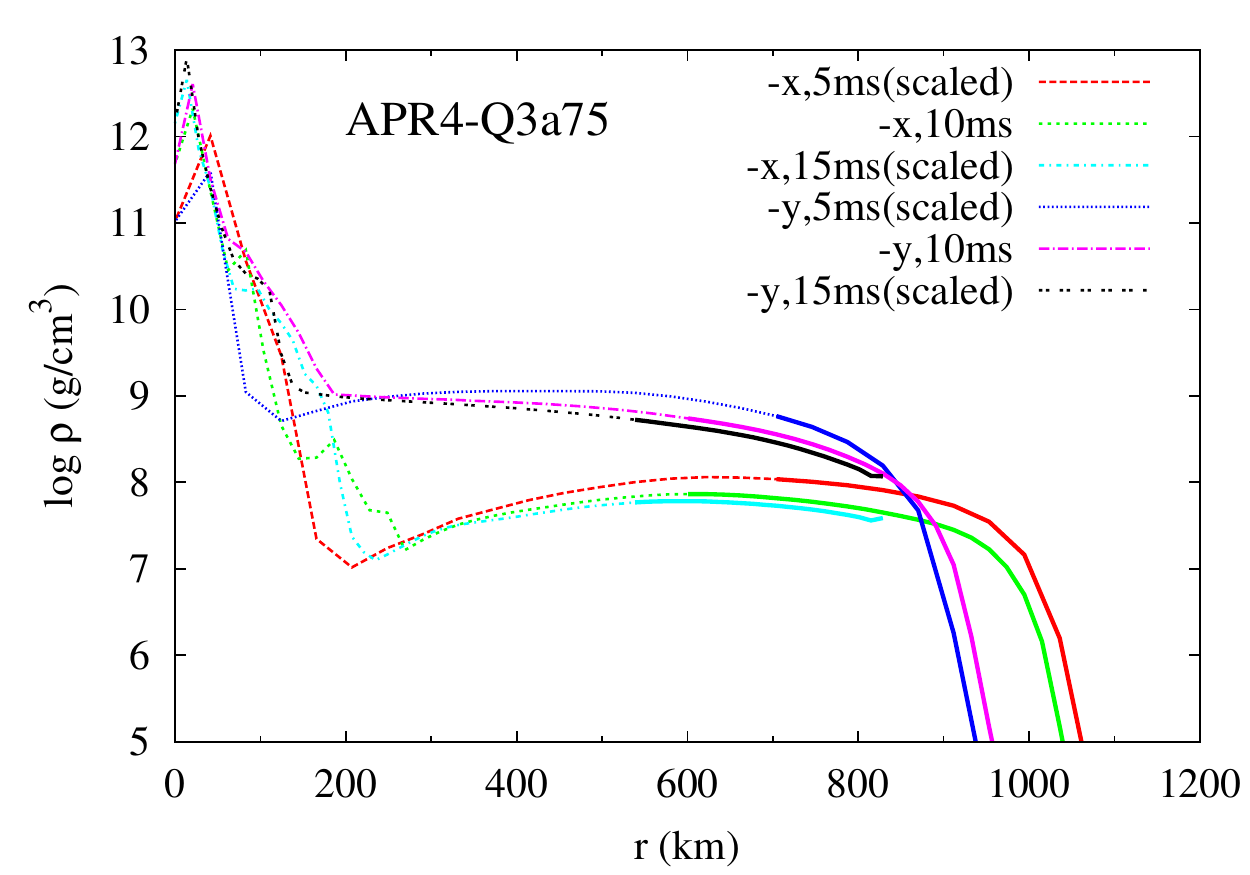} \\
  \includegraphics[width=.95\linewidth]{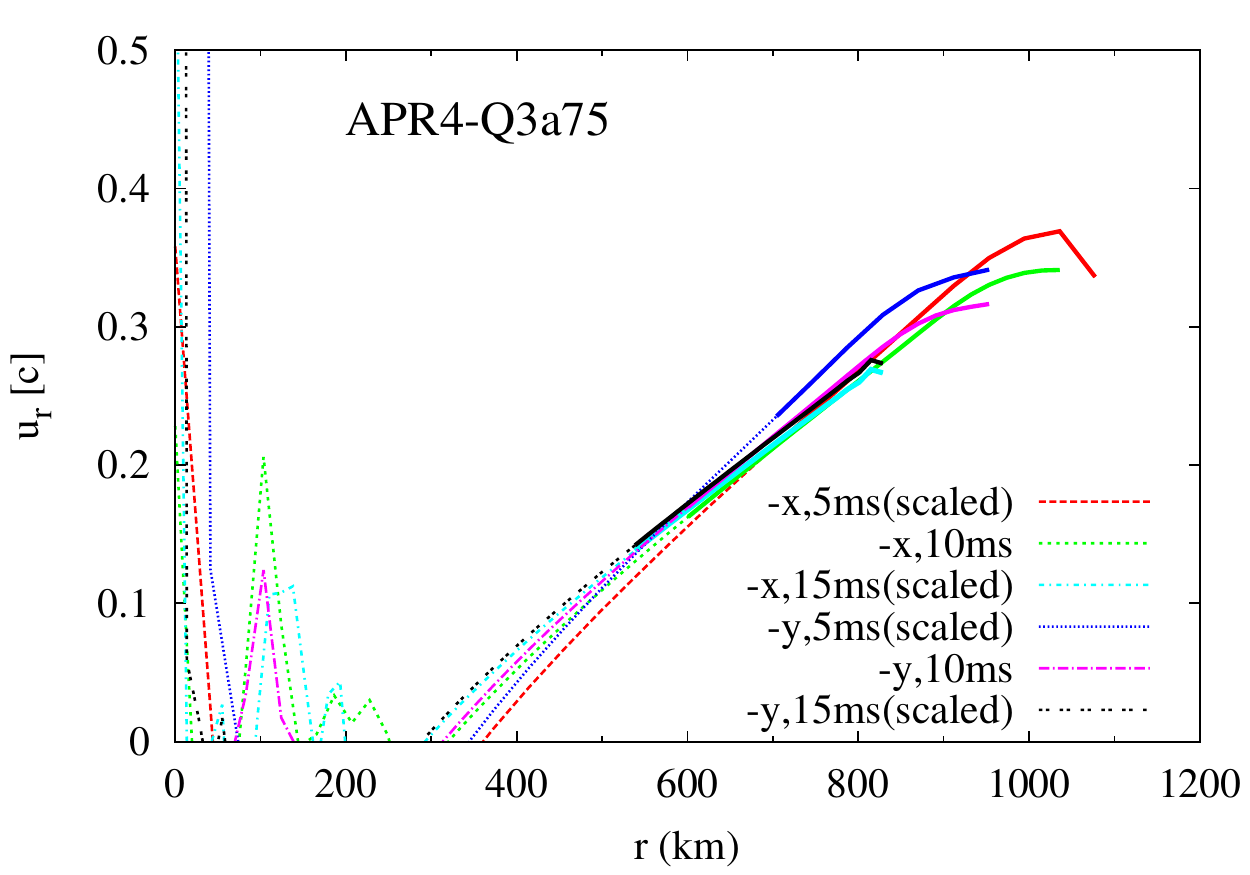}
 \end{tabular}
 \caption{Rest-mass density on a logarithmic scale (top) and radial
 velocity (bottom) along the $-x$ and $-y$ directions at 5, 10, and
 \SI{15}{\ms} after the onset of merger for APR4-Q3a75. Solid and dashed
 portions of each curve denote unbound and bound material,
 respectively. Assuming homologous expansion, the radius is multiplied
 by 2 for the \SI{5}{\ms} profile and divided by 1.5 for the
 \SI{15}{\ms} profile, so that they can be compared directly with that
 at \SI{10}{\ms}. Similarly, the density is divided by $2^3$ for the
 \SI{5}{\ms} profile and multiplied by $(1.5)^3$ for the \SI{15}{\ms}
 profile. The velocity is not scaled and is truncated at $\rho =
 \SI{e5}{\gram\per\cubic\cm}$ to avoid showing artificial atmospheres
 accumulated near ejecta surfaces. Truncation of profiles at $\approx
 \SI{850}{\km}$ for \SI{15}{\ms} is due to escape of material from the
 second largest domain, outside which hydrodynamic evolution equations
 are not solved.} \label{fig:scaleeqrho}
\end{figure}

The ejecta evolves in an approximately homologous manner. That is, the
velocity of each fluid element is kept approximately constant, and its
position and density evolve according to the free-expansion law,
\begin{equation}
 r \propto t \; , \; \rho \propto t^{-3} . \label{eq:homexp}
\end{equation}
Figure \ref{fig:scaleeqrho} shows rest-mass density and velocity
profiles at 5, 10, and \SI{15}{\ms} after the onset of merger in the
$-x$ and $-y$ directions of APR4-Q3a75. In these plots, the radius and
rest-mass density are scaled according to Eq.~\eqref{eq:homexp} so that
those at 5 and \SI{15}{\ms} can be compared directly to those at
\SI{10}{\ms}. Both the density and velocity profiles overlap
approximately among different time slices after the scaling, and the
agreement is particularly good between 10 and \SI{15}{\ms}. These facts
imply that homologous expansion is achieved at the late phase. We also
observe approximate homologous expansion for other models, but the
deviation is slightly more severe for a larger value of $Q$ at a fixed
time (i.e., \SI{10}{\ms}) due probably to stronger residual
gravitational binding.

\subsubsection{Polar direction} \label{sec:res_struc_z}

\begin{figure}
 \begin{tabular}{c}
  \includegraphics[width=.95\linewidth]{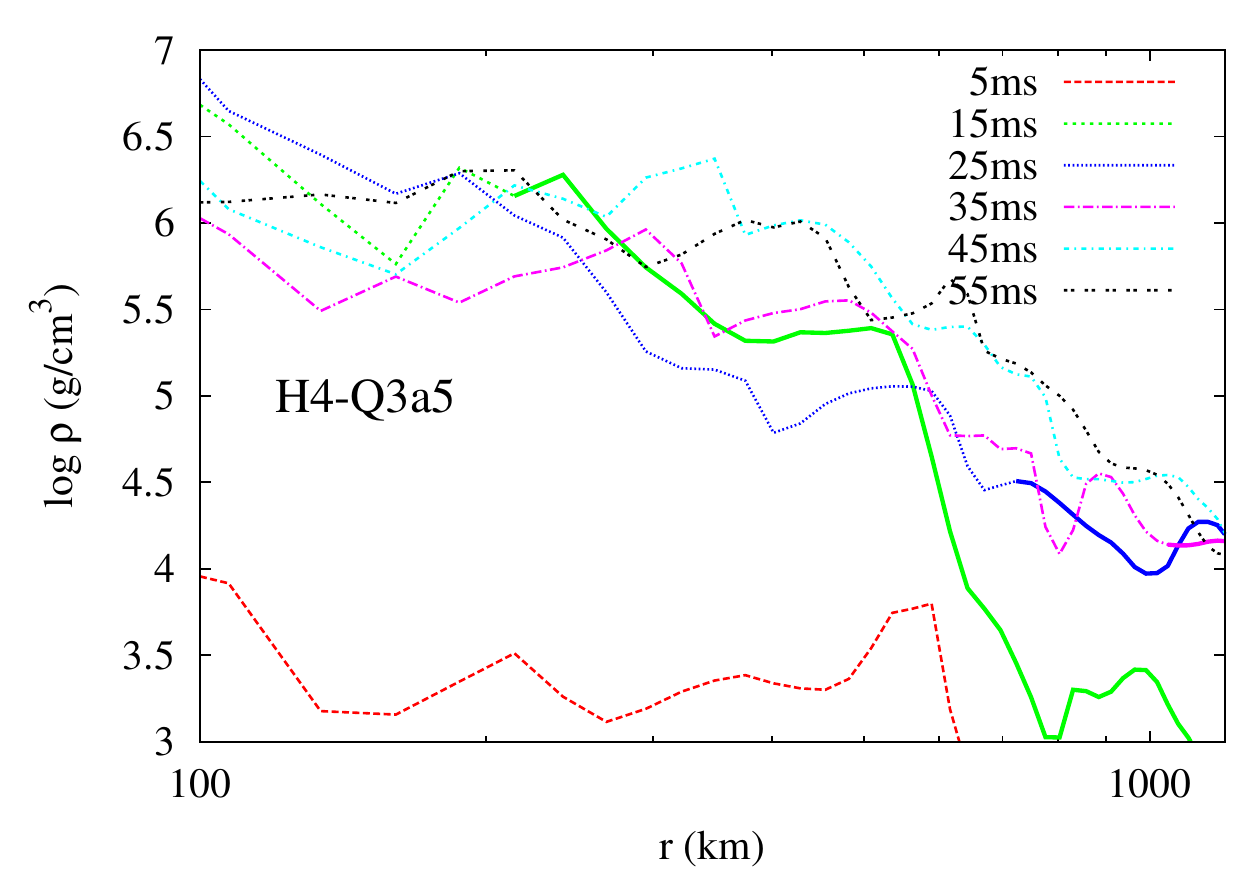} \\
  \includegraphics[width=.95\linewidth]{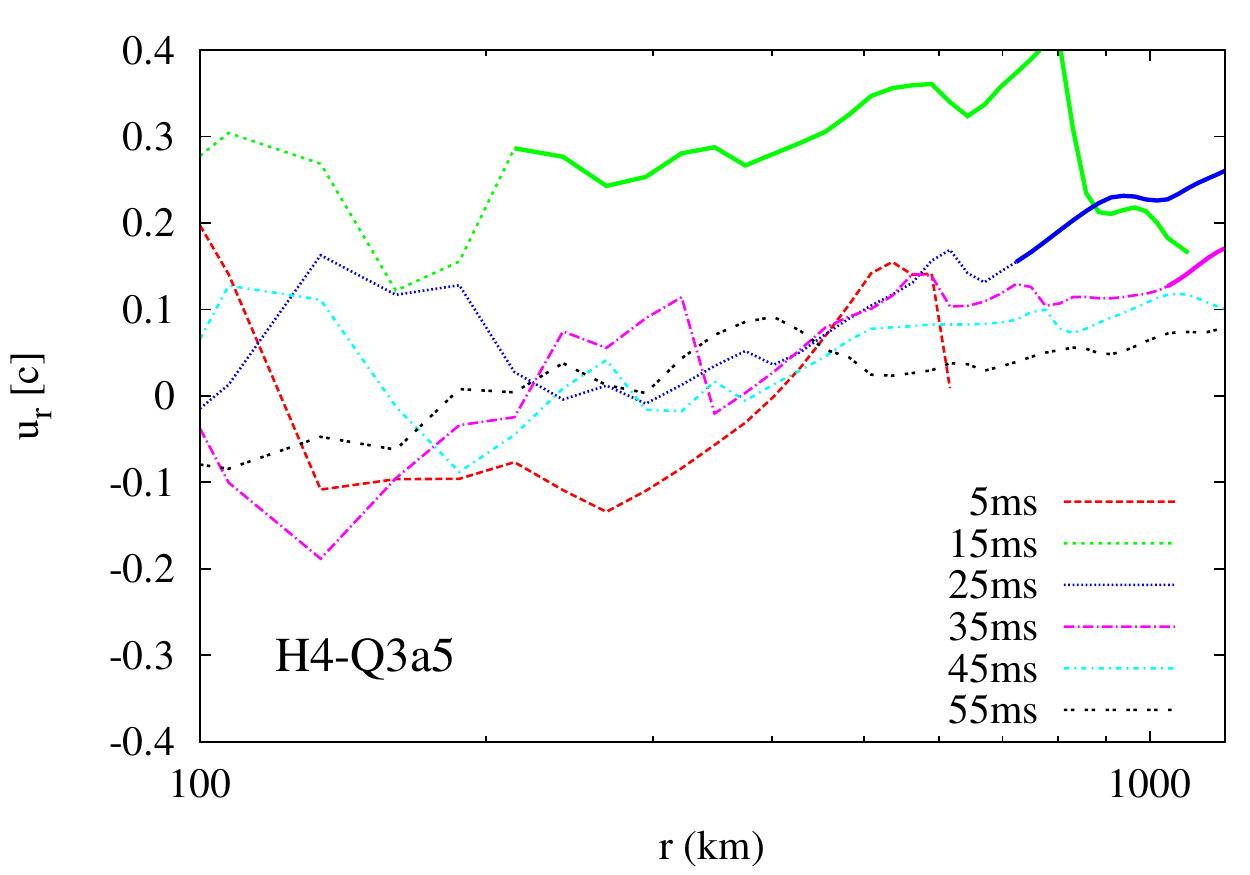}
 \end{tabular}
 \caption{Rest-mass density in the logarithmic scale (top) and radial
 velocity (bottom) along the $z$ axis at several time slices for
 H4-Q3a5. Distances are also shown in the logarithmic scale. The time is
 given as that after the onset of merger, $t_\mathrm{merge}$. Solid and
 dashed portions of each curve denote unbound and bound material,
 respectively. The velocity of material with $\rho \le
 \SI{e3}{\gram\per\cm\cubic}$ is not shown.} \label{fig:profzq3}
\end{figure}

Figure \ref{fig:profzq3} shows rest-mass density and velocity profiles
along the $z$ axis for H4-Q3a5. Because our purpose is to study the
formation of an envelope, profiles at several time slices are shown
together without scalings. At \SI{5}{\ms} after the onset of merger, no
unbound material is found, and the rest-mass density is very low
everywhere. This is because the tidal torque does not eject material
toward the polar region. Material is pushed significantly toward the
polar region only after the shock heating in the disk region sets
in. This is reflected in the increase of the rest-mass density for $t -
t_\mathrm{merge} \gtrsim \SI{10}{\ms}$. Unbound material is ejected from
the disk with $v \approx 0.3c$ in the beginning and is beyond a radius
of \SI{1000}{\km} by $\approx \SI{35}{\ms}$ for this particular model.

A long-lived envelope is formed following the shock-driven disk
outflow. The velocity of envelope material is smaller than the typical
ejecta velocity, and in particular, the radial velocity of bound
material falls below $0.1 c$ at \SI{55}{\ms}. This suggests that the
envelope is in an approximately stationary state at this time. Indeed,
the rest-mass density profiles do not change very much from 25 to
\SI{55}{\ms}. The profile may be approximated by a power law, $\rho
\propto r^{-p_\mathrm{env}}$, with its index $p_\mathrm{env} \approx
2$--3. The magnitude of the rest-mass density implies that the total
mass of the envelope formed after the merger of H4-Q3a5 is much smaller
than that formed after binary neutron star mergers
\cite{hotokezaka_kkosst2013,nagakura_hssi2014}. This could be
advantageous for a hypothetical jet to overcome a baryon loading
problem, but it will not be easy to obtain a collimated jet in the
absence of a heavy envelope. Firm conclusions to the jet propagation
require an extensive study of disk winds.

\begin{figure}
 \begin{tabular}{c}
  \includegraphics[width=.95\linewidth]{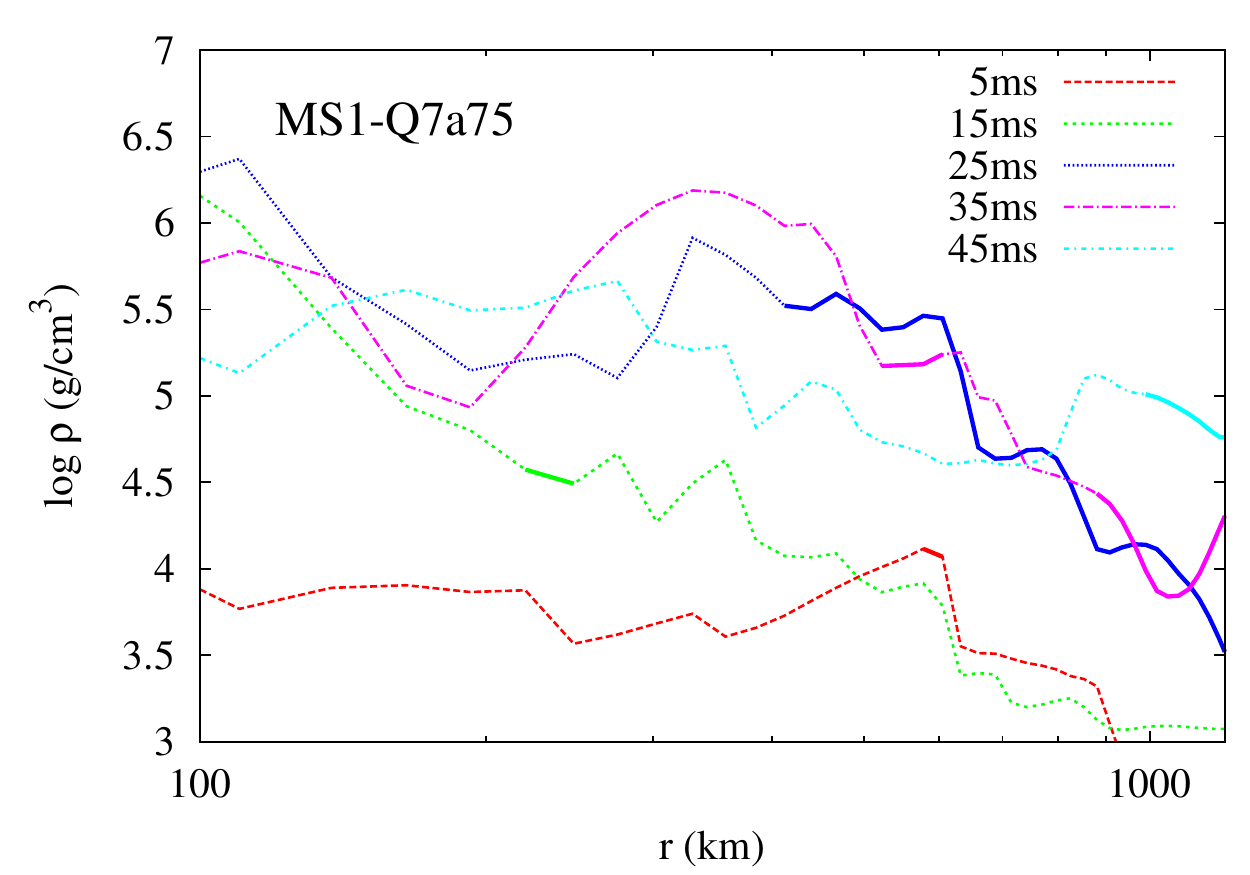} \\
  \includegraphics[width=.95\linewidth]{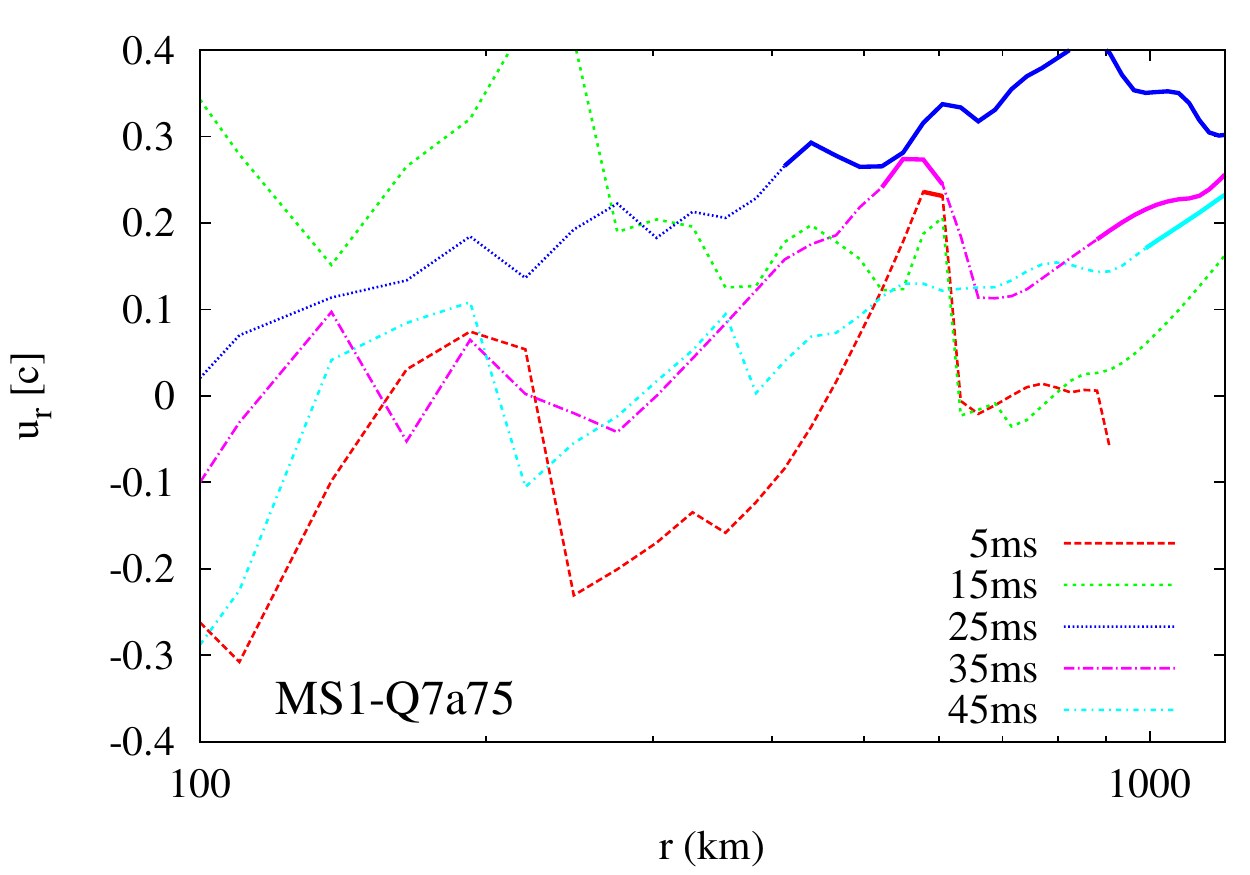}
 \end{tabular}
 \caption{The same as Fig.~\ref{fig:profzq3} but for MS1-Q7a75.}
 \label{fig:profzq7}
\end{figure}

It takes a long time for the remnant of a high-mass-ratio binary merger
to develop a long-lived envelope in the polar region. Figure
\ref{fig:profzq7} shows rest-mass density and velocity profiles along
the $z$ axis for MS1-Q7a75. In this model, the ejecta generated by the
disk are beyond \SI{1000}{\km} only for $t - t_\mathrm{merge} \gtrsim
\SI{45}{\ms}$, and material behind it exhibits significantly more time
variability than that of H4-Q3a5. The velocity profile with $v \gtrsim
0.1c$ also indicates significant time variability. It can, however,
still be seen that the rest-mass density of the envelope is comparable
to that of H4-Q3a5 (Fig.~\ref{fig:profzq3}). Thus, we may safely
conclude that the mass of the envelope formed after the merger is much
smaller for black hole-neutron star binaries than for binary neutron
stars unless (or possibly even if) binary parameters are extreme as far
as the dynamical processes are concerned.

\subsubsection{Velocity distribution} \label{sec:res_struc_v}

\begin{figure}
 \includegraphics[width=.95\linewidth]{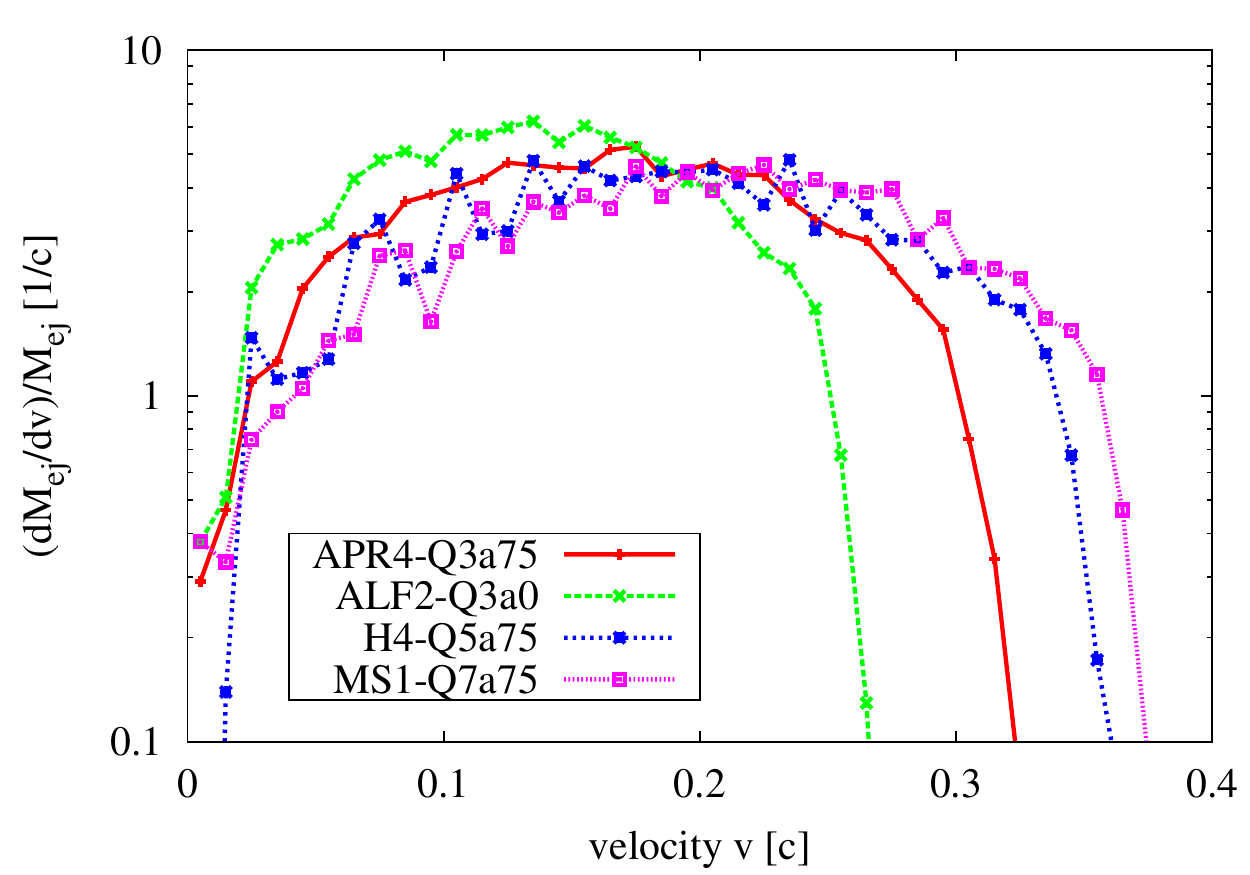} \caption{Velocity
 distribution normalized by the ejecta mass measured at \SI{10}{\ms}
 after the onset of merger for selected models. The velocity is defined
 as $\sqrt{1 - 1/(- u_t)^2}$. We use $dM_\mathrm{ej} / dv$ rather than
 its integration over a finite velocity interval, because the former
 quantity is independent of binning.}  \label{fig:veldistrib}
\end{figure}

Figure \ref{fig:veldistrib} shows the velocity distributions of
dynamical ejecta normalized by the ejecta mass, $M_\mathrm{ej}$,
measured at \SI{10}{\ms} after the onset of merger for selected
models. Namely, integrating each distribution over the velocity returns
unity. They are derived by analyzing unbound material on the equatorial
plane as described in Sec.~\ref{sec:setup_diag_ej}, and we checked that
estimation at different time slices gives very similar results.

All the models exhibit a relatively flat distribution with a cutoff at
low and high velocities rather than, say, a power-law distribution. This
agrees semiquantitatively with previous results obtained in Newtonian
simulations \cite{rosswog_pn2013}. This distribution implies that the
density structure of ejecta can be approximated by $\rho \propto v^{-2}
\propto r^{-2}$ within the range between lower and higher cutoff
velocities, because the free-expansion law, Eq.~\eqref{eq:homexp}, gives
$dM/dv \propto \rho v^2$. This observation is largely consistent with
the spatial profile shown in Fig.~\ref{fig:profeqrho}.

The velocity distribution is shifted toward larger velocities when the
ejecta mass is larger (see the top panel of Fig.~\ref{fig:evol} for
visual comparisons). We also find that the distribution tends to be
shifted toward larger velocities when the mass ratio, $Q$, is
larger. This is consistent with the observations of $M_\mathrm{ej} /
M_\mathrm{bd}$ and $v_\mathrm{ave}$ in Sec.~\ref{sec:res_char_dep},
where the dynamical ejecta from a higher-mass-ratio binary is seen to be
energetic. Previous numerical-relativity simulations also found this
tendency \cite{foucart_etal2014}.

\subsection{Fallback} \label{sec:res_fb}

\begin{figure}
 \includegraphics[width=.95\linewidth]{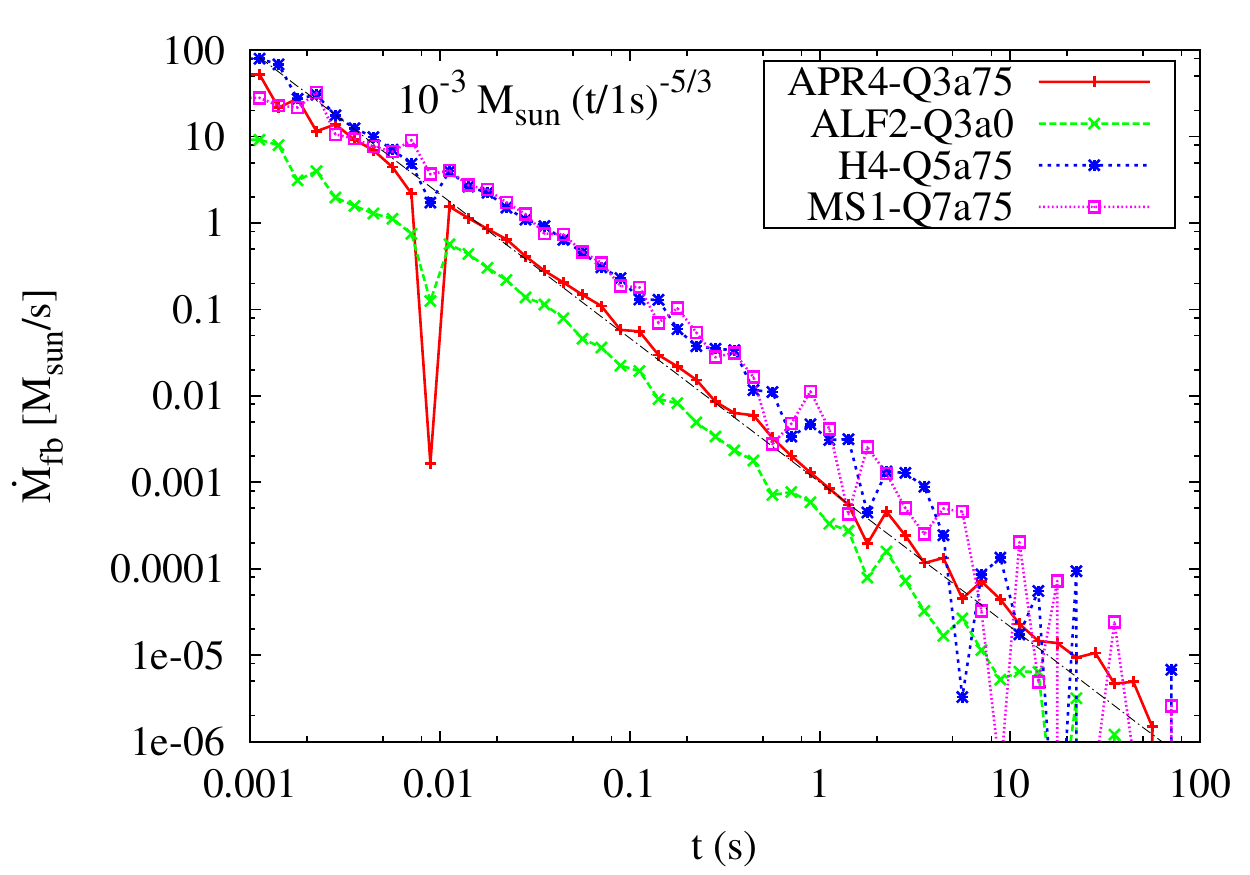} \caption{Fallback rate
 measured by analyzing matter profiles at \SI{10}{\ms} after the onset
 of merger for selected models. A power law $ \num{e-3}
 M_\odot~\si{\per\second} ( t / \SI{1}{\second} )^{-5/3}$ is also
 plotted (black dot-dashed line) as an eye guide. Apparent deviation
 from the power law at $t \gtrsim$ \SI{1}{\second} is ascribed to the
 limited number of grid data, and the power law is recovered if we
 compute $\dot{M}_\mathrm{fb}$ using a wide time interval.}
 \label{fig:fallback}
\end{figure}

The fallback rate as a function of time is found to obey a power law
with the index $-5/3$ irrespective of the models. Figure
\ref{fig:fallback} shows fallback rates determined by the method
described in Sec.~\ref{sec:setup_diag_fb} analyzing matter profiles at
\SI{10}{\ms} after the onset of merger for the selected models. Aside
from statistical fluctuations due to the limited number of grid data,
overall behavior is consistent with the structureless power law,
$\dot{M}_\mathrm{fb} \propto t^{-5/3}$, and no significant time
evolution is found when we compute $\dot{M}_\mathrm{fb}$ at different
time slices. This power-law fallback rate is known to be achieved after
the tidal disruption of main sequence stars by supermassive black holes
\cite{rees1988,phinney1989}. The same power law is found for black
hole-neutron star binaries in Newtonian simulations
\cite{rosswog2007,rosswog_pn2013} and is also reported in a
numerical-relativity simulation for a single binary model with the
$\Gamma = 2$ polytrope \cite{chawla_abllmn2010}. Our results confirm
their findings for a wide range of binary parameters in numerical
relativity. Nuclear interaction neglected in this study may not be
important, because Newtonian studies show that nucleosynthesis in the
nuclear statistical equilibrium does not modify the power-law behavior
\cite{rosswog2007,rosswog_pn2013} and \textit{r}-process heating can
modify it only on rare occasions \cite{metzger_aqm2010}.

This power-law behavior implies that the mass spectrum with respect to
specific energy takes a constant profile, i.e., $dM_\mathrm{fb} /
d\tilde{E} = \mathrm{const}$. The usual reasoning behind the power-law
index $-5/3$ is the combination of $dM_\mathrm{fb} / dP_\mathrm{fb} = (
dM_\mathrm{fb} / d\tilde{E} ) ( d\tilde{E} / dP_\mathrm{fb} )$, the
Keplerian relation $P_\mathrm{fb} \propto a_\mathrm{fb}^{3/2} \propto (
- \tilde{E} )^{-3/2}$, and the assumption that $dM_\mathrm{fb} /
d\tilde{E}$ is constant. The first and second relations are
universal. The third assumption is verified for the tidal disruption of
main sequence stars by various hydrodynamic simulations (e.g.,
Ref.~\cite{evans_kochanek1989}) and is pointed out to be more
appropriate for a stiffer polytrope due to stronger shock interaction
\cite{lodato_kp2009}. Because the neutron-star self-gravity cannot be
neglected and shocks do not appear to play a significant role in energy
redistribution for a neutron star disrupted by a stellar-mass black
hole, the reason for constant energy distribution is nontrivial and may
be worth future investigation.

Although the overall magnitude of the power law is not computed very
accurately by our approximate estimation method, we may safely conclude
that the fallback rates span $\dot{M}_\mathrm{fb} \sim
\num{e-4}$--$\num{e-2} M_\odot~\si{\per\second} ( t / \SI{1}{\second}
)^{-5/3}$ when substantial mass ejection occurs. Because the periapsis
distance of the fallback material is found to agree approximately with
the radius at which the neutron star is disrupted, the material will
join the accretion disk before reaching the periapsis. Thus, the
black-hole accretion rate and electromagnetic luminosity could be
smaller than the fallback rate (see
Refs.~\cite{rossi_begelman2009,lee_rl2009} for relevant discussions).

In this analysis, the center of mass is always assumed to be located at
the coordinate origin. This is not justified in a rigorous manner,
because the remnant black hole-disk system acquires a substantial
velocity of $O(100)$ \si{\km\per\second} by two mechanisms. One is
backreaction from the anisotropic mass ejection
\cite{rosswog_dtp2000,kyutoku_is2013}, and the other is recoil due to
the anisotropic gravitational-wave emission \cite{fitchett1983}. We will
describe the former and latter in Secs.~\ref{sec:res_diskbh} and
\ref{sec:res_gw}, respectively.

\subsection{Remnant disk and black hole} \label{sec:res_diskbh}

Because remnant disks and black holes are thoroughly investigated in
previous work \cite{shibata_taniguchi2011}, we describe their properties
only briefly. The amount of mass outside the apparent horizon,
$M_{r>r_\mathrm{AH}}$, is shown in Table \ref{table:result} and is
discussed in Sec.~\ref{sec:res_char_dep}. Typical accretion time scales
due to purely hydrodynamic processes are estimated to be
30--\SI{300}{\ms} when measured at $\approx \SI{10}{\ms}$ after the
onset of merger irrespective of the models. We do not go into details of
accretion dynamics, expecting that realistic behavior will be determined
by unincorporated physics like neutrino processes and
magnetohydrodynamics.

\begin{figure*}
 \includegraphics[width=.95\linewidth]{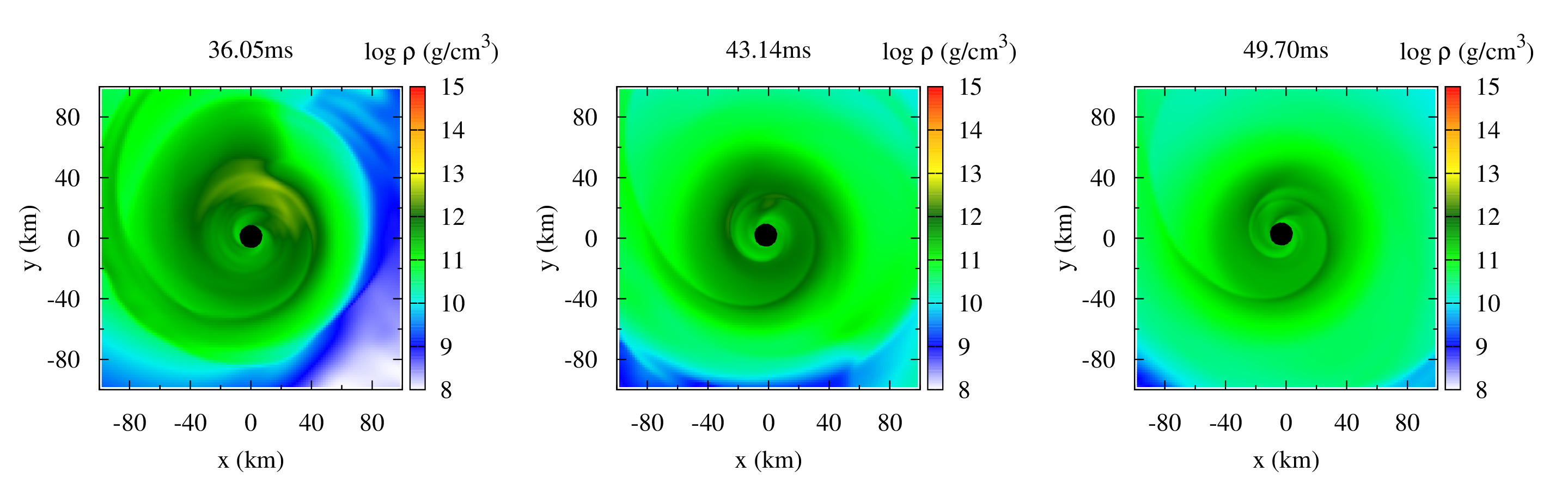} \caption{The same as
 Fig.~\ref{fig:snap_merger} (H4-Q5a75) but at different time slices.}
 \label{fig:snap_disk}
\end{figure*}

One feature of remnant disks overlooked in our previous studies is the
existence of standing spiral accretion shocks. Figure
\ref{fig:snap_disk} shows rest-mass density profiles on the equatorial
plane in the central region at different time slices for H4-Q5a75. This
figure (see also the right panel of Fig.~\ref{fig:snap_merger}) shows
that sharp spirals extending to the apparent horizon stay in
approximately the same location over \SI{10}{\ms} without exhibiting any
rotation. Similar structures are found for most of the models as long as
the remnant disk is appreciable, and we find no cases in which this
spiral structure disappears by the end of simulations, which is at
$\gtrsim \SI{50}{\ms}$ after the onset of merger for the longest
runs. The standing spiral shocks appear to be formed as a trace of
self-collision of tidal tails rather than as a result of disk
instability. This spiral structure should serve to dissipate the angular
momentum of disk material and enhance mass accretion by the remnant
black hole.

\begin{table*}
 \caption{Characteristic physical quantities associated with the remnant
 black hole measured at \SI{10}{\ms} after the merger and with
 gravitational waves for our fiducial, $N=60$ runs. $M_\mathrm{BH,f}$
 and $\chi_\mathrm{f}$ are the mass and dimensionless spin parameter,
 respectively, of the remnant black hole. $V_\mathrm{ej}$ and
 $V_\mathrm{GW}$ are the magnitude of velocities imparted to the remnant
 black hole-disk system due to the ejecta backreaction and
 gravitational-wave recoil, respectively. $\Delta E_\mathrm{GW}$ and
 $\Delta J_\mathrm{GW}$ are the energy and angular momentum,
 respectively, radiated by gravitational waves.}
 \begin{tabular}{c|ccc|ccc} \hline
  ~Model~ & ~$M_\mathrm{BH,f} [M_\odot]$~ & ~$\chi_\mathrm{f}$~ &
  ~$V_\mathrm{ej}$ (\si{\km\per\second})~ & ~$V_\mathrm{GW}$
  (\si{\km\per\second})~ & ~$\Delta E_\mathrm{GW} [M_\odot]$~ & ~$\Delta
  J_\mathrm{GW} [M_\odot^2]$~ \\
  \hline \hline
  APR4-Q3a75 & 5.07 & 0.87 & 100 & 90 & \num{9.3e-2} & 5.6 \\
  ALF2-Q3a75 & 5.02 & 0.86 & 500 & 60 & \num{6.3e-2} & 4.6 \\
  H4-Q3a75 & 4.99 & 0.88 & 500 & 60 & \num{4.9e-2} & 4.0 \\
  MS1-Q3a75 & 4.97 & 0.88 & 800 & 20 & \num{4.0e-2} & 3.5 \\
  \hline
  APR4-Q3a5 & 5.17 & 0.77 & 20 & 70 & \num{9.2e-2} & 5.3 \\
  ALF2-Q3a5 & 5.10 & 0.76 & 300 & 70 & \num{6.1e-2} & 4.3 \\
  H4-Q3a5 & 5.07 & 0.76 & 300 & 50 & \num{4.8e-2} & 3.7 \\
  MS1-Q3a5 & 5.05 & 0.75 & 600 & 50 & \num{3.9e-2} & 3.3 \\
  \hline
  APR4-Q3a0 & 5.26 & 0.55 & $<1$ & 60 & \num{8.2e-2} & 4.3 \\
  ALF2-Q3a0 & 5.26 & 0.56 & 20 & 30 & \num{6.1e-2} & 3.8 \\
  H4-Q3a0 & 5.20 & 0.55 & 70 & 40 & \num{4.6e-2} & 3.2 \\
  MS1-Q3a0 & 5.16 & 0.53 & 200 & 40 & \num{3.6e-2} & 2.8 \\
  \hline
  APR4-Q5a75 & 7.80 & 0.85 & 30 & 20 & 0.16 & 10 \\
  ALF2-Q5a75 & 7.69 & 0.83 & 400 & 40 & 0.11 & 9.1 \\
  H4-Q5a75 & 7.65 & 0.83 & 400 & 70 & \num{9.0e-2} & 8.0 \\
  MS1-Q5a75 & 7.62 & 0.83 & 700 & 50 & \num{7.5e-2} & 7.2 \\
  \hline
  APR4-Q5a5 & 7.90 & 0.71 & $<1$ & 30 & 0.13 & 8.8 \\
  ALF2-Q5a5 & 7.89 & 0.71 & 30 & 30 & 0.11 & 8.3 \\
  H4-Q5a5 & 7.81 & 0.70 & 200 & 50 & \num{8.9e-2} & 7.4 \\
  MS1-Q5a5 & 7.74 & 0.68 & 400 & 50 & \num{7.2e-2} & 6.7 \\
  \hline
  APR4-Q7a75 & 10.5 & 0.83 & $<1$ & 40 & 0.17 & 14 \\
  ALF2-Q7a75 & 10.5 & 0.83 & 40 & 30 & 0.16 & 13 \\
  H4-Q7a75 & 10.4 & 0.82 & 200 & 40 & 0.13 & 12 \\
  MS1-Q7a75 & 10.3 & 0.81 & 400 & 30 & 0.11 & 11 \\
  \hline
  APR4-Q7a5 & 10.6 & 0.67 & $<1$ & 30 & 0.12 & 10 \\
  ALF2-Q7a5 & 10.6 & 0.67 & $<1$ & 30 & 0.12 & 11 \\
  H4-Q7a5 & 10.6 & 0.67 & 6 & 20 & 0.11 & 10 \\
  MS1-Q7a5 & 10.6 & 0.67 & 30 & 20 & 0.10 & 9.9 \\
  \hline
 \end{tabular}
 \label{table:gresult}
\end{table*}

The mass and dimensionless spin parameter of the remnant black holes at
\SI{10}{\ms} after the onset of merger are listed in Table
\ref{table:gresult}. They are consistent with our previous results for
models with comparable binary parameters
\cite{kyutoku_st2010,kyutoku_st2010e,kyutoku_ost2011}. After the
measurement, the dimensionless spin parameters increase by up to
$\approx 0.01$ due to long-term accretion depending on the models. Thus,
the values of $\chi_\mathrm{f}$ shown in Table \ref{table:gresult} may
be regarded as the lower limits of hypothetical final configurations
which will be achieved by purely hydrodynamic processes.

The remnant black hole-disk system including fallback material receives
a recoil velocity due to the backreaction of anisotropic mass ejection
\cite{rosswog_dtp2000,kyutoku_is2013}, which we call the ejecta kick
velocity. The ejecta kick velocity, $V_\mathrm{ej}$, is estimated by
linear-momentum conservation as
\begin{align}
 V_\mathrm{ej} & \approx \frac{P_\mathrm{ej}}{m_0} \notag \\
 & = \SI{555}{\km\per\second} \; \left( \frac{P_\mathrm{ej}}{0.01
 M_\odot} \right) \left( \frac{Q+1}{4} \right)^{-1} \label{eq:ejkick} ,
\end{align}
where $M_\mathrm{NS} = 1.35 M_\odot$ is assumed. For simplicity, the
mass of the remnant black hole-disk system is approximated by $m_0$ in
this expression, neglecting energy loss to the ejecta and to
gravitational waves. The former is $\lesssim 0.02 m_0$, and the latter
is $\lesssim 0.03 m_0$ for the cases considered here, where the energy
radiated during the very early inspiral phase that existed before our
initial condition, $m_0 - M_0$, is also taken into account. Because the
ejecta mass is large only when tidal disruption occurs at a distant
orbit and gravitational radiation is not very strong, the sum of both
does not exceed $0.03 m_0$.

Values of the ejecta kick velocity for each model are presented in Table
\ref{table:gresult}. This table shows that $V_\mathrm{ej}$ can be
several hundreds of \si{\km\per\second} when mass ejection is efficient
and easily dominates kick velocities due to the gravitational radiation
reaction, $V_\mathrm{GW}$, which we discuss in Sec.~\ref{sec:res_gw}.

\subsection{Gravitational waves} \label{sec:res_gw}

Gravitational waves from black hole-neutron star binaries are thoroughly
investigated in our previous work
\cite{kyutoku_st2010,kyutoku_st2010e,kyutoku_ost2011}, and derived
waveforms are used to construct phenomenological models aiming at data
analysis \cite{lackey_ksbf2012,pannarale_bks2013,lackey_ksbf2014}. In
the following, we instead discuss integrated or instantaneous properties
of gravitational waves.

The energy, linear momentum, and angular momentum carried away by
gravitational waves are presented in Table \ref{table:gresult}. While
the energy $\Delta E_\mathrm{GW}$ and the angular momentum $\Delta
J_\mathrm{GW}$ are presented as they are, the magnitude of linear
momentum $\Delta P_\mathrm{GW}$ is shown instead as the velocity
imparted to the remnant black hole-disk system including fallback
material,
\begin{equation}
 V_\mathrm{GW} \approx \frac{\Delta P_\mathrm{GW}}{m_0} ,
\end{equation}
where we adopt $m_0$ as in Eq.~\eqref{eq:ejkick}. We call
$V_\mathrm{GW}$ the gravitational-wave kick velocity. Although the
accuracy in computing $\Delta P_\mathrm{GW}$ is not very high due to
mode couplings, we do not find $V_\mathrm{GW}$ larger than
\SI{100}{\km\per\second} for the models considered in this
study. Broadly speaking, the ejecta kick velocity, $V_\mathrm{ej}$,
dominates the gravitational-wave kick velocity, $V_\mathrm{GW}$, when
$M_\mathrm{ej} \gtrsim 0.01 M_\odot$.

\begin{figure}
 \includegraphics[width=.95\linewidth]{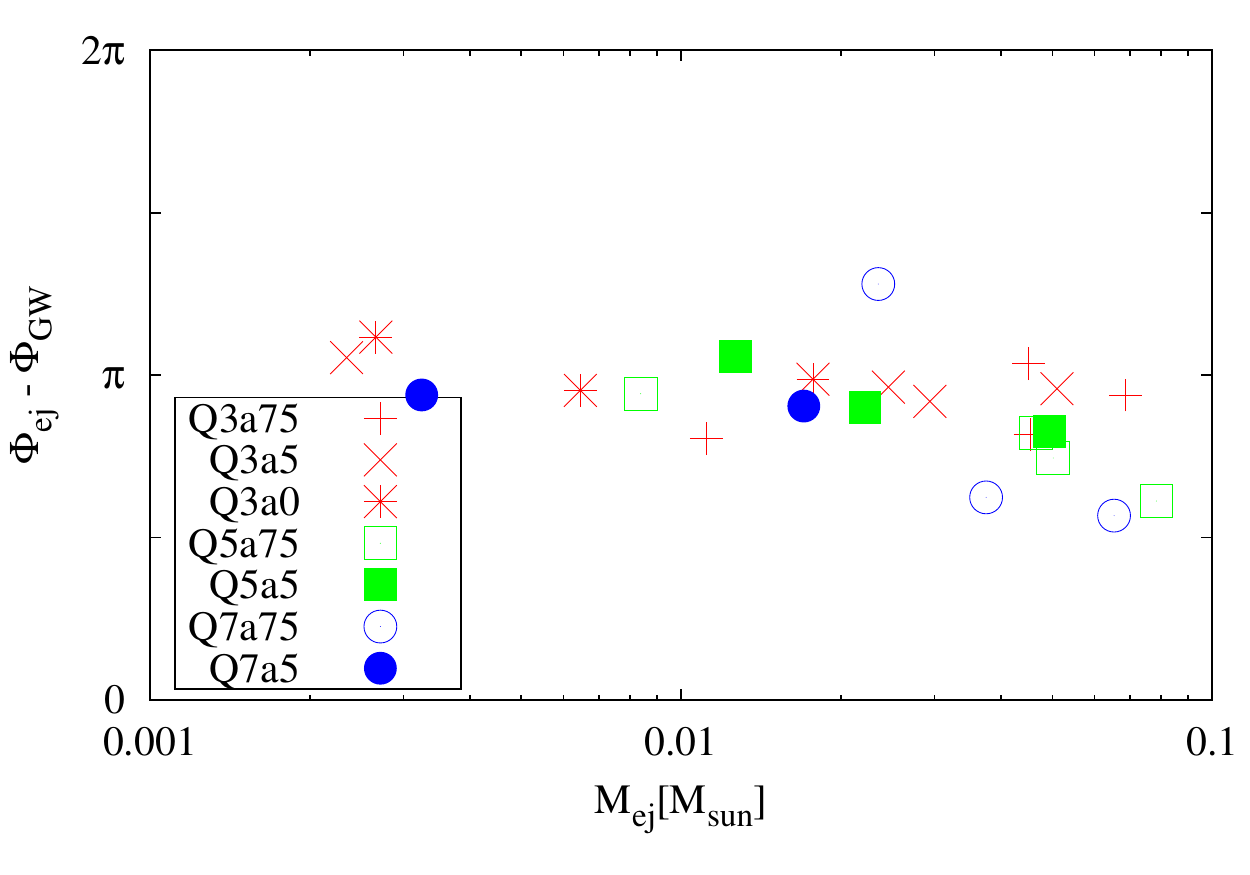} \caption{Difference
 between the angle of ejecta linear momentum, $\Phi_\mathrm{ej}$, and of
 gravitational-wave linear momentum, $\Phi_\mathrm{GW}$, vs the ejecta
 mass, $M_\mathrm{ej}$. We restrict the range to $M_\mathrm{ej} \ge
 0.001 M_\odot$.} \label{fig:angle}
\end{figure}

The ejecta kick velocity and gravitational-wave kick velocity partially
cancel out each other, because their angles $\Phi_\mathrm{ej}$ and
$\Phi_\mathrm{GW}$ point in approximately opposite directions. Figure
\ref{fig:angle} shows the difference between them, $\Phi_\mathrm{ej} -
\Phi_\mathrm{GW}$, vs $M_\mathrm{ej}$. The differences cluster around
$\pi$ irrespective of the model parameters, and this means that ejecta
and gravitational waves carry linear momenta in opposite
directions. This tendency does not depend on grid resolutions. While the
origin of anticorrelation is nontrivial, it is reasonable that
$\Phi_\mathrm{ej} - \Phi_\mathrm{GW}$ prefers a specific value, because
both dynamical mass ejection and linear-momentum emission are determined
primarily by merger dynamics including tidal disruption. The largest
velocity in the coalescence event is achieved by the plunge motion of
material promptly swallowed by the black hole after the tidal
disruption, and the plunge should emit the linear momentum efficiently
in its direction due to the large velocity (see
Refs.~\cite{wiseman1992,blanchet_qw2005} for relevant discussions). A
possible explanation of the anticorrelation between $\Phi_\mathrm{ej}$
and $\Phi_\mathrm{GW}$ is that the linear momentum is emitted right
after the tidal disruption primarily in the direction of the plunge
motion, which should be opposite to the ejecta motion. This
anticorrelation implies that the realistic value of the remnant velocity
is given approximately by $| V_\mathrm{ej} - V_\mathrm{GW} |$.

\begin{figure}
 \includegraphics[width=.95\linewidth]{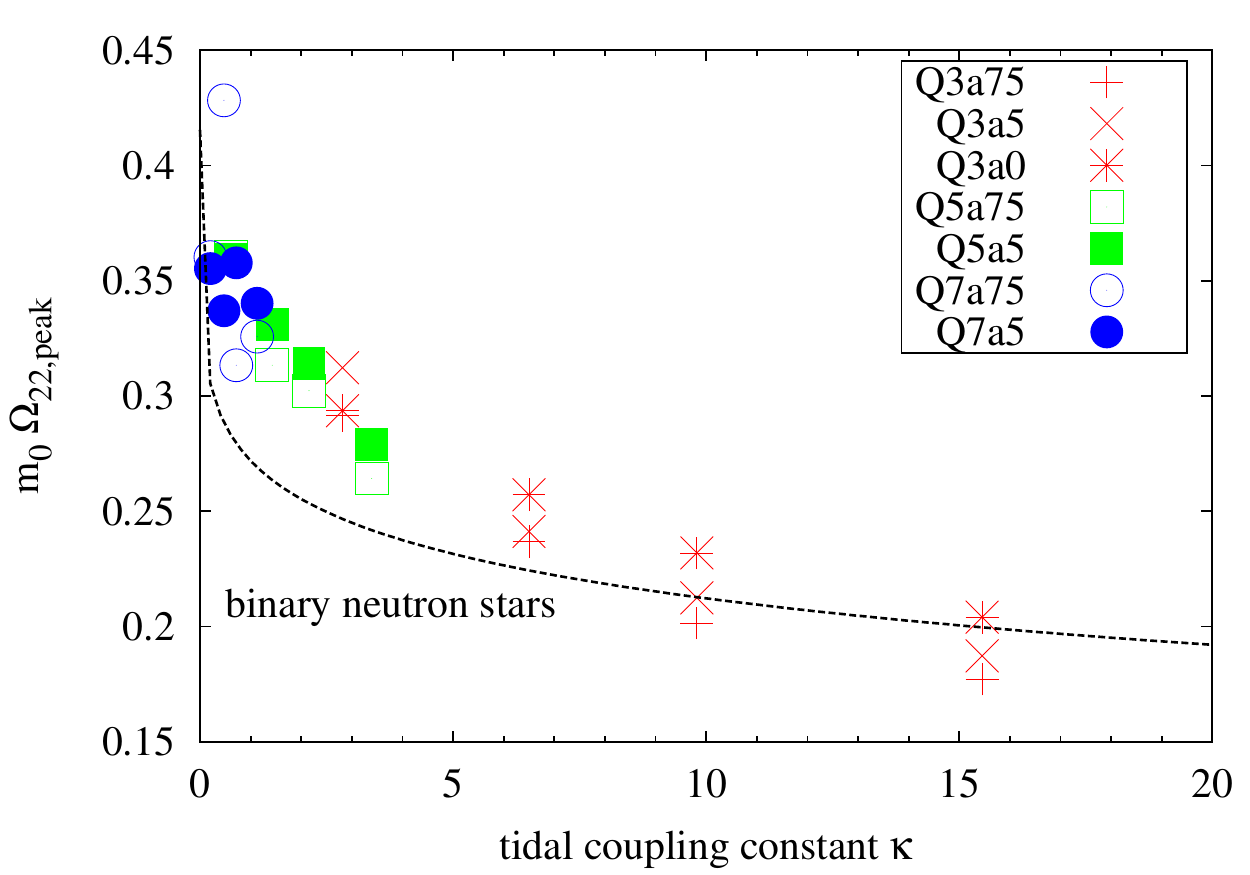} \caption{Dimensionless
 gravitational-wave frequency of $(2,2)$ mode gravitational waves, $m_0
 \Omega_{22,\mathrm{peak}}$, at the amplitude peak as a function of the
 tidal coupling constant $\kappa$ defined by
 Eq.~\eqref{eq:coupling}. The dashed line is a fit obtained from binary
 neutron star simulations due to Ref.~\cite{read_bcfgkmrst2013}.}
 \label{fig:gw}
\end{figure}

Finally, we comment on the possible existence of a tight correlation
between the strength of tidal effects and gravitational-wave frequency
at the maximum amplitude, which is suggested to exist for binary neutron
stars \cite{read_bcfgkmrst2013,bernuzzi_etal2014}. Figure \ref{fig:gw}
shows a dimensionless gravitational-wave frequency of the $(2,2)$ mode
at the maximum amplitude, $m_0 \Omega_{22,\mathrm{peak}}$, as a function
of a tidal coupling constant,
\begin{equation}
 \kappa = \frac{2Q}{(1+Q)^5} \frac{k}{\mathcal{C}^5} =
  \frac{3Q}{(1+Q)^5} \Lambda , \label{eq:coupling}
\end{equation}
adapted to black hole-neutron star binaries \cite{damour_nagar2010} (see
also Ref.~\cite{vines_fh2011}). This figure suggests the existence of
relations independent of the mass ratio and equation of state. If the
correlations are tight, it implies that the finite-size effect in the
black hole-neutron star binary merger is described fairly well by the
quadrupolar tidal deformability up to tidal disruption. These relations
depend on the black-hole spin, and our results suggest that $m_0
\Omega_{22,\mathrm{peak}}$ is smaller for a larger black-hole spin. This
agrees with Ref.~\cite{bernuzzi_etal2014}. The same value of $\chi$ may
not be compared directly among different mass ratios except for
nonspinning cases, and effective spin parameters weighted by the mass
ratio such as $\chi [ 1 + 3/(4Q) ] Q^2 / (1+Q)^2$ (see, e.g.,
Refs.~\cite{kidder_ww1993,damour2001}) will be more appropriate. If such
correlations are confirmed accurately by future simulations, they would
help to extract neutron-star equations of state without detailed
analysis of the phase evolution, just as knowledge of the cutoff
frequency would do
\cite{vallisneri2000,shibata_kyt2009,shibata_kyt2009e,kyutoku_st2010,kyutoku_st2010e,kyutoku_ost2011}.

Because the gravitational-wave amplitude peaks during the rapid increase
of the frequency, the error of $m_0 \Omega_{22,\mathrm{peak}}$ is not
very small. Typical errors are estimated to be $\approx 5\%$ due to
eccentricities,\footnote{This is estimated as twice the eccentricity in
the inspiral phase, because it is difficult to isolate the eccentricity
contribution during the merger phase.} $\approx 10\%$ due to the finite
resolution, and $\approx 5\%$ due to the gravitational-wave extraction
method like extraction radii (even with the extrapolation). Hence, the
total error may be $\approx 20\%$ in the worst case.

A relation satisfied by nonspinning black hole-neutron star binaries (if
really exists) is not necessarily the same as that by binary neutron
stars, because the merger dynamics is very different. We include a
fitting curve derived from binary neutron star simulations
\cite{read_bcfgkmrst2013} in Fig.~\ref{fig:gw}. We cannot determine
whether the relations are different or not from the current data by two
reasons. One is the numerical error associated with each simulation. The
other is the fact that tidal coupling constants, $\kappa$, spanned by
binary neutron star simulations are much larger than those by black
hole-neutron star binary simulations, and thus the extrapolated
relations cannot be seriously trusted. Specifically, the relation
derived in Ref.~\cite{read_bcfgkmrst2013} is obtained by fitting results
of simulations with $26 \le \kappa \le 440$, none of which overlaps with
that in our current simulations. It may be worth future investigation to
test whether relations are distinct between binary neutron stars and
nonspinning black hole-neutron star binaries.

\section{Electromagnetic counterpart} \label{sec:cp}

In this section, we discuss expected characteristics of electromagnetic
counterparts based on the properties of dynamical ejecta derived by our
simulations. We focus primarily on the effect of anisotropy, which is
characterized by the opening angle in the equatorial plane,
$\varphi_\mathrm{ej}$, and in the meridional plane,
$\theta_\mathrm{ej}$, on the macronova/kilonova
\cite{li_paczynski1998,kulkarni2005,metzger_mdqaktnpz2010,kisaka_it2015}
and synchrotron radio emission \cite{nakar_piran2011,takami_ki2014}. A
concise summary of the main results derived in this section is found in
Ref.~\cite{kyutoku_is2013}, in which other aspects of the ejecta like
gravitational-wave memory emission and cosmic-ray acceleration are also
discussed.

For simplicity, we adopt slightly different notations for ejecta
quantities in this section from those in other sections. Specifically,
we denote the ejecta mass by $M$ instead of $M_\mathrm{ej}$. The opening
angles are denoted by $\theta$ and $\varphi$ instead of
$\theta_\mathrm{ej}$ and $\varphi_\mathrm{ej}$, respectively. Recall
that $\theta ( = \theta_\mathrm{ej} )$ is defined as the half-opening
angle taking the equatorial symmetry into account, and a full sphere
corresponds to $\theta = \pi / 2$ and $\varphi = 2 \pi$. We also adopt
short-hand notations $M_{-2} \equiv M / ( 0.01 M_\odot )$, $\theta_i
\equiv \theta / (1/5)$, and $\varphi_i \equiv \varphi / \pi$. We recover
the speed of light, $c$, everywhere.

\subsection{Macronova/kilonova} \label{sec:cp_mk}

The macronova/kilonova is quasithermal radiation from the ejecta heated
by the decay of unstable \textit{r}-process elements. The dynamical
ejecta from mergers of black hole-neutron star binaries will be composed
primarily of neutrons as discussed in Sec.~\ref{sec:res_ej_case}, and
then \textit{r}-process elements should be synthesized
\cite{lattimer_schramm1974,lattimer_schramm1976}. After the neutrons are
exhausted within a few seconds, $\beta$-decay and fission of unstable
\textit{r}-process elements heat the ejecta.\footnote{Some of the energy
liberated in the $\beta$-decay does not contribute to the heating
because of the energy deposited to neutrinos and $\gamma$-ray photons
\cite{metzger_mdqaktnpz2010}. The latter does not escape freely in the
early stage of the ejecta evolution and contribute to the ejecta
heatup.}  The heated ejecta emit radiation primarily in red-optical and
infrared bands on a day-to-month time scale \cite{tanaka_hkwkss2014},
where a bunch of Doppler-broadened lines associated with the complicated
energy-level structure of \textit{r}-process elements blanket the
emission in blue-optical and ultraviolet bands
\cite{kasen_bb2013,tanaka_hotokezaka2013}.

\subsubsection{Analytic model} \label{sec:cp_mk_model}

Qualitative features of the macronova/kilonova from the anisotropic
ejecta can be understood by modifying the prototypical model for
spherical ejecta proposed in Ref.~\cite{li_paczynski1998}. In this
section, we introduce short-hand notations $V_{-1} \equiv V / ( 0.1 c )$
for the surface velocity, $\kappa_1 \equiv \kappa / (
\SI{10}{\per\gram\cm\squared} )$ for the opacity, and $f_{-6} \equiv f /
\num{e-6}$ for the heating efficiency. The precise meaning of these
quantities is explained in the following.

We approximate the hydrodynamic evolution of the ejecta by the free
expansion of a uniform-density truncated sphere characterized by the
opening angles $\theta$ and $\varphi$. The radius of the (truncated)
sphere is given by $R (t) = Vt$ using the surface velocity $V$, and thus
the rest-mass density of the ejecta is
\begin{equation}
 \rho (t) = \frac{3 \pi M}{4 \theta \varphi V^3 t^3} .
\end{equation}
In this uniform-density free-expansion model, the surface velocity is
related to the average velocity of the ejecta defined by
Eq.~\eqref{eq:vave} via $V = \sqrt{5/3} v_\mathrm{ave} \approx 1.3
v_\mathrm{ave}$. Because the ejecta material is expected to be
radiation-dominated in the relevant epoch due to \textit{r}-process
heating, the internal energy density $u$ is related to the pressure $P$
and temperature $T$ by $u = 3P = aT^4$, where $a$ is the radiation
constant. The time evolution of the internal energy density is derived
by the first law of thermodynamics as
\begin{equation}
 t^3 \frac{du}{dt} + 4 t^2 u = \frac{3 \pi M}{4 \theta \varphi V^3}
  \dot{\varepsilon} - \frac{3 \pi}{4 \theta \varphi V^3} L ,
\end{equation}
where $\dot{\varepsilon}$ is the specific heating rate and $L$ is the
luminosity. Time-dependent quantities are $u$, $\dot{\varepsilon}$, and
$L$.

We assume that the specific heating rate is given by a power law
$\dot{\varepsilon} (t) = f c^2 / t$ parametrized by the heating
efficiency $f$ in the same manner as the spherical model
\cite{li_paczynski1998}. An appropriate value of the heating efficiency,
$f$, will depend significantly on the electron fraction (the number of
electrons per baryon) \cite{rosswog_katp2014,wanajo_snkks2014}. The
uncertainty is particularly high when fission is an important heating
source rather than $\beta$-decay of elements near the stability line
\cite{wanajo_snkks2014}. In this study, we take the fiducial value of
$f$ to be \num{e-6} following Ref.~\cite{wanajo_snkks2014}.

We give the luminosity by a diffusion approximation in a similar manner
to the spherical model \cite{li_paczynski1998} but assuming geometry
adapted to anisotropic mass ejection. The assumption is that the
radiation is emitted not from the truncated spherical surface but from
the cross section of truncation. In the language of our simulations,
photons from the anisotropic ejecta are assumed to escape mainly into
the $\pm z$ directions, and the emitting surfaces are taken to be those
observed from the $\pm z$ direction like ones depicted in
Fig.~\ref{fig:snap_eqvar}. The temperature gradient $dT/dr$ relevant to
the diffusion flux is approximated by $\approx T / ( \theta R )$ rather
than $\approx T/R$ of the spherical ejecta under this assumption. Thus,
the flux may be given by
\begin{equation}
 F (t) \approx \frac{\sigma_\mathrm{SB} T^4}{\kappa \rho \theta R} ,
  \label{eq:mkflux}
\end{equation}
where $\kappa$ is the opacity and $\sigma_\mathrm{SB}$ is the
Stefan--Boltzmann constant. In this estimation, a factor of order unity
is neglected in exactly the same manner as in
Ref.~\cite{li_paczynski1998}. The emitting area is then given
approximately by $2 \times \varphi R^2 / 2 = \varphi R^2$, where the
first ``2'' stands for two emitting surfaces at $+z$ and $-z$, and
therefore the bolometric luminosity may be given by
\begin{equation}
 L (t) \approx \frac{\varphi^2 V^4 c}{3 \pi \kappa M} t^4 u .
\end{equation}
This expression does not reduce to that for the spherical ejecta even if
we adopt $\theta = \pi / 2$ and $\varphi = 2 \pi$ because of different
assumptions. The neglected truncated spherical surface has the area $( 4
/ \pi ) \theta \varphi R^2$, and thus the luminosity may be
underestimated by a fraction of $( 4 / \pi ) \theta \approx
30\%$. Although this term can be included with no difficulty, we omit
this contribution so that the parameter dependence becomes clear.

The value of opacity, $\kappa$, is highly uncertain due to our
incomplete knowledge of \textit{r}-process elements and their line
features \cite{kasen_bb2013,tanaka_hotokezaka2013}. Although the
realistic opacity of \textit{r}-process elements is safely assumed to be
dominated by various bound-bound transition lines in optical and
ultraviolet wavelengths, no complete line list exists so far. In this
study, we take the fiducial value of $\kappa$ to be
\SI{10}{\per\gram\cm\squared}, because this approximately reproduces
results obtained by radiation transfer simulations performed adopting
currently available line lists \cite{tanaka_hotokezaka2013}. The gray
approximation adopted in this model is not realistic and limits the
predictability of the spectra.

The thermodynamic evolution equation can be solved analytically. For
this purpose, it is convenient to cast the equation into a dimensionless
form. First, we normalize the surface velocity by the speed of light as
$\beta_s \equiv V/c$ in the usual manner. Next, we define a
characteristic time scale by the condition that the optical depth of a
characteristic path becomes unity, $\kappa \rho \theta R = 1$, and this
gives a critical time of the onset of transparency,
\begin{equation}
 t_c = \sqrt{\frac{3 \pi \kappa M}{4 \varphi V^2}} \; .
\end{equation}
Finally, a characteristic internal energy density can be defined by
\begin{equation}
 u_c = f \rho ( t_c ) c^2 = \sqrt{\frac{4 \varphi f^2 c^4}{3 \pi
  \kappa^3 \theta^2 M}} \; .
\end{equation}
Introducing dimensionless variables $\tilde{t} = t / t_c$ and $\tilde{u}
= u / u_c$, we obtain the dimensionless evolution equation,
\begin{equation}
 \frac{d\tilde{u}}{d\tilde{t}} + \left( \frac{4}{\tilde{t}} + \frac{3
				  \pi \tilde{t}}{16 \theta \beta_s}
				 \right) \tilde{u} =
 \frac{1}{\tilde{t}^4} ,
\end{equation}
which has an analytic solution
\begin{equation}
 \tilde{u} (t) = \frac{C}{\tilde{t}^4} \exp \left( - \frac{3 \pi
					     \tilde{t}^2}{32 \theta
					     \beta_s} \right) +
 \sqrt{\frac{32 \theta \beta_s}{3 \pi}} \frac{1}{\tilde{t}^4} Y \left(
								 \sqrt{\frac{3
								 \pi}{32
								 \theta
								 \beta_s}}
								 \tilde{t}
								\right)
 ,
\end{equation}
where $C$ is the integration constant and $Y$ is Dawson's integral
defined by
\begin{equation}
 Y (x) \equiv e^{-x^2} \int_0^x e^{s^2} ds .
\end{equation}
Because the initial internal energy of the ejecta is rapidly lost due to
the adiabatic cooling,\footnote{Further energy injection could modify
the thermodynamic evolution via different $\dot{\varepsilon} (t)$
\cite{kisaka_it2015}.} we may safely set the integration constant, $C$,
to be zero. The key issue which allows us to derive this analytic
solution for nonspherical ejecta is that the temporal dependence of each
term (adiabatic cooling, radiative cooling, and heating) is not affected
by the geometry in our model.

The peak time, peak bolometric luminosity, and effective temperature at
the peak time can be estimated using this solution. Note that Dawson's
integral takes the maximum value $y_p \approx 0.54$ at $x_p \approx
0.92$. The peak time is
\begin{align}
 t_\mathrm{peak} & = x_p \sqrt{\frac{8 \kappa \theta M}{\varphi V c}}
 \notag \\
 & = \SI{11}{day} \; \kappa_1^{1/2} M_{-2}^{1/2} V_{-1}^{-1/2}
 \theta_i^{1/2} \varphi_i^{-1/2} .
\end{align}
The peak bolometric luminosity is
\begin{align}
 L_\mathrm{peak} & = y_p \sqrt{\frac{\varphi f^2 M V c^5}{2 \theta
 \kappa}} \notag \\
 & = \SI{1.8e40}{erg.cm^{-3}} \; f_{-6} \kappa_1^{-1/2} M_{-2}^{1/2}
 V_{-1}^{1/2} \theta_i^{-1/2} \varphi_i^{1/2} .
\end{align}
The effective temperature is defined from the diffusion flux,
Eq.~\eqref{eq:mkflux}, by $T_\mathrm{eff} \equiv ( F /
\sigma_\mathrm{SB} )^{1/4}$, and its value at the peak time is
\begin{align}
 T_\mathrm{peak} & = \frac{y_p^{1/4}}{x_p^{1/2}} \left( \frac{\varphi
 f^2 c^5}{8 a^2 \theta^3 \kappa^3 M V} \right)^{1/8} \notag \\
 & = \SI{1900}{\kelvin} \; f_{-6}^{1/4} \kappa_1^{-3/8} M_{-2}^{-1/8}
 V_{-1}^{-1/8} \theta_i^{-3/8} \varphi_i^{1/8} .
\end{align}
These expressions share the same parameter dependence as those derived
in Ref.~\cite{kyutoku_is2013} by random-walk arguments. This indicates
that the parameter dependence is robust as far as similar assumptions
are adopted.

In the range of opening angles observed in our numerical simulations,
the macronova/kilonova from black hole-neutron star binaries tends to
peak slightly earlier with slightly higher bolometric luminosity than
that from the spherical ejecta for given values of other
parameters. This tendency is also observed in radiation transfer
simulations \cite{tanaka_hkwkss2014}. When the opening angle in the
meridional plane, $\theta$, is small, the peak time becomes early and
the peak bolometric luminosity increases. The reason for this is that
photons can escape easily from the ejecta when $\theta$ is
small. Specifically, the optical depth $\kappa \rho \theta R$ at
$t_\mathrm{peak}$ is proportional to $\theta^{-1}$ and independent of
$\varphi$. When the opening angle in the equatorial plane, $\varphi$, is
small, the peak time becomes late and the peak bolometric luminosity
decreases. The reason is that a small value of $\varphi$ increases the
rest-mass density, optical depth, and characteristic time scales. The
dependence of luminosity may be understood by the constancy of
$L_\mathrm{peak} t_\mathrm{peak}$ for both cases. The combined effects
of these two angles tend to prefer the slightly earlier peak with
slightly brighter emission. When the ejecta becomes transparent, the
bolometric luminosity does not depend on the geometry, because we simply
have $L = \dot{\varepsilon} M$ even within this model derived with the
diffusion approximation.\footnote{It is probable that the photons are
depleted when the free-free emission becomes inefficient
\cite{kulkarni2005}, and this effect is not taken into account in the
current model.}

At a given time, the material temperature $T$ and effective temperature
$T_\mathrm{eff}$ are higher for the anisotropic ejecta than for the
spherical one due to different geometry. In typical situations, the
material temperature, $T$, is higher by about a factor of 2, and this
agrees approximately with the result of
Ref.~\cite{tanaka_hkwkss2014}. The reason of the high temperature is
that the decay heat of unstable \textit{r}-process elements is deposited
to a small volume for a given mass and velocity of the
ejecta. Accordingly, the effective temperature, $T_\mathrm{eff}$, is
higher by $\approx 30\%$--50\%. Even this amount of difference could
have a significant effect on the observed flux [not to be confused with
the diffusion flux, Eq.~\eqref{eq:mkflux}, which is trivially related to
$T_\mathrm{eff}$] in optical and near-infrared bands, because the
typical value of $T_\mathrm{peak}$ is in the infrared band. Thus, a
small increase of $T_\mathrm{eff}$ could enhance the flux at optical and
near-infrared bands. In fact, we see that absolute/apparent magnitudes
increase by 1--2 in these bands if we assume a perfect blackbody
spectrum. We do not, however, regard this amount of increase as very
quantitative, because realistic spectra will be very different from the
blackbody. The high temperature could also affect possible dust
formation \cite{takami_ni2014}.

The dependence of the peak quantities on $f$, $\kappa$, $M$, and $V$ is
the same as that of the spherical model \cite{li_paczynski1998}. A large
value of $f$ increases the luminosity, a large value of $\kappa$ delays
the peak and decreases the luminosity, a large value of $M$ delays the
peak and increases the luminosity, and a large value of $V$ hastens the
peak and increases the luminosity. Among these parameters, the ejecta
mass can become much larger for black hole-neutron star binaries than
for binary neutron stars \cite{hotokezaka_kkosst2013}, and higher
luminosity could be achieved \cite{hotokezaka_ktkssw2013}. In fact, this
difference may dominate corrections due to the opening angles. Typical
velocities of the ejecta cannot be very different. The heating
efficiency and opacity could change reflecting different compositions of
the ejecta, but we do not discuss them in this study.

\subsubsection{Directional dependence, line, and polarization}
\label{sec:cp_mk_misc}

We briefly discuss possible aspects of the macronova/kilonova from
anisotropic ejecta that cannot be captured in the analytic model
developed above. An obvious outcome of the anisotropy is the directional
dependence (see Refs.~\cite{roberts_klr2011,grossman_krp2014} for binary
neutron stars). Emission should be brighter when observed from the
direction perpendicular to the equatorial plane than in the equatorial
plane. Specifically, the flux will be larger by $1 / \theta \approx
3$--5 near $t_\mathrm{peak}$ for the former situation. Accordingly, the
light curves will exhibit different evolution near $t_\mathrm{peak}$ and
become indistinguishable after the entire ejecta becomes
transparent. These behaviors are observed in radiation transfer
simulations \cite{tanaka_hkwkss2014}. Followup observations of
electromagnetic counterparts will benefit from this directional
dependence, because gravitational waves are emitted most strongly in the
direction perpendicular to the equatorial plane, in which the
macronova/kilonova will be the brightest. That is, observed binaries
will be biased toward the brightest direction of the macronova/kilonova.

A chance to observe spectral lines associated with \textit{r}-process
elements will be better for black hole-neutron star binaries than for
binary neutron stars. In any case, it will be very challenging to
observe such lines from the macronova/kilonova, because a bunch of lines
are expected to be significantly blended due to Doppler broadening in
the ejecta with a large surface velocity. This broadening may be
mitigated in the direction perpendicular to the equatorial plane,
because the expansion velocity is smaller by a factor of $\theta$ than
in other directions and spherical cases. Furthermore, the emission is
expected to be the brightest in this direction. Thus, the
macronova/kilonova associated with black hole-neutron star binaries
would deserve detailed spectroscopic observations to seek a
(serendipitous) strong and isolated line (see also
Ref.~\cite{kasen_bb2013} for relevant discussions).

Potential diagnostics of the anisotropic geometry is polarization
induced by electron scattering, but the polarization degree is not
likely to be high for the macronova/kilonova. If the optical depth to
electron scattering is sufficiently high and lines do not contribute to
depolarization significantly, the linear polarization observed from the
equatorial plane would be 4\%--5\% because of the highly deformed
photosphere \cite{Hoflich1991}. However, the number density of free
electrons will be much smaller in the ejecta composed of
\textit{r}-process elements than in, e.g., the supernova ejecta near the
peak luminosity. While the \textit{r}-process elements have the mass
number $\gtrsim 100$, the ionization degree will not be particularly
high around the peak of the macronova/kilonova
\cite{kasen_bb2013,tanaka_hotokezaka2013}. Hence, the opacity for the
electron scattering will be lower by about 3 orders of magnitude than
that for bound-bound transitions if $\kappa =
\SI{10}{\per\gram\cm\squared}$ is an appropriate representative. The
optical depth to electron scattering will be only $O(10^{-2})$ near
$t_\mathrm{peak}$ when the total optical depth is at most $O(10)$, and
therefore the polarization degree may be reduced by a similar factor. In
addition, interaction with lines will further depolarize the radiation
\cite{jeffery1989}.

\subsection{Synchrotron radio emission} \label{sec:cp_sync}

Nonthermal radiation such as synchrotron emission is expected to arise
from blast waves formed between the ejecta and ambient interstellar
medium in a similar manner to the supernova remnant and gamma-ray burst
afterglow \cite{nakar_piran2011,takami_ki2014}. Subrelativistic blast
waves will develop as the ejecta sweeps the interstellar medium, and the
kinetic energy of the ejecta is converted to postshock internal
energy. A fraction of the internal energy at the forward shock will be
converted to energy of nonthermal electrons assembled from the
interstellar medium and of amplified magnetic fields. The accelerated
electrons will radiate synchrotron emission in a magnetized environment,
and the emission would be observed in radio bands
\cite{nakar_piran2011,piran_nr2013} and possibly in optical, x-ray, and
$\gamma$-ray bands \cite{takami_ki2014}.

\subsubsection{Radiation at the deceleration time}
\label{sec:cp_sync_dec}

The most luminous emission is expected when the ejecta begins to be
decelerated significantly, and the deceleration time $t_\mathrm{dec}$
depends on the ejecta geometry for a given mass and velocity of the
ejecta. We describe the synchrotron radio emission expected at
$t_\mathrm{dec}$ adopting a simplified version of the nonrelativistic
model developed in Ref.~\cite{takami_ki2014} (see also
Ref.~\cite{ioka_meszaros2005}). We do not, however, attempt to model the
time evolution in this study, because the lateral expansion should
become important after the deceleration time for the anisotropic
ejecta. While the late-time evolution of the spherical ejecta will be
described reasonably by Sedov--Taylor's self-similar solution as in
Ref.~\cite{piran_nr2013}, it is difficult to formulate the lateral
expansion of the anisotropic ejecta in a simple manner. We introduce
short-hand notations $v_{-1} \equiv v / (0.1c)$ for the ejecta velocity,
$n_0 \equiv n / ( \SI{1}{\per\cm\cubic} )$ for the ambient number
density, $\epsilon_{\mathrm{e},-1} \equiv \epsilon_\mathrm{e} / 0.1$ for
the fraction of postshock internal energy given to nonthermal electrons,
$\epsilon_{B,-1} \equiv \epsilon_B / 0.1$ for the fraction of postshock
internal energy given to magnetic fields, and $D_2 \equiv D / (
\SI{100}{Mpc} )$ for the distance from the observer to the site of
binary coalescence. We also introduce the fraction $\eta$ of accelerated
electrons and power-law index $p$ of the Lorentz factor distribution.

The ejecta is decelerated significantly when the mass comparable to its
own is assembled from the interstellar medium. The deceleration radius
of the anisotropic ejecta is given by
\begin{align}
 R_\mathrm{dec} & = \left( \frac{3 \pi M}{4 \theta \varphi n
 m_\mathrm{p}} \right)^{1/3} \notag \\
 & = \SI{1.1}{pc} \; n_0^{-1/3} M_{-2}^{1/3} \theta_i^{-1/3}
 \varphi_i^{-1/3} , \label{eq:rdec}
\end{align}
where $m_\mathrm{p}$ is the proton mass and $n$ is the number density of
the ambient medium. Modeling the ejecta by a single-velocity shell with
$v$, the corresponding deceleration time is
\begin{align}
 t_\mathrm{dec} & = \frac{R_\mathrm{dec}}{v} \notag \\
 & = \SI{38}{year} \; n_0^{-1/3} M_{-2}^{1/3} v_{-1}^{-1}
 \theta_i^{-1/3} \varphi_i^{-1/3} .
\end{align}
Here, the ejecta velocity, $v$, may be identified with the average
velocity of the ejecta, $v_\mathrm{ave}$, defined by
Eq.~\eqref{eq:vave}. The value of the ambient density, $n$, should vary
by orders of magnitude depending on the location of the binary
coalescence, and we normalize it by the typical value of the Galactic
disk, \SI{1}{\per\cm\cubed}, following previous work
\cite{nakar_piran2011,piran_nr2013,takami_ki2014}. These expressions
reduce to values for the spherical ejecta when we set $\theta = \pi / 2$
and $\varphi = 2 \pi$.

If the typical opening angles observed in numerical simulations are kept
until the deceleration time, $R_\mathrm{dec}$ and $t_\mathrm{dec}$ are
larger by a factor of 2--3 than those for the spherical ejecta for given
values of the other parameters. The reason for this is that only a
limited fraction of the volume inside $R_\mathrm{dec}$ is swept by the
anisotropic ejecta. Whereas the ejecta will approach a spherical state
to some extent before $t_\mathrm{dec}$ \cite{rosswog_katp2014}, the
synchrotron radio emission from black hole-neutron star binaries will be
a longer-lasting event than that for binary neutron stars.

The geometry does not modify the Lorentz factor distribution of
nonthermal electrons and magnetic fields. The number of assembled
electrons at $t_\mathrm{dec}$ is given by
\begin{align}
 N_\mathrm{e,tot} & = \frac{M}{m_\mathrm{p}} \notag \\
 & = \num{1.2e55} \; M_{-2} .
\end{align}
Assuming that a fraction $\eta \le 1$ of these electrons is accelerated
to power-law distribution of the Lorentz factor $\gamma_\mathrm{e}$ with
the index $p>2$ as
\begin{equation}
 \frac{d N_\mathrm{e} ( \gamma_\mathrm{e} )}{d \gamma_\mathrm{e}}
  \propto \gamma_\mathrm{e}^{-p} \; ( \gamma_m < \gamma_\mathrm{e} ) ,
\end{equation}
the minimum Lorentz factor may be derived from the number and energy of
accelerated electrons as
\begin{align}
 \gamma_m & = \frac{\epsilon_\mathrm{e}}{\eta} \frac{p-2}{p-1}
 \frac{m_\mathrm{p}}{m_\mathrm{e}} \frac{(v/c)^2}{2} \notag \\
 & = 0.92 \; g(p) \eta^{-1} \epsilon_{\mathrm{e},-1} v_{-1}^{2} ,
\end{align}
where $m_\mathrm{e}$ is the electron mass, $\epsilon_\mathrm{e}$ is the
fraction of the postshock internal energy given to the accelerated
nonthermal electrons, and $g(p) \equiv (p-2)/(p-1)$. Care must be taken
in applying this equation to subrelativistic blast waves, because
$\gamma_m$ can fall below unity and become unphysical, particularly when
the ejecta is significantly decelerated (not considered here). The
strength of magnetic fields is given by
\begin{align}
 B & = \sqrt{9 \pi \epsilon_B n m_\mathrm{p}} v \notag \\
 & = \SI{6.5}{mGauss} \; \epsilon_{B,-1}^{1/2} n_0^{1/2} v_{-1} ,
\end{align}
where $\epsilon_B$ is the fraction of postshock internal energy
converted to magnetic fields.

Parameters characterizing microphysics, $p$, $\eta$ ,
$\epsilon_\mathrm{e}$, and $\epsilon_B$, are all uncertain. Following
Ref.~\cite{piran_nr2013}, we normalize $\epsilon_\mathrm{e}$ and
$\epsilon_B$ by 0.1 and take the fiducial value of $p$ to be 2.5, which
gives $g(p) = 1/3$. Typical values of $p$ observed in nonrelativistic
blast waves may be 2.5--3 \cite{chevalier_fransson2006}. The fiducial
value of $\eta$ is set to be unity. Detailed spectroscopic observations
of nonthermal radiation could determine these parameters in principle
\cite{takami_ki2014}.

Quantities characterizing the instantaneous spectrum are estimated as
follows. The synchrotron frequency of an electron with
$\gamma_\mathrm{e}$ is defined by $\nu_\mathrm{e} ( \gamma_\mathrm{e} )
\equiv q B \gamma_\mathrm{e}^2 / ( 2 \pi m_\mathrm{e} c )$, where $q$ is
the elementary charge, and the power of the electron is by $P_\mathrm{e}
( \gamma_\mathrm{e} ) \equiv \sigma_\mathrm{T} c B^2 \gamma_\mathrm{e}^2
/ ( 6 \pi )$, where $\sigma_\mathrm{T}$ is the Thomson cross
section. The specific flux from a single electron at its peak frequency,
$\nu_\mathrm{e}$, is estimated to be $P_\nu \approx P_\mathrm{e} /
\nu_\mathrm{e} = \sigma_\mathrm{T} m_\mathrm{e} c^2 B / (3q)$
independently from the electron Lorentz factor. The characteristic
frequency of the electron distribution corresponding to $\gamma_m$ is
given by
\begin{equation}
 \nu_m = \SI{1.5e4}{\hertz} \; g(p)^2 \eta^{-2}
  \epsilon_{\mathrm{e},-1}^2 \epsilon_{B,-1}^{1/2} n_0^{1/2} v_{-1}^5 .
\end{equation}
An unabsorbed specific flux, i.e., a hypothetical specific flux in the
absence of self-absorption, at $\nu_m$ is estimated to be
\begin{align}
 F_{\nu,m} & = \frac{\eta N_\mathrm{e,tot} P_\nu}{4 \pi D^2} \notag \\
 & = \SI{2.4}{Jy} \; \eta \epsilon_{B,-1}^{1/2} n_0^{1/2} M_{-2} v_{-1}
 D_2^{-2} ,
\end{align}
where $D$ is the distance from the observer to the site of binary
coalescence. If we neglect synchrotron self-absorption and cooling, the
specific flux is given by $F_{\nu,m} ( \nu / \nu_m )^{1/3}$ below
$\nu_m$ and $F_{\nu,m} ( \nu / \nu_m )^{-(p-1)/2}$ above $\nu_m$. We
show below that the self-absorption could suppress the radio spectrum,
while the cooling is not important in the radio band.

The self-absorption frequency $\nu_a$ may be obtained approximately by
comparing the hypothetical unabsorbed flux with the blackbody flux in
the Rayleigh regime \cite{ioka_meszaros2005}. The blackbody flux at
$\nu_m$ is given by
\begin{equation}
 F_{\nu,\mathrm{BB}} = 2 \pi \nu_m^2 \gamma_m m_\mathrm{e}
  \frac{A_\mathrm{em}}{4 \pi D^2} ,
\end{equation}
where $A_\mathrm{em}$ is the blackbody emitting area. While this should
be $4 \pi R_\mathrm{dec}^2$ for the spherical ejecta, we take
$A_\mathrm{em} = \varphi R_\mathrm{dec}^2 \propto \theta^{2/3}
\varphi^{-1/3}$ for the anisotropic ejecta in a similar manner to
Sec.~\ref{sec:cp_mk_model}. It is readily found that $\nu_m < \nu_a$
when $F_{\nu,\mathrm{BB}} < F_{\nu,m}$, and vice versa. The case that
$\nu_m < \nu_a$ is typical for subrelativistic blast waves, and the
self-absorption frequency is defined as the frequency at which the
synchrotron and blackbody fluxes are equal. Specifically, we obtain
\begin{align}
 \nu_a & = \left( \frac{F_{\nu,m}}{F_{\nu,\mathrm{BB}}}
 \right)^{2/(4+p)} \nu_m \notag \\
 & = \SI{3.9e7}{\hertz} \; \eta^{-2(p-2)/(4+p)} \notag \\
 & \times \epsilon_{\mathrm{e},-1}^{2(p-1)/(4+p)}
 \epsilon_{B,-1}^{(2+p)/[2(4+p)]} n_0^{(14+3p)/[6(4+p)]} \notag \\
& \times M_{-2}^{2/[3(4+p)]} v_{-1}^{(5p-2)/(4+p)} \theta_i^{4/[3(4+p)]}
 \varphi_i^{-2/[3(4+p)]} ,
\end{align}
where the prefactor is given for $p=2.5$ and varies by a factor of 2
within $2.1 < p < 3$.\footnote{More precisely, the prefactor $(
\num{6.0e11} )^{2/(4+p)} \times \SI{1.5e4}{\hertz} \;
g(p)^{2(p-1)/(4+p)}$ is applicable to all the values of $p>2$.} The
dependence on the opening angles is inherited from the emitting area,
$A_\mathrm{em}$, and $\nu_a$ is smaller by a few tens of percent for the
anisotropic ejecta than for the spherical one. The self-absorption
frequency can increase to $\sim \SI{1}{\GHz}$ in a plausible parameter
range, and thus the self-absorption could be important at low-frequency
radio bands for such cases. When $\nu_a < \nu_m$, we instead obtain
$\nu_a = ( F_{\nu,m} / F_{\nu,\mathrm{BB}} )^{3/5} \nu_m$.

The cooling Lorentz factor $\gamma_c$ at the deceleration time, above
which the radiative energy loss plays a significant role, is estimated
by the condition $\gamma_c m_\mathrm{e} c^2 = P ( \gamma_c )
t_\mathrm{dec}$, and we obtain
\begin{align}
 \gamma_c & = \frac{6 \pi m_\mathrm{e} c}{\sigma_\mathrm{T} B^2
 t_\mathrm{dec}} \notag \\
 & = \num{1.5e4} \; \epsilon_{B,-1}^{-1} n_0^{-2/3} M_{-2}^{-1/3}
 v_{-1}^{-1} \theta_i^{1/3} \varphi_i^{1/3} .
\end{align}
The corresponding cooling frequency is given by
\begin{equation}
 \nu_c = \SI{4.3e12}{\hertz} \; \epsilon_{B,-1}^{-3/2} n_0^{-5/6}
  M_{-2}^{-2/3} v_{-1}^{-1} \theta_i^{2/3} \varphi_i^{2/3} .
\end{equation}
Although the cooling frequency decreases by a factor of several for the
anisotropic ejecta due to the long deceleration time, this could affect
the radio spectrum at high frequency only in a limited parameter range.

Finally, the instantaneous spectrum for $\nu_m < \nu_a < \nu_c$ is given
by
\begin{equation}
 \frac{F_\nu}{F_{\nu,m}} =
  \begin{cases}
   ( \nu_a / \nu_m )^{-(p+4)/2} ( \nu / \nu_m )^2 & ( \nu < \nu_m) \\
   ( \nu_a / \nu_m )^{-(p-1)/2} ( \nu / \nu_a )^{5/2} & ( \nu_m \le \nu
   < \nu_a) \\
   ( \nu / \nu_m )^{-(p-1)/2} & ( \nu_a \le \nu < \nu_c ) \\
   ( \nu_c / \nu_m )^{-(p-1)/2} ( \nu / \nu_c )^{-p/2} & ( \nu_c \le \nu
   )
  \end{cases}
  .
\end{equation}
The third segment is the most relevant to radio observations, and it
would be useful to reexpress the spectrum in this range as
\begin{align}
 F_\nu & = \SI{0.12}{mJy} \; \left( \frac{\nu}{\SI{1}{\GHz}}
 \right)^{-(p-1)/2} \notag \\
 & \times \eta^{1-p} \epsilon_{\mathrm{e},-1}^{p-1}
 \epsilon_{B,-1}^{(p+1)/4} n_0^{(p+1)/4} M_{-2} v_{-1}^{(5p-3)/2}
 D_2^{-2} ,
\end{align}
where the prefactor is for $p=2.5$ and decreases by a factor of 40 as
$p$ increases from 2.1 to 3.\footnote{The prefactor $( \num{6.5e4}
)^{(1-p)/2} \times \SI{24}{Jy} \; g(p)^{p-1}$ is applicable to all the
values of $p$.} This expression indicates that the emission associated
with the massive ejecta from black hole-neutron star binaries will be
bright.

\subsubsection{Proper motion} \label{sec:cp_sync_prop}

Aside from the expansion, the anisotropic ejecta from black hole-neutron
star binaries exhibits center-of-mass motion, and thus the proper motion
of radio images could be observed \cite{kyutoku_is2013}. The
characteristic distance of the center-of-mass motion may be given
approximately by $R_\mathrm{com} = v_\mathrm{ej} t_\mathrm{dec} =
R_\mathrm{dec} ( v_\mathrm{ej} / v_\mathrm{ave} )$. The projected
distance on the celestial sphere should be smaller by a factor of
$\approx 2$ due to the angular average, whereas the observational bias
due to the directional dependence of gravitational radiation should
mitigate this decrease. The expected amount of projected travel
distances is $O(1)$ pc [see Eq.~\eqref{eq:rdec}], and we expect the
radio image of the anisotropic ejecta to move $O(1)$ milliarcsecond
during its bright emission for a event at $O(100)$ Mpc. This amount of
proper motion could be resolved by current radio instruments depending
on the parameters and observed frequency \cite{taylor_fbk2004} and could
help to distinguish black hole-neutron star binaries from binary neutron
stars only by electromagnetic counterparts.

\section{Summary} \label{sec:summary}

We performed numerical-relativity simulations of black hole-neutron star
binary mergers to study dynamical mass ejection. The mass ratio,
black-hole spin, and neutron-star equation of state were systematically
varied to reveal the dependence of the ejecta properties on these
parameters. We found that dynamical mass ejection is driven primarily by
the tidal torque exerted from black holes to elongated neutron stars,
and this process progresses over $\approx \SI{2}{\ms}$ after the onset
of merger. The dynamical ejecta is concentrated around the equatorial
plane with a half opening angle of \ang{10}--\ang{20} and sweeps out
about a half of the plane, except for cases that mass ejection is
inefficient. Because of this anisotropy, the ejecta carries a bulk
linear momentum, and thus the remnant black hole-disk system receives an
ejecta kick velocity due to the backreaction.

The ejecta mass can be as large as $\sim 0.1 M_\odot$, and the average
velocity of the ejecta defined from the kinetic energy is typically
0.2--$0.3 c$. Dynamical mass ejection tends to become efficient when the
neutron-star compactness is small, the mass ratio is small, and/or the
black-hole spin is large. The dependence of ejecta properties on the
compactness, however, is not as simple as that of the total mass
remaining outside the apparent horizon. This suggests that not only the
compactness but also detailed properties of the equation of state
influence the ejecta properties significantly. Furthermore, the
dependence on the mass ratio is not always monotonic. The ratio of the
ejecta mass to the bound mass is large when the mass ratio is large, and
the average velocity of the ejecta is also large for such cases. These
suggest that the dynamical ejecta from higher-mass-ratio binaries is
more energetic for a given ejecta mass.

We also found that the bound envelope along the polar axis of the
central remnant is not as heavy as that for binary neutron star mergers
as far as the dynamical processes are concerned. This would be
advantageous for a hypothetical gamma-ray burst jet to overcome the
baryon loading problem, while how to collimate it is uncertain in the
absence of a heavy envelope. Fallback rates of bound material obey the
canonical $-5/3$ power law. The remnant disk exhibits a standing spiral
shock structure, which enhances the mass accretion.

Because the gravitational-wave kick velocity imparted to the remnant
does not exceed \SI{100}{\km\per\second} for our models, the ejecta kick
velocity dominates motion of the remnant. We found that ejecta and
gravitational waves usually carry the linear momentum in the opposite
direction, and thus these two kick velocities would partially cancel
out. Tight correlations between the gravitational-wave frequency at the
maximum amplitude and tidal coupling constant were suggested to exist in
a similar manner to that found for binary neutron stars. The relations
for black hole-neutron star binaries depend on the black-hole spin.

Properties of electromagnetic counterparts were discussed based on the
results of numerical simulations focusing on the effect of ejecta
anisotropy. An analytic model of the macronova/kilonova shows that both
the material and effective temperatures become high for the anisotropic
ejecta from black hole-neutron star binaries. We also found that the
peak time is slightly early and the peak bolometric luminosity is
slightly high for the typical ejecta opening angles. The synchrotron
radio emission is long lasting for the anisotropic ejecta, and the
proper motion of the radio images could also be observed. The most
significant difference from electromagnetic counterparts associated with
binary neutron stars would come from different ejecta masses for both
emission models.

\begin{acknowledgments}
 Koutarou Kyutoku is grateful to John L. Friedman, Kenta Hotokezaka,
 Kazumi Kashiyama, Kyohei Kawaguchi, Kenta Kiuchi, Richard
 O'Shaughnessy, and Masaomi Tanaka for valuable discussions. Koutarou
 Kyutoku is also grateful to Tom Downes for the management of the Jacobi
 cluster at University of Wisconsin--Milwaukee, which is supported by
 NSF Awards No.~PHY-0923409 and No.~PHY-1104371. This work is supported
 by JSPS KAKENHI Grant-in-Aid for Scientific Research (Grants
 No.~24244028, No.~26247042, No.~26287051, No.~26400267), JSPS KAKENHI
 Grant-in-Aid for Specially Promoted Research (Grant No.~24000004), and
 MEXT KAKENHI Grant-in-Aid for Scientific Research on Innovative Areas
 (Grant No.~24103006). Koutarou Kyutoku is supported by a JSPS
 Postdoctoral Fellowship for Research Abroad.
\end{acknowledgments}

\appendix

\section{Convergence and uncertainty} \label{app:err}

Ejecta are only a fraction of the material remaining outside the horizon,
which itself is only a fraction of a neutron star. Therefore, quantities
associated with the ejecta could entail large fractional
errors. Furthermore, our numerical simulations have various parameters
both physical and unphysical. In this appendix, we estimate errors and
uncertainties in our computations. We also discuss seemingly spurious
high-velocity ejecta found in Sec.~\ref{sec:res_ej_case}.

\subsection{Convergence with respect to the grid resolution}
\label{app:err_resol}

\begin{figure}
 \includegraphics[width=.95\linewidth]{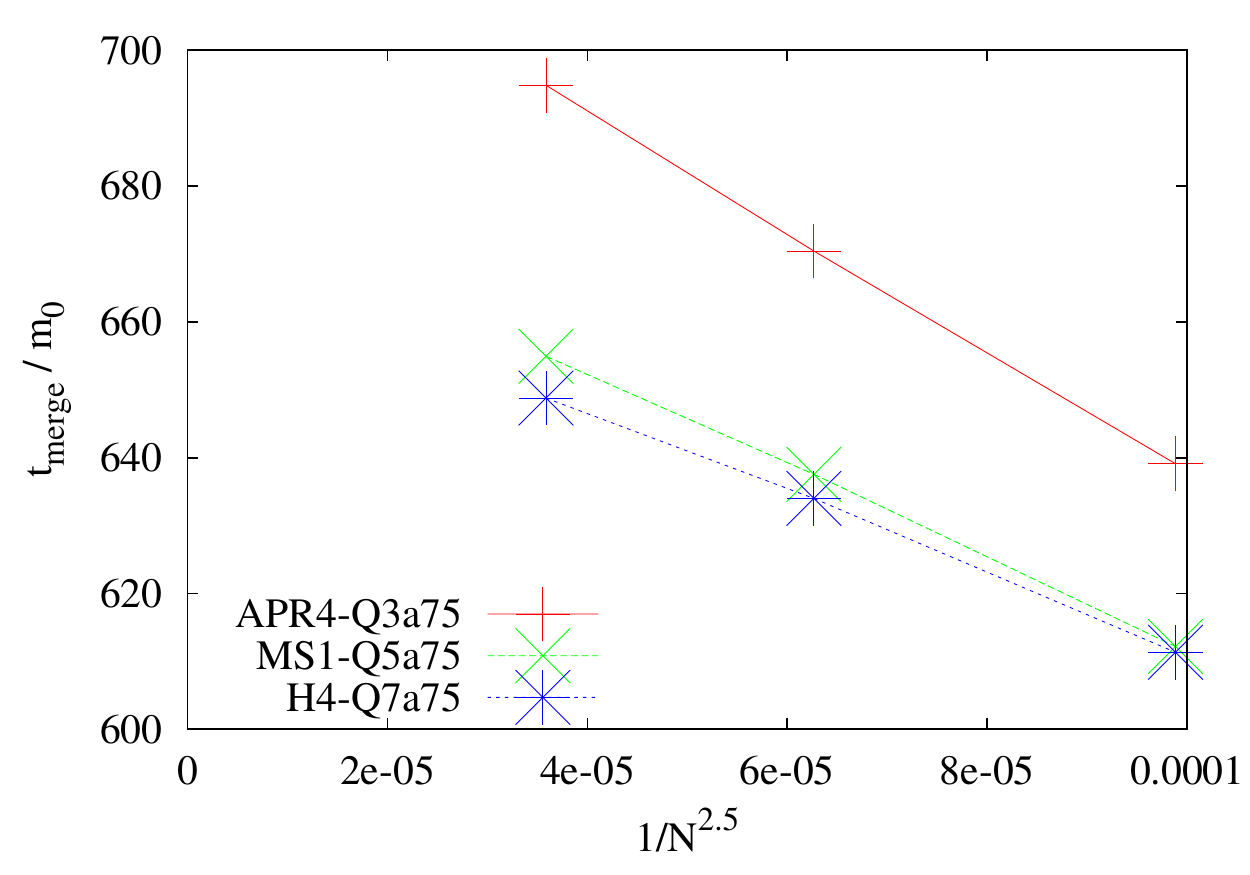} \caption{Merger time
 normalized by the total mass, $t_\mathrm{merge} / m_0$, vs grid
 resolutions for selected models. We assume a hypothetical convergence
 order 2.5 for all the models. Actual numerical data for APR4-Q3a75,
 MS1-Q5a75, and H4-Q7a75 show convergence orders 2.2, 2.9, and 3.1,
 respectively.} \label{fig:tmerge}
\end{figure}

Finite grid resolutions are obvious sources of errors. First of all, we
demonstrate that reasonable convergence behavior is observed in our
numerical simulations. Figure \ref{fig:tmerge} shows the merger time,
$t_\mathrm{merge}$, as a function of grid resolutions represented by $N$
(see Sec.~\ref{sec:setup_sim} for the definition) for selected
models. The exact convergence order estimated from numerical data varies
among models and typically lies between 2 and 3. Taking the different
accuracies for different parts of our code {\small SACRA}
\cite{yamamoto_st2008} into account, the observed behavior is
reasonable.

\begin{table*}
 \caption{The same as Table \ref{table:result} but for runs with
 different grid resolutions for selected models. The fiducial grid
 resolution adopted in the body text is $N=60$.}
 \begin{tabular}{c|ccccccc} \hline
  ~$N$~ & ~$M_{r>r_\mathrm{AH}} [M_\odot]$~ & ~$M_\mathrm{bd}
  [M_\odot]$~ & ~$M_\mathrm{ej} [M_\odot]$~ & ~~$T_\mathrm{ej}$
  (\si{erg})~~ & ~~$P_\mathrm{ej} [M_\odot]$~~ & ~~$v_\mathrm{ave}$~~ &
  ~~$v_\mathrm{ej}$~~ \\
  \hline \hline
  \multicolumn{8}{c}{APR4-Q3a75} \\
  \hline
  60 & 0.194 & 0.182 & 0.0111 & \num{5.48e50} & \num{2.08e-3} & 0.235 &
			      0.187 \\
  48 & 0.210 & 0.201 & 0.0087 & \num{3.98e50} & \num{1.61e-3} & 0.227 &
			      0.186 \\
  40 & 0.206 & 0.198 & 0.0079 & \num{3.40e50} & \num{1.41e-3} & 0.219 &
			      0.178 \\
  \hline
  \multicolumn{8}{c}{APR4-Q5a75} \\
  \hline
  60 & 0.068 & 0.059 & 0.0084 & \num{4.77e50} & \num{8.30e-4} & 0.252 &
			      0.099 \\
  48 & 0.070 & 0.063 & 0.0067 & \num{3.60e50} & \num{7.87e-4} & 0.246 &
			      0.118 \\
  40 & 0.074 & 0.069 & 0.0047 & \num{2.43e50} & \num{6.51e-4} & 0.239 &
			      0.137 \\
  \hline
  \multicolumn{8}{c}{MS1-Q5a75} \\
  \hline
  60 & 0.356 & 0.277 & 0.0785 & \num{5.55e51} & \num{1.80e-2} & 0.281 &
			      0.223 \\
  48 & 0.361 & 0.282 & 0.0795 & \num{5.26e51} & \num{1.80e-2} & 0.272 &
			      0.226 \\
  40 & 0.370 & 0.290 & 0.0797 & \num{5.28e51} & \num{1.77e-2} & 0.272 &
			      0.222 \\
  \hline
  \multicolumn{8}{c}{H4-Q7a75} \\
  \hline
  60 & 0.194 & 0.157 & 0.0375 & \num{2.78e51} & \num{7.19e-3} & 0.288 &
			      0.192 \\
  48 & 0.207 & 0.171 & 0.0360 & \num{2.64e51} & \num{6.89e-3} & 0.287 &
			      0.191 \\
  40 & 0.214 & 0.179 & 0.0341 & \num{2.41e51} & \num{6.64e-3} & 0.281 &
			      0.195 \\
  \hline
 \end{tabular}
 \label{table:convresult}
\end{table*}

Table \ref{table:convresult} compares characteristic quantities among
different grid resolutions for selected models. It is evident that these
quantities are not always monotonic with respect to the grid
resolution. Such behavior is frequently seen in hydrodynamic quantities,
which severely suffer low convergence order when shock waves
exist. Relative errors are smaller for $M_{r>r_\mathrm{AH}}$ than for
$M_\mathrm{ej}$, and this suggests that the accurate determination of
boundaries separating bound and unbound material is an important but
difficult task. If we assume the first-order convergence between $N=60$
and 48 results, the worst-case error with $N=60$ results are $\approx
30\%$, 40\%, and 100\% for $M_{r>r_\mathrm{AH}}$, $M_\mathrm{bd}$, and
$M_\mathrm{ej}$, respectively. The accuracy of $M_\mathrm{ej}$ is
especially low when the ejecta mass is as small as $M_\mathrm{ej}
\lesssim 0.01 M_\odot$, and the error decreases to $\lesssim 20\%$ for
more massive ejecta. It is reasonable that the relative error is large
for the small mass ejecta, where the absolute error is always estimated
to be $\approx 0.005$--$0.01 M_\odot$ for $M_\mathrm{ej}$.

\subsection{Effect of an artificial atmosphere} \label{app:err_atm}

An artificial atmosphere affects the ejecta properties. Some portion of
the atmosphere happens to satisfy the unbound criterion, $u_t < -1$, as
a result of hydrodynamic interaction, and this error spuriously
increases the amount of the ejecta. At the same time, the atmosphere
decelerates physical material ejected from neutron stars, and this error
spuriously decreases the amount of ejecta. Low atmospheric density will
mitigate both these errors. Another source of error is a steep density
gradient at the neutron-star surface, which induces spurious shock
heating in numerical simulations and helps the material to become
unbound. Although this error will be suppressed as grid resolutions are
improved to resolve the stellar surface accurately, lowering the
atmospheric density at a fixed resolution does not always suppress it,
because the shock could become strong.

\begin{table*}
 \caption{The same as Table \ref{table:result} but with different values
 of $f_\mathrm{at}$ for APR4-Q3a75 with $N=60$ and 40. The fiducial
 value of $f_\mathrm{at}$ adopted in the body text is \num{e-12}.}
 \begin{tabular}{c|ccccccc} \hline
  ~$f_\mathrm{at}$~ & ~$M_{r>r_\mathrm{AH}} [M_\odot]$~ &
  ~$M_\mathrm{bd} [M_\odot]$~ & ~$M_\mathrm{ej} [M_\odot]$~ &
  ~~$T_\mathrm{ej}$ (\si{erg})~~ & ~~$P_\mathrm{ej} [M_\odot]$~~ &
  ~~$v_\mathrm{ave}$~~ & ~~$v_\mathrm{ej}$~~ \\
  \hline \hline
  \multicolumn{8}{c}{$N=60$} \\
  \hline
  \num{e-11} & 0.196 & 0.185 & 0.0111 & \num{5.45e50} & \num{2.07e-3} &
			  0.235 & 0.186 \\
  \num{e-12} & 0.194 & 0.182 & 0.0111 & \num{5.48e50} & \num{2.08e-3} &
			  0.235 & 0.187 \\
  \num{e-13} & 0.194 & 0.183 & 0.0112 & \num{5.51e50} & \num{2.08e-3} &
			  0.235 & 0.186 \\
  \hline
  \multicolumn{8}{c}{$N=40$} \\
  \hline
  \num{e-10} & 0.207 & 0.199 & 0.0080 & \num{3.44e50} & \num{1.43e-3} &
			  0.219 & 0.178 \\
  \num{e-11} & 0.209 & 0.201 & 0.0078 & \num{3.33e50} & \num{1.38e-3} &
			  0.219 & 0.177 \\
  \num{e-12} & 0.206 & 0.198 & 0.0079 & \num{3.40e50} & \num{1.41e-3} &
			  0.219 & 0.178 \\
  \hline
 \end{tabular}
 \label{table:diffatm}
\end{table*}

Table \ref{table:diffatm} compares characteristic quantities obtained
with different values of $f_\mathrm{at}$. We find that ejecta quantities
like $M_\mathrm{ej}$, $T_\mathrm{ej}$, and $P_\mathrm{ej}$ increase by
$\approx 0.5\%$ for $N=60$ when the atmospheric density is decreased by
an order of magnitude, while the corresponding change (either increase
or decrease) is $\approx 1\%$--2\% for $N=40$. Although the dominant
mechanism responsible for the error is not certain, this suggests that
the error associated with the artificial atmosphere will decrease
significantly as grid resolutions are improved probably due to the
suppression of spurious shocks at the stellar surface. We also find a
similar amount of variations when we change the value of the atmospheric
power-law index $n_\mathrm{at}$ from 3 to 2.

\subsection{Effect of thermal correction $\Gamma_\mathrm{th}$}
\label{app:err_gamth}

Dynamical mass ejection is expected to be governed basically by
zero-temperature equations of state, because shock heating does not play
a significant role. However, the ejecta properties depend weakly on the
finite-temperature part of equations of state due to the spurious shock
heating. Thus, the dependence of results on $\Gamma_\mathrm{th}$ also
requires investigation.

\begin{table*}
 \caption{The same as Table \ref{table:result} but for runs with
 different values of $\Gamma_\mathrm{th}$ for selected models. The
 fiducial value of $\Gamma_\mathrm{th}$ adopted in the body text is
 1.8. All the simulations are performed with $N=60$.}
 \begin{tabular}{c|ccccccc} \hline
  ~$\Gamma_\mathrm{th}$~ & ~$M_{r>r_\mathrm{AH}} [M_\odot]$~ &
  ~$M_\mathrm{bd} [M_\odot]$~ & ~$M_\mathrm{ej} [M_\odot]$~ &
  ~~$T_\mathrm{ej}$ (\si{erg})~~ & ~~$P_\mathrm{ej} [M_\odot]$~~ &
  ~~$v_\mathrm{ave}$~~ & ~~$v_\mathrm{ej}$~~ \\
  \hline \hline
  \multicolumn{8}{c}{APR4-Q3a75} \\
  \hline
  1.6 & 0.196 & 0.184 & 0.0116 & \num{5.95e50} & \num{2.13e-3} & 0.240 &
			      0.184 \\
  1.8 & 0.194 & 0.182 & 0.0111 & \num{5.48e50} & \num{2.08e-3} & 0.235 &
			      0.187 \\
  2.0 & 0.192 & 0.184 & 0.0082 & \num{3.69e50} & \num{1.51e-3} & 0.225 &
			      0.184 \\
  \hline
  \multicolumn{8}{c}{H4-Q3a75} \\
  \hline
  1.6 & 0.326 & 0.280 & 0.0455 & \num{2.41e51} & \num{9.34e-3} & 0.243 &
			      0.205 \\
  1.8 & 0.331 & 0.285 & 0.0454 & \num{2.30e51} & \num{9.22e-3} & 0.238 &
			      0.203 \\
  2.0 & 0.334 & 0.287 & 0.0464 & \num{2.35e51} & \num{9.28e-3} & 0.238 &
			      0.200 \\
  \hline
  \multicolumn{8}{c}{APR4-Q5a75} \\
  \hline
  1.6 & 0.064 & 0.053 & 0.0108 & \num{6.75e50} & \num{8.82e-4} & 0.265 &
			      0.082 \\
  1.8 & 0.068 & 0.059 & 0.0084 & \num{4.77e50} & \num{8.30e-4} & 0.252 &
			      0.099 \\
  2.0 & 0.072 & 0.063 & 0.0089 & \num{5.08e50} & \num{9.80e-4} & 0.253 &
			      0.110 \\
  \hline
  \multicolumn{8}{c}{H4-Q5a75} \\
  \hline
  1.6 & 0.314 & 0.261 & 0.0529 & \num{3.74e51} & \num{1.14e-2} & 0.281 &
			      0.216 \\
  1.8 & 0.316 & 0.266 & 0.0502 & \num{3.33e51} & \num{1.10e-2} & 0.272 &
			      0.220 \\
  2.0 & 0.320 & 0.269 & 0.0516 & \num{3.34e51} & \num{1.13e-2} & 0.269 &
			      0.220 \\
  \hline
 \end{tabular}
 \label{table:gamthresult}
\end{table*}

Table \ref{table:gamthresult} compares characteristic quantities
obtained with different values of $\Gamma_\mathrm{th}$ for selected
models. Results do not depend monotonically on the value of
$\Gamma_\mathrm{th}$. This fact suggests that the effect of
$\Gamma_\mathrm{th}$ is not very physical, and the difference is
ascribed to numerical errors such as spurious shock heating. We checked
that the differences develop during dynamical mass ejection which
progresses over $\approx \SI{2}{\ms}$ after the onset of merger, and
late-time physical shock heating in the disk region does not introduce
significant differences. The difference of results among different
values of $\Gamma_\mathrm{th}$ is as large as $\approx 20\%$ when
$M_\mathrm{ej} \lesssim 0.01 M_\odot$ and tends to become small when the
mass ejection is efficient. We regard the difference observed in Table
\ref{table:gamthresult} as an estimate of systematic uncertainty, which
could converge as grid resolutions are improved.

\subsection{Effect of different initial separations as a substitute for
  the eccentricity} \label{app:err_sep}

Ejecta properties computed in our simulations deviate from those of
hypothetical genuinely circular mergers due to unphysical eccentricities
inherent in initial data. Specifically, different orbital and
approaching velocities at the tidal disruption lead to deviations of
characteristic quantities on the order of the eccentricity. Although
this error can be eliminated by iterative eccentricity reduction
\cite{foucart_kpt2008,kyutoku_st2014}, it is demanding to reduce the
eccentricities for all the models considered in this study.

\begin{table*}
 \caption{The same as Table \ref{table:result} but with a different
 initial angular velocity, $m_0 \Omega_0$, for APR4-Q3a75. The fiducial
 value of $m_0 \Omega$ adopted in the body text is 0.036. All the
 simulations are performed with $N=60$.}
 \begin{tabular}{c|ccccccc} \hline
  ~$m_0 \Omega_0$~ & ~$M_{r>r_\mathrm{AH}} [M_\odot]$~ & ~$M_\mathrm{bd}
  [M_\odot]$~ & ~$M_\mathrm{ej} [M_\odot]$~ & ~~$T_\mathrm{ej}$
  (\si{erg})~~ & ~~$P_\mathrm{ej} [M_\odot]$~~ & ~~$v_\mathrm{ave}$~~ &
  ~~$v_\mathrm{ej}$~~ \\
  \hline \hline
  0.036 & 0.194 & 0.182 & 0.0111 & \num{5.48e50} & \num{2.08e-3} & 0.235
			  & 0.187 \\
  0.034 & 0.206 & 0.194 & 0.0112 & \num{5.59e50} & \num{2.11e-3} & 0.236
			  & 0.188 \\
  0.032 & 0.203 & 0.192 & 0.0109 & \num{5.38e50} & \num{2.02e-3} & 0.235
			  & 0.186 \\
  0.030 & 0.205 & 0.194 & 0.0114 & \num{5.64e50} & \num{2.12e-3} & 0.236
			  & 0.186 \\
  \hline
 \end{tabular}
 \label{table:diffsep}
\end{table*}

To estimate systematic errors associated with unphysical eccentricities,
we instead compare results obtained by models with different values of
$m_0 \Omega_0$ with $e \sim 0.01$--0.02 for APR4-Q3a75. These models
should merge at a different true anomaly (angle measured from the
periapsis), admitting that it is very difficult to quantify this
statement in numerical simulations. Therefore, the results will give us
an idea of errors associated with eccentricities. As shown in Table
\ref{table:diffsep}, ejecta quantities like $M_\mathrm{ej}$,
$T_\mathrm{ej}$, and $P_\mathrm{ej}$ fluctuate within $\pm 2.5\%$ as
expected from the value of the eccentricity. Because the
increase/decrease of a single ejecta quantity is accompanied by that of
the others, derived quantities like $v_\mathrm{ave}$ and $v_\mathrm{ej}$
are relatively robust with respect to the unphysical eccentricity.

\subsection{Comment on seemingly spurious high-velocity ejecta}
\label{app:err_hv}

In Sec.~\ref{sec:res_ej_case}, a small amount of unbound material is
found to be ejected with a large velocity. We regard this component as
an artifact, because the amount of high-velocity ejecta does not
converge even approximately with respect to grid resolutions, admitting
that ejecta cannot be decomposed unambiguously into physical and
unphysical components unless reliable extrapolation to the continuum
limit is performed. We speculate that this high-velocity ejecta is
created by the artificial atmosphere and finite grid resolutions. They
induce unphysical shocks at the stellar surface during the inspiral
phase, and tenuous material continuously flows out from the inner edge
of the neutron star. This artificial outflow is accumulated around the
black hole and forms a small unphysical disk during the inspiral phase,
whereas some of this disk may be supplied by the neutron star in a
physical manner after the onset of mass shedding. A fraction of this
unphysical disk is ejected impulsively during the merger (due possibly
to the tidal torque exerted by the neutron star) with a large velocity
reflecting the large escape velocity of black holes, i.e., the speed of
light.

\section{Analytic estimate of ejecta opening angle}
\label{app:ejang}

As we discussed in Sec.~\ref{sec:res_ej_var}, the ejecta geometry may be
characterized by the opening angle in the meridional plane,
$\varphi_\mathrm{ej}$, and that in the meridional plane,
$\theta_\mathrm{ej}$. Before looking at numerical results, it is
instructive to estimate these angles by analytic arguments for
comparisons. These estimates help us to distinguish between expected and
unexpected features.

Allowing more than one revolution, the opening angle of the dynamical
ejecta in the equatorial plane should be given by
\begin{equation}
 \varphi_\mathrm{ej} \approx 2 \pi \frac{t_\mathrm{td}}{P_\mathrm{td}} ,
\end{equation}
where $t_\mathrm{td}$ is the time scale of tidal disruption and
$P_\mathrm{td}$ is the orbital period at the tidal disruption radius,
$r_\mathrm{td}$ (see Eq.~\eqref{eq:msradius}). On one hand,
$t_\mathrm{td}$ may be given approximately by the sound crossing time
$t_\mathrm{sc}$ of the neutron star as
\begin{equation}
 t_\mathrm{td} \approx t_\mathrm{sc} \propto \frac{1}{\sqrt{\bar{\rho}}}
  \; ,
\end{equation}
where $\bar{\rho}$ is the average stellar rest-mass density, which is
determined by the equation of state. On the other hand, $P_\mathrm{td}$
should be given by
\begin{equation}
 P_\mathrm{td} \approx 2 \pi \sqrt{\frac{r_\mathrm{td}^3}{m_0}} \propto
  \sqrt{\frac{Q}{(1+Q) \bar{\rho}}} \; ,
\end{equation}
where Eq.~\eqref{eq:msradius} is used and spin-induced corrections are
temporarily neglected. This suggests that the dependence of
$\varphi_\mathrm{ej}$ on the equation of state is weak, because
$\bar{\rho}$ cancels. This expression also suggests that
$\varphi_\mathrm{ej}$ is smaller for a larger mass ratio, but the
expected change is less than 10\% between $Q=3$ and 7. Prograde
black-hole spins will decrease $\varphi_\mathrm{ej}$, because the
orbital frequency around a Kerr black hole is given by
\cite{bardeen_pt1972}
\begin{equation}
 \Omega_\mathrm{K} = \frac{\sqrt{M_\mathrm{BH}}}{r^{3/2} + \chi
  M_\mathrm{BH}^{3/2}} ,
\end{equation}
and thus $P_\mathrm{td}$ increases as $\chi$ increases.

The opening angle of the dynamical ejecta in the meridional plane,
$\theta_\mathrm{ej}$, is determined by the ratio of the velocity
perpendicular to the orbital plane, $v_\perp$, to that in the equatorial
direction, $v_\parallel$, as $\theta_\mathrm{ej} \approx \arctan (
v_\perp / v_\parallel ) \approx v_\perp / v_\parallel$. This value
should be given by the ratio of the neutron-star radius perpendicular to
the orbital plane to the tidal disruption radius, $r_\mathrm{td}$ [see
Eq.~\eqref{eq:msradius}]. Thus, the dependence of $\theta_\mathrm{ej}$
on the equation of state will be weak again, because both $v_\parallel$
and $v_\perp$ should scale linearly with $R_\mathrm{NS}$. Dependence on
the mass ratio is expected to be $\theta_\mathrm{ej} \propto Q^{-1/3}$,
inherited from $r_\mathrm{td}$, but the expected change is only 25\%
between $Q=3$ and 7. The spin will not modify the value of
$\theta_\mathrm{ej}$.

\end{document}